\documentclass[11pt,a4paper]{article}

\usepackage[utf8]{inputenc}
\usepackage{graphicx}
\usepackage{caption} 
\usepackage{subcaption} % for subfigures
\usepackage[colorlinks]{hyperref} % for hyperlinks (to section, citations, etc.)
\usepackage{booktabs} % for better tables
\usepackage{multirow} % for merging rows
\usepackage{tabularx} 
\usepackage[margin=2cm]{geometry}
\usepackage{authblk} % for author affiliations
\usepackage{physics} % for some physics notation (differentials, etc.)
\usepackage{amsmath} % for math
\usepackage{amssymb} % for math
\usepackage{siunitx} % for scientific notation
\usepackage[usenames,dvipsnames]{xcolor} % for coloring text
\usepackage{adjustbox} % for rotating tables
\usepackage{bbm} % for indicator function
\usepackage[sorting=none,backend=bibtex]{biblatex}
\newcommand{\sref}[1]{Sec.~\ref{#1}}
\newcommand{\fref}[1]{Fig.~\ref{#1}}

\newcommand{\tref}[1]{Tab.~\ref{#1}}
\newcommand{\eref}[1]{Eq.~\ref{#1}}

\makeatletter
\renewcommand\paragraph{\@startsection{paragraph}{4}{\z@}%
 {-2.5ex\@plus -1ex \@minus -.25ex}%
 {1.25ex \@plus .25ex}%
 {\normalfont\normalsize\bfseries}}
\makeatother
\title{Ecological Data Reveal Imbalances in Collision Avoidance Due to Groups' Social Interaction}

\author[1]{Adrien Gregorj}
\author[1,2]{Zeynep Y\"ucel}
\author[1,2,3]{Francesco Zanlugo}
\author[2,4]{Takayuki Kanda}
\affil[1]{Okayama University, Okayama, Japan}
\affil[2]{Advanced Telecommunication Research Institute International, Kyoto, Japan}
\affil[3]{Osaka International Professional University, Osaka, Japan}
\affil[4]{Kyoto University, Kyoto, Japan}
\date{\today}

\begin{document}
\maketitle

\begin{abstract}
    The relative dynamics in collision avoidance between individual pedestrians and dyads has been recently studied, and it was shown that individuals may intrude and disrupt those dyads that are not socially interacting, but not the ones that are engaged in interaction. Building on this, our current study examines how much each party contributes to collision avoidance in the absolute sense, i.e.\ by measuring deviations from their intended paths.

    Our findings suggest that individuals prioritise trajectory efficiency in undisturbed situations, but prioritise safety when encountering dyads, by deviating more from their intended path. Not socially interacting dyads present a similar behavior, although their trajectories appear to be even more efficient than those of individuals in undisturbed situations, and their deviations during encounters less pronounced. On the other hand, socially interacting dyads are not very efficient in undisturbed situations, and their behavior is mostly unaffected by encounters.

    These results strongly suggest that group dynamics affects in two ways the behavior of pedestrians, namely it has a dynamical and a social effect. The dynamical effect, i.e.\ the necessity to keep a spatial vicinity to one's partner, stabilises their trajectory, while the social one, by redirecting one's attention to the partner, decreases the ability to focus on the external environment, and thus leads to reduced efficiency and safety.

    Another finding concerns the tendency of individuals to avoid in a more prominent way the interacting dyads as compared to non-interacting ones. This suggests that individuals not only observe their surroundings to anticipate the future paths of others, but may also assess others' contribution to collision avoidance.

    An impact parameter analysis reveals that collision risk influences path deviations in pedestrian encounters. For individuals, larger behavioral differences between low and high interaction levels of the dyad occur both when the collision risk is high and during less critical encounters. For dyads, the deviation differences between low and high interaction levels are most pronounced when the individual is on course to pass close to the dyad.

\end{abstract}
\section{Introduction}
\label{sec:introduction}

%\subsection{
%Research on human walking motion:
%A brief overview 
%}

Human walking motion has been a research subject in various disciplines, each with distinct scopes and objectives.
Crowd dynamics primarily studies collective human movement~\cite{corbetta2023physics}, often with the goal of ensuring safety in large-scale event management (e.g.\ Olympic games, Hajj)~\cite{baqui2019pedpiv}, prevention of crowd accidents (e.g.\ during festivals)~\cite{vanumu2020human}, and evacuation planning (e.g.\ during fire, terror attacks etc.)~\cite{ronchi2021developing}, among others.
On the other hand, architecture and urban design extend this focus to everyday life settings~\cite{gehl2001life}, prioritizing safety and comfort. This includes architectural designs for enhancing business in malls~\cite{goss1993magic}, landscape designs~\cite{francis2012creating,braham2019spatial} for recreational spaces, and urban infrastructure planning for efficient travel and minimal detours~\cite{calthorpe1993next}.
In addition, robotics directs its attention towards individuals~\cite{bartneck2020human} with the objective of achieving natural and comfortable motion at the actuator and/or navigation levels.
Applications range from assisting human limbs~\cite{zhang2017human} to seamless integration into human crowds~\cite{zhou2022human} for services such as museum guiding or companionship~\cite{shiomi2014towards}.
Furthermore, kinesiology examines motion from a biomechanical standpoint, aiming to explain the fundamentals of motor control~\cite{patla2006any}. In contrast, cognitive science predominantly explores the neurophysiological perspective, with a specific focus on such aspects as perception, visuo-motor coordination, and embodied cognition~\cite{wilson2013embodied}.
%Apart from these, human walking motion has been studied in many other disciplines such as rehabilitation, traffic engineering~\cite{kothari2021human}, psychology~\cite{reicher2001psychology}, epidemiology~\cite{sajjadi2021social}, ethology, computer vision~\cite{solera2015socially}, ergonomics etc.
%\subsection{
%The scenario studied in this study
%}
Despite its brevity,
% and limited focus on a small number of representative domains and scenarios, 
we believe that the above overview already provides a glimpse into the multifaceted nature of research on human walking motion and the vast variety of scopes, objectives and methodologies.

In this study, we choose to view human walking motion in its simplest form, namely as the most fundamental means of transportation~\cite{fruin1992designing} and as an essential activity for an independent life and daily tasks (e.g.\ trips to the market, the bank etc.).
% the below does not fit well with the flow
In such settings, individuals maneuver among fellow pedestrians and obstacles with a focus on safety and fluidity, a concept known as collision avoidance. These localised interactions are thought to give rise to self-organization within a local-to-global framework~\cite{rio2018local}.
For capturing the properties of this sort of navigation behaviour, a large variety of computational models have been proposed~\cite{lohner2010modeling}, where the early models were inspired by physics and used repulsive forces to reproduce collision avoidance~\cite{helbing1995social}. While these models have been successful in generating coherent patterns at the collective level and have greatly contributed to our understanding of pedestrian motion~\cite{helbing1995social}, they often fall short in accurately capturing the realistic attributes of human trajectories~\cite{rio2018local}.
% too strong remark?

In that respect, this study will focus on human walking behavior in an ordinary public space aiming to unveil nuanced patterns in decision-making and path selection. Specifically, we will focus on collision avoidance between individual pedestrians and social groups of two people. By dissecting such encounters according to the group's level of engagement in social interaction, we aim to illustrate the emergence of an intriguing pattern, namely a degree of involvement in collision avoidance and possibly an understanding, by participants not absorbed in social interaction, of the others' involvement.

In a previous study, we studied dyad-individual collision avoidance in a relative sense —specifically, in terms of how closely they approach each other during collision avoidance~\cite{gregorj2023social}— and demonstrated that individuals tend to intrude and disrupt a dyad when its members are not socially interacting, but they refrain from doing so when the dyad is socially interacting or appears to have a strong social bond (e.g.\ couples). Additionally, we have shown that the collision avoidance dynamics between the dyad and individual can be modeled using an ``interaction potential'', obtained through an analogy with the two-body scattering problem, increasing rapidly as the distance decreases, and depending on the level of interaction or the nature of the social relation of the dyad.

In the current study, we aim to build upon our previous research by examining individual-dyad collision avoidance in an absolute sense, i.e.\ quantifying each peer's contribution to collision avoidance based on their deviations from intended paths. Firstly, we will illustrate, using ecological data, that social interaction within groups leads to less efficient path adjustments. Specifically, when groups move without other pedestrians nearby, their deviations from a straight path increase with higher levels of interaction, resulting in longer and less economical trajectories. Conversely, when groups encounter individuals, their deviations decrease with higher interaction levels, indicating reduced responsiveness in collision avoidance.

On the other hand, individual pedestrians clearly modify their behavior in presence of dyads. Additionally, individuals show sensitivity to the level of interaction, albeit not significantly so.
This result, when combined with the statistically significant result reported in~\cite{gregorj2023social}, namely that relative collision avoidance is stronger between individuals and socially interacting dyads with respect to the one between
individuals and not socially interacting dyads, suggests that pedestrians anticipate the diminished involvement of groups in collision avoidance and adjust their deviations from intended paths.

Additionally, we will investigate the role of collision risk in shaping path deviations during pedestrian encounters. The concept of impact parameter, borrowed from physics and used in~\cite{gregorj2023social}, will be employed to quantify the collision risk between individuals and dyads. We will show that collision risk influences path deviations in pedestrian encounters, with larger behavioral differences between low and high interaction levels of the dyad occurring both when the collision risk is high and during less critical encounters. The particularly straight trajectories of individuals encountering non-interacting dyads echoes the findings of~\cite{gregorj2023social}, suggesting that are more susceptible to pass through such dyads. We will also demonstrate that the deviation differences between low and high interaction levels are most pronounced when the individual is on course to pass close to the dyad.

Furthermore, a comparison between the behavior of individuals,  dyads with and without social interaction shows that group dynamics presents both a dynamical and a social effect, suggesting that this aspect should be introduced also in crowd models to properly asses the presence of groups and their effect on overall dynamics.

The adaptive behavior observed suggests a nuanced understanding of social cues and an ability to adjust behavior accordingly. Moreover, it sheds light on the complexity of human interaction within crowded environments, where individuals must constantly explore path choices/adjustments and negotiate space alongside others. Understanding these dynamics not only enhances our comprehension of pedestrian behavior but also has implications for urban planning, crowd management, and the design of intelligent systems aimed at facilitating smooth and safe movement in public spaces.

In the following section, we will frame our work and the scenario in focus and explain related literature concerning the factors relevant for shaping human motion in those settings.

\section{Background}
\label{sec:Background}

%\subsection{
%Community ambulation and the extrinsic factors acting on human walking motion
%}

The concept of ``community ambulation'' as described in the literature refers to individuals' ability to move independently in public spaces~\cite{bhojwani2022impact}\footnote{A more precise definition given in~\cite{shumway2002environmental} specifies that community ambulation entails walking a defined distance (e.g., 800~m) and navigating stairs without assistance. }.
Community ambulation, by definition, necessitates the capacity to integrate walking with a diverse range of demands arising from the dynamic nature of the surrounding public environment. Patla et al.\ precisely identify factors contributing to this integration as walking distance and speed, ambient conditions, physical load, terrain variations, postural transitions, traffic density, and attentional demands~\cite{patla1999dimensions}.
% (e.g.\ coping with distractors or dual-tasking). 

In order to meet these demands, individuals must constantly navigate decision-making processes while engaged in ongoing activities~\cite{cisek2014challenges}. These embodied decisions necessitate rapid and continuous processing of multi-modal sensory information~\cite{gibson2014ecological} as well as evaluating all possibilities in parallel, culminating in the execution of the anticipated optimal choice. Although there is ongoing debate regarding the exact mechanisms underlying these decision-making processes~\cite{busemeyer1993decision,cisek2007cortical}, the complexity of this cognitive process is widely accepted.
%for explaining embodied decisons, there exist various accounts such as the cognitive and ecological perspectives~\cite{busemeyer1993decision},cisek2007cortical}.
%\subsection{
%Our data set and the extrinsic factors of community ambulation in such environment
%}

We consider a typical urban setting to be characterised by low to medium density, plain geometry, and even terrain. Such an environment is often populated by individuals from diverse backgrounds, spanning various ages, occupations, and other demographic factors. In this study, as an adequate example of urban environment, we use the DIAMOR pedestrian trajectory data set~\cite{zanlungo2014potential} (see~\sref{sec:data set} for details).
The use of ecological data is essential in our study, as it captures real-world complexities and nuances inherent in human behaviour within urban environments. Unlike simulated or laboratory-based data, ecological data reflect the genuine interactions and responses of individuals navigating through authentic urban settings. By leveraging such data, we hope to better understand the intricacies of pedestrian dynamics, including the influence of social interactions, and individual behaviours.

Even though there is an inherent lack of control over the factors influencing navigation choices of pedestrians in this naturalistic environment~\cite{mobbs2021promises}, we can assess extrinsic factors in accordance with Patla and Shumway-Cook's guidelines~\cite{patla1999dimensions}. Regarding walking speed, the environment imposes no significant constraints, as evidenced by empirical velocity distributions that align with ranges observed in other ecological studies on community ambulation~\cite{zanlungo2014potential,costa2010interpersonal} (see \sref{sec:metadata}). Analyzing a sufficiently large sample size allows for the proper representation of interpersonal variations. Regarding postural transitions,  the absence of curves, slopes, or stairs in the environment indicates minimal need for significant adjustments, a conclusion further supported by video recordings. In terms of physical load, analysis of the same data reveals that pedestrians predominantly carry either no load or a seemingly light handbag or backpack, typical for urban travel\footnote{Human annotations of video data show that pedestrians mainly consist of students, workers, or shoppers commuting to schools, workplaces, or commercial centers.}.

Furthermore, the terrain and ambient conditions within the data set remain constant, whereas density undergoes minute changes (see \sref{sec:metadata} and \sref{sec:data preparation}). Regarding distance, as detailed in \sref{sec:data preparation}, specific preprocessing steps are employed to ensure the comparability of path lengths across pedestrians\footnote{However, we cannot comment on how long the pedestrians might have walked before entering our observation space.}. This leaves us with the last extrinsic factor of community ambulation, namely, attentional demands, which may actually be quite challenging to assess.

While navigating in public environments, pedestrians need to allocate attentional resources for the observation of their surroundings to identify potential hazards (e.g.\ stairs, static obstacles) and remain vigilant to changes~\cite{cutting1995we}. Nevertheless, the underlying principles of anticipatory locomotor adaptations used to circumvent such challenges is not completely understood~\cite{gerin2005negotiation}. Shumway-Cook et al.\ identify three key factors influencing attentional demands in community ambulation: familiarity with the trip location, environmental distractions, and the presence or absence of travel companions~\cite{shumway2002environmental}. In our case, information regarding pedestrians' familiarity with the environment is unavailable. Nevertheless, given the straightforward nature of the corridor shown in~\fref{fig:occupancy_grid_diamor}, it is unlikely that familiarity with such an environment would significantly affect route finding. Additionally, with no visual or auditory distractions such as shops or music present, environmental distractors are expected to be minimal and consistent across all observed pedestrians. This allows us to isolate the effect of the \textit{presence or absence of travel companions} and analyze its specific impact.

The distinction between physical crowds, comprised of individuals in close spatial proximity but lacking a shared psychological identity, and psychological crowds, characterised by a collective sense of identity or purpose among its members, has been introduced in~\cite{templeton2015mindless,vonsivers2016modelling}. Pedestrians observed in the DIAMOR data set are likely to belong to the former category, as they do not appear to have a collective identity. Nevertheless, our data set  contains instances of small social \textit{groups} and \textit{individuals}. Here, by groups we refer to 2 or more pedestrians travelling together towards a shared destination and engaged in a social relationship~\cite{mcphail1982using, bugental2000acquisition}. Conversely, pedestrians who are not part of a group are termed individuals\footnote{It is important to acknowledge that while members of a group are individuals, in the context of this research, their group affiliation is considered to significantly influence their navigation and, thus, they are identified by this aspect of their social dynamics throughout the study.}. Additionally, we use the term \textit{interaction}, in a specific context, referring exclusively to social interaction characterised by verbal communication, gestures, gaze, physical contact, etc., among members of a group. In this respect, we emphasise that we strictly separate social interaction from collision avoidance.

Note that while individuals need to allocate a certain portion of their attentional resources for the inspection of their surroundings and collision avoidance, they do not need to allocate any resources for coordinating their motion with a partner or for social interaction. Consequently, we posit that cognitive load is likely to be lower for individuals, as well as relatively consistent across different individuals.

On the other hand, groups are likely to experience higher mental workload compared to individuals, since they need to invest effort in managing group's internal dynamics as well as social interaction. Concerning the former, group members need to maintain a cohesive pace and orientation with their partners while ensuring group consistency~\cite{zanlungo2015spatial,yucel2013deciphering}. In addition, if they also carry out social interaction, they need to ensure its smoothness through practices like mutual gaze (on a common target or as eye contact), facial expressions, hand gestures, head/body pose, back-channeling etc.  Furthermore, the level of effort may vary depending on factors such as group size, hierarchy intricacy, and the level of group members' engagement in social interaction~\cite{yucel2018modeling}. Therefore, we cannot simply assume comparable attentional demands for all groups.

At this point, in order to dissociate the attentional demands of groups from their size and hierarchy, we choose to focus on two-people groups, known as dyads. We argue that this approach is not oversimplified, as the majority of groups in crowds  consist of two people~\cite{schultz2014group} and larger groups are shown to break down into sub-groups of two or three people, making dyads a fundamental building block of crowds~\cite{costa2010interpersonal,zanlungo2013walking}. In addition, by breaking down these dyads into sub-categories according to their level of involvement in social interaction~\cite{yucel2018modeling}, we can also address the non-uniformity of attentional demands and approximate the gradation of mental workload.

%\subsection{
%Allocation of attention in community ambulation
%}

Regardless of whether they travel as an individual or as part of a group, to ensure safe community ambulation, humans need to attentively monitor the continuously changing environment. To that end, among the five sensory information channels, the visual one emerges as being the most important.
% followed by auditory (and to a certain extent by tactile) senses. 
% Nevertheless, humans' attentional resources are limited and shared by all sensory information channels~\cite{gerin2005negotiation}, which implies that paying attention to one channel competes for available attentional resources with others and dual-tasking or distraction may lead to lower walking speeds and larger clearances\footnote{Due to the fact that the cognitive demand increases with the additional task, and the decrease in the amount of available resources to plan collision avoidance is argues to be balanced by devoting a larger space to make trajectory adjustments.}, or falls~\cite{gerin2005negotiation, murakami2021mutual,beauchet2009stops}. 
Generally speaking, humans have high quality perceptual access to the world~\cite{gibson2014ecological}, although not flawless~\cite{marr2010vision}. In community ambulation, such visual information is processed online~\cite{patla2006any} under the influence of visual conspicuity~\cite{itti2000saliency, gibson2014ecological} and task demands~\cite{torralba2006contextual}.
% indiv visual
The principal role of visual channel in locomotion is providing an understanding of location of the self, the goal, and the environment (e.g.\ dimensions, terrain features etc.), which are essential in the control of adaptive locomotion~\cite{patla1997understanding} as demonstrated by studies contrasting open loop obstacle avoidance to full visual sampling~\cite{patla2006any}.

Studies~\cite{patla2007gaze,saeedpour-parizi2021target} have demonstrated pedestrians' reliance on visual cues for efficient path planning around static obstacles. This process entails utilizing visual information to locate the target destination and assess the surrounding path area. Additionally, pedestrians evaluate the necessary magnitude of the deviation to navigate obstacles effectively, ensuring a seamless progression towards their intended destination.

In addition to static obstacles, pedestrians must also be aware of moving targets (e.g.\ other pedestrians, bicycles, cars etc.) and anticipate potential collisions, which present greater challenges compared to stationary obstacles due to their momentum and unpredictable motion~\cite{cinelli2007travel}. In this study, we focus on human-human collision avoidance and thus consider as moving obstacles the other pedestrians in the environment.

Previous research on gaze analysis of pedestrians encountering other moving people has revealed several noteworthy findings. It was observed that gaze fixations occur at consistent frequencies regardless of increases in pedestrian density, albeit with a narrower scanning range of the street~\cite{berton2020eye-gaze}. This suggests that pedestrians tend to focus more on people directly in front of them, particularly those in closer proximity. Moreover, pedestrians in the central and right positions were fixated at greater distances compared to those on the left, indicating a modulation of gaze behaviour based on the location and direction of pedestrians in the community environment~\cite{joshi2021gaze}. Additionally, studies have found that during encounters, pedestrians tend to focus around the chest of the oncoming pedestrian~\cite{arai2017analysis}, and initially scanning various parts of their body, providing evidence that body motion cues serve as a significant source of visual information during such situations~\cite{kitazawa2010pedestrian,murakami2022spontaneous}. Hessels et al.\ also concluded that walkers tend to look at different body parts based on the behaviour of the other pedestrian, with more attention directed towards individuals who address or direct themselves towards them, with the exception of pairs engaged in conversation, who are likely to be looked at despite not being directed towards the observer~\cite{hessels2020task}.

%~\cite{kitazawa2010pedestrian,joshi2021gaze,berton2020eye-gaze}

Pedestrians use visual information not only for anticipating others' path, but also for estimating certain personal features~\cite{dabbs1975beauty,van2009exploiting}. Some person-specific information which can be observed by visual inspection (even without detailed scrutiny) involve height, gender, and group relation.
Uninstructed pedestrians circumventing around standing people are found to maintain a greater distance from males than females, from groups than individuals, and from an attractive person than an unattractive one~\cite{dabbs1975beauty}. In addition, role dependent strategies in collision avoidance (i.e.\ who passes first and who gives way) are shown to depend on gender and height~\cite{van2009exploiting}\footnote{Knorr et al.\ failed to observe any significance on gender and height, possibly due to the lack of cognitive load and oblique (i.e.\ not frontal) crossing situation ~\cite{knorr2016influence}.}. Other person-specific information which cannot be directly observed but which can be estimated visually involve age, personality and mood. Interestingly, it has been shown that people can estimate such traits as domination, boldness, easygoingness, happiness, youthfulness with quite high agreement, even by watching only walkers' point light displays~\cite{montepare1988impressions, johansson1975visual}. To the best of our knowledge, no study has explored whether pedestrians perceive social interactions within the groups they encounter. However, given our findings, we believe it is probable that they do notice such interactions.

In addition, eye gaze is shown to serve not only for  collecting  information from one's surroundings, but also for delivering information about one's intended path to others~\cite{hessels2020task, nummenmaa2009ll}.
In particular, Nummenmaa et al.~\cite{nummenmaa2009ll} interestingly shed light on the communicative efficacy of ``gaze aversion'' and demonstrated  that diverting gaze to one side can effectively signal a planned path to others, suggesting steering in that direction to avoid collision, thereby prompting the other pedestrian to maneuver in the opposite direction.
Generally speaking, studies on the Theory of Mind Model (ToMM) suggest that eye gaze plays a crucial role in understanding others' intentions~\cite{baron1985does}, and the absence of gaze alternation can result in a failure to anticipate others' attentional focus and intentions in static scenarios, referred to as ``mind blindness''~\cite{baron1997mindblindness}.
In addition to gaze direction, head orientation, known to indicate focus of attention~\cite{yucel2013joint}, serves as a cue for path selection, while body orientation is associated with a ``potential to move''~\cite{zhou2022human}.
Yet, such findings imply that the brain, responsible for orchestrating swift and successful behavior in such settings, relies not only on spontaneous sensory inputs but also on internal representations of surrounding pedestrians to fulfill this role~\cite{wilson2013embodied}.

On the other hand, the auditory channel can be considered as the second most important sensory channel during community ambulation. %Although not likely to provide as much information as the visual one, 
Although dual-task studies suggest that environmental sounds (e.g.\ construction noise, music etc.) are likely to be distractive rather than supportive~\cite{bhojwani2022impact}, holistically speaking, the audio channel plays a significant role in shaping the overall perception of the surroundings. Specifically, task load on the auditory channel has been shown to interfere with locomotion, leading to lower walking speed or collisions, and a preference for maintaining a larger personal space~\cite{shumway2002environmental,gerin2005negotiation}\footnote{The concept of personal space can be interpreted as a zone pedestrians maintain for themselves and others, ensuring a comfortable distance~\cite{hall1963system}. More recently, researchers have proposed an alternative interpretation of personal space as a buffer zone that enables people to perceive risks and plan trajectory adaptations~\cite{gerin2005negotiation,fruin1971designing}. Some recent works advocate for a combined approach, incorporating both a social influence field and a collision avoidance field~\cite{zhou2022human}}.

Furthermore, in dual-task scenarios, individuals tend to move at lower speeds, and the increased workload on the auditory channel may lead to changes in visual scanning patterns. Specifically, eye gaze is seen to be directed more frequently towards pedestrians posing a higher risk of collision, and fixations are  observed more commonly on upper body segments, possibly since the upper body is more informative regarding future walking directions~\cite{bhojwani2022impact,zhou2022human}.

In addition to securing a certain degree of awareness about the environment, group members may also need to observe their partners for ensuring group cohesion and possibly for orchestrating social interactions. This monitoring process often involves utilizing primarily visual and auditory channels, akin to a dual-task mobility scenario.
% Social interaction therefore is likely to consume a significant part of attentional resources, since it requires visual information (e.g.\ keeping the partner in visual field or making a gaze contact with him/her,) as well as audio (e.g.\ conversation) and tactile information (i.e.\ physical contact). 
Specifically, social interaction in a group is likely to require visual resources for tracking focus of attention (e.g.\ gaze on partner or a mutual gaze on a target) and elicitation of emotions (e.g.\ facial expression, gestures) etc.~\cite{kendon1990conducting}. In addition, audio channel is essential for comprehension of speech, detection of adjustments on pitch, loudness, intonation etc. of the interlocutor, as well as turn management (signaling turn taking, holding, giving or skipping), communicating feedback (e.g.\ agreement, surprise etc.) or acknowledgement (i.e.\ back-channeling).
Note that since humans have limited attentional resources and group members may already need to devote part of those to the afore-mentioned commitments towards their partners, they are likely to be left with less resources for navigation planning than individuals\footnote{One may also expect them to require a larger personal space to make the trajectory adjustments necessary for collision avoidance~\cite{gerin2005negotiation}.}.
Intuitively, assessing the likelihood of collision with moving obstacles, such as other pedestrians, is more challenging compared to navigating around stationary obstacles~\cite{cutting1995we} and requires an anticipation of others' trajectory.
%To anticipate such risks and react safely, pedestrians continuously collect information from other surrounding pedestrians, and similarly from static or other moving obtacles. 
At this point, it is highly likely that humans consider other pedestrians around them not simply as \textit{obstacles that can move}, but as humans like themselves~\cite{baron1985does}. Therefore, they possibly also monitor and assess whether others are engaged in the same visual exploration and this meta-awareness can play a crucial role in their own planning and, in turn, visual sampling.

The current study aims to contribute beyond existing research by showcasing how social interaction affects the ability of pedestrians to focus on the surrounding environment, with possible effects on efficiency and safety. Furthermore, it attempts to demonstrate that humans possess awareness not only of others' planned paths inferred from subtle actions, but also of the absence of such indications. Specifically, we believe individuals are capable of recognizing the lack of these subtle communicative actions or understanding that they are not directed towards an implication of a planned path (but, for instance, addressed at a travel companion accompanying their social interaction). In such cases, they proactively anticipate a limited contribution to collision avoidance from those pedestrians and take compensatory measures on their part.

\section{Methods}

\subsection{Data set}
\label{sec:data set}

The data set utilised in this study is the DIAMOR data set~\cite{zanlungo2014potential}, which was previously employed in ~\cite{glas2014automatic,yucel2018modeling} for the purpose of group recognition and pedestrian dynamics modeling.
The recordings were made in an underground pedestrian street network located in a commercial district of Osaka, Japan. Specifically, the data set comprises recordings from two straight corridors within this street network, and our focus centers on one of these. With several train stations, business centers, and shopping malls accessible from the recording location, there is a diversity in pedestrian profiles. The recording area is roughly 200~m$^2$ and allows continuous tracking along approximately 50~m and the recording spans eight hours in a weekday.

We would like to highlight that this data set is collected from uninstructed pedestrians in their ecological environment, which we consider a valuable asset for observing naturalistic behaviour\footnote{Experimentation has been reviewed and approved by ATR ethics board with document number 10-502-1. Posters explaining that an experiment concerning pedestrian tracking was being held were present in the environment. The data are publicly available~\cite{Diamor_dataset} and contain anonymous trajectories derived from range data~\cite{glas2009laser}.}.
Notably, studies involving non-human animals have revealed intriguing distinctions and qualitative differences in behaviour between constrained tasks and uncontrolled settings~\cite{rosenberg2021mice,calhoun2019unsupervised,stringer2019spontaneous}. Similarly, in human studies, the phenomenon of modifying of one's behaviour in response to the awareness of being observed has even been given a name, the ``Hawthorne effect''\footnote{Although it is now largely agreed that such effect was less significant than originally thought in the scenario from which it takes its name.}~\cite{sedgwick2015understanding,mccambridge2014systematic}. This effect has been observed in various settings, including when assessing the quality of care provided by trained practitioners~\cite{leonard2010using}, or the energy awareness of consumers~\cite{schwartz2013hawthorne}. In the specific context of human locomotion, Farhan et al. have shown that observed participants exhibit lower variability in gait parameters~\cite{farhan2023variability}, and Friesen et al. have demonstrated that locomotion parameters (e.g.\ speed, step length) were impacted by the number of researchers present in the room~\cite{friesen2020all,farhan2023variability}. In this respect, the ecological data studied in the upcoming sections are considered to be largely devoid of experimental or behavioural bias, or subconscious alterations in behaviour, or at least minimally affected by such factors
\footnote{
    Note that with the above arguments, we by no means intend to assert that the outcomes derived from traditional, meticulously controlled experimental paradigms are inaccurate or invalid. These approaches, proficient at dissecting intricate behaviours into their individual components, have significantly contributed to our understanding of the fundamental processes that govern behaviour. Nonetheless, their reductionist approach is likely to constrain their capacity to elucidate naturalistic behaviour comprehensively, since pristine experiences are more of an exception than the norm in real-world settings and it is ideal to study humans in naturalistic settings for ultimately explaining their real-world behaviour~\cite{mobbs2021promises}.
}.
%in virtual ecologies and controlled settings, realism is questionable
%not possible to deploy such sensors due to practical, economical or ethical reasons, and thus our data obtained with only environmental sensors and does not involve the location or orientation of limbs, eye gaze etc.

\begin{figure}[htb]
    \centering
    \includegraphics[width=0.8\textwidth]{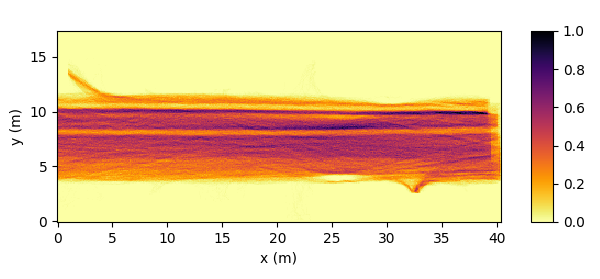}
    \caption{The normalised cumulative density map for the DIAMOR data set. It is obtained by dividing the recording area into a grid of 10~cm $\times$ 10~cm cells and counting the number of pedestrians that have been in each cell at any point in time. The counts are then normalised by dividing by the maximum count in the grid. Darker areas indicate higher pedestrian density.
    }
    \label{fig:occupancy_grid_diamor}
\end{figure}

The data include both depth and video information. The depth information is utilised to derive the trajectories of pedestrians~\cite{glas2009laser}, which can be freely downloaded from~\cite{Diamor_dataset}. As a result of this tracking process, we obtain the normalised cumulative density map shown in \fref{fig:occupancy_grid_diamor} (refer also to~\cite{gregorj2023social}). The map is obtained by dividing the recording area into a grid of 10~cm $\times$ 10~cm cells and counting the number of pedestrians that have been in each cell at any point in time. The counts are then normalised by dividing by the maximum count in the grid. Darker areas indicate higher pedestrian density.

The video data served as the basis for establishing the ground truth regarding dyads and the intensity of interaction. To assess errors arising from coding fatigue and coder bias, each relation (belonging to a group and intensity) was labeled by two different coders. In the first stage of coding, coders observed walking patterns, age, gender, and clothing etc.\ to determine which pedestrians formed a group. In the second stage, focusing solely on pedestrians labeled as dyads in the first stage, coders assessed the intensity of interaction. This reduced the amount of data each coder had to view, thus enhancing coding efficiency.

Coders were asked to label the intensity of interaction for pedestrian groups using a 0-3 subjective scale. To avoid biasing the assessment, only the resolution (i.e.\ the number of interaction levels) was predefined (four levels), with no guidelines provided on what constituted weak, mild, or strong interaction intensity. Instead, coders carried out free-viewing, watching three hours of video footage of groups to intuitively grasp variations in interaction intensity, before the actual coding task began.

The agreement between coders for group relation labeling was evaluated using Cohen's $\kappa$ coefficient, resulting in a high value of $\kappa$ = 0.96, indicating strong agreement~\cite{fleiss2003statistical}. For interaction intensity labeling, the reliability was assessed using Krippendorff's $\alpha$ coefficient, yielding a value of $\alpha$ = 0.67, which is usually considered sufficiently high~\cite{krippendorff2004reliability}.

\subsection{Metadata}
\label{sec:metadata}
\begin{figure}[htb]
    \centering
    \begin{subfigure}[t]{0.45\textwidth}
        \centering
        \includegraphics[width=\textwidth]{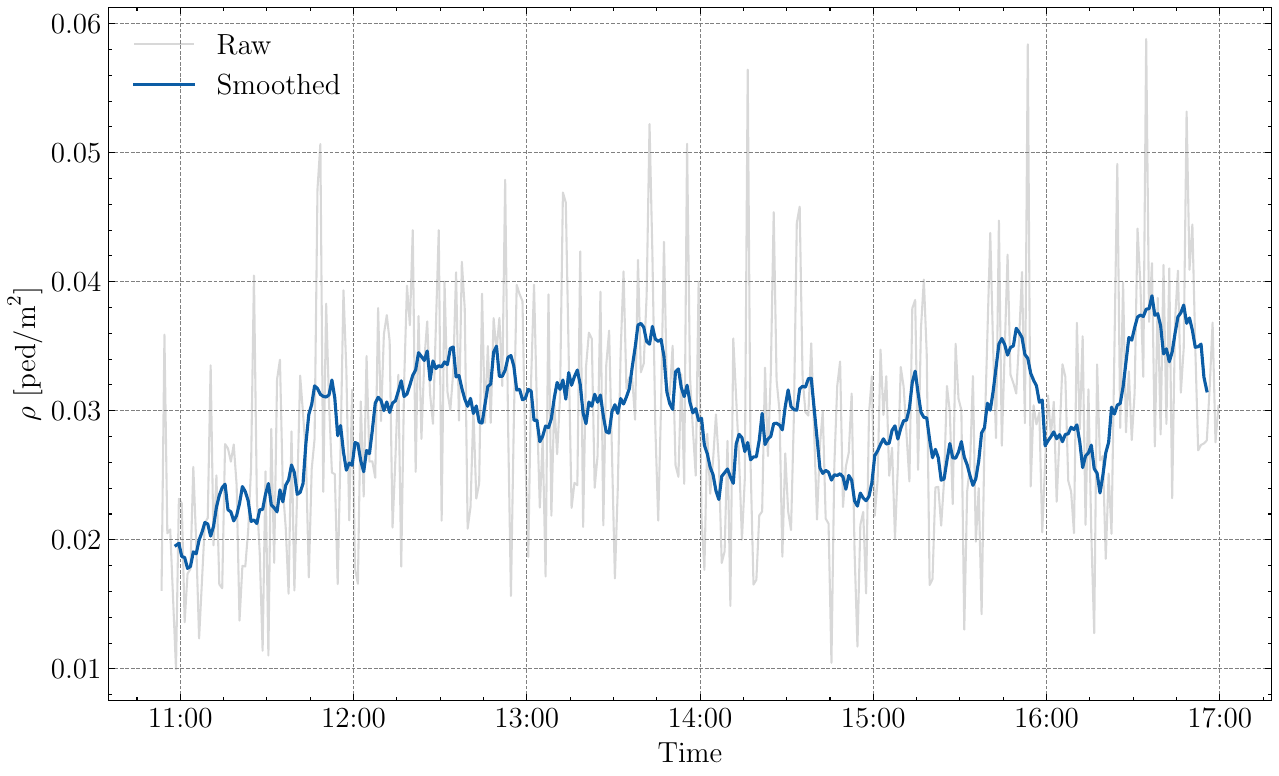}
        \caption{}
        \label{fig:density}
    \end{subfigure}
    \begin{subfigure}[t]{0.45\textwidth}
        \centering
        \includegraphics[width=\textwidth]{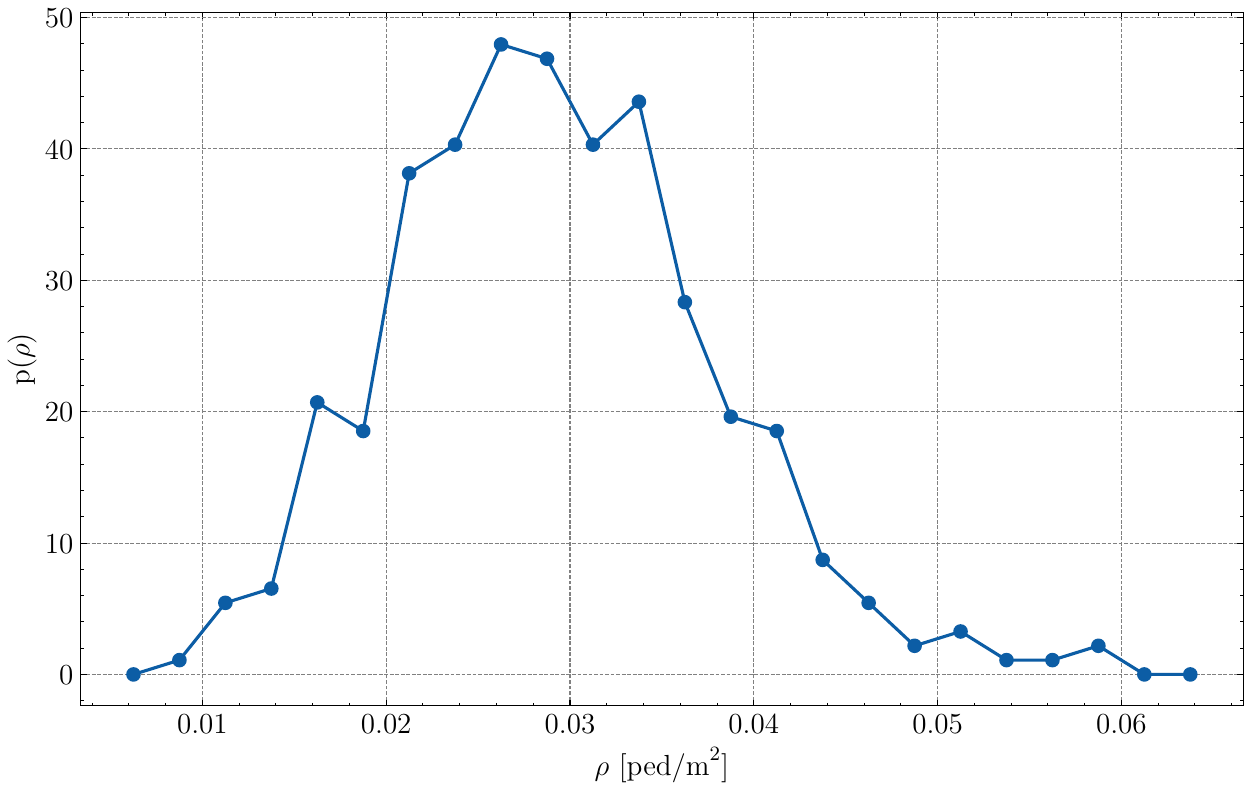}
        \caption{}
        \label{fig:density_histogram}
    \end{subfigure}
    \caption{Density of pedestrians in the DIAMOR data set. (a) Density of pedestrians in the recording area over time. (b) Histogram of the density of pedestrians. Density is calculated as the number of pedestrians in the recording area during a 1~min time window, divided by the area of the recording area.}
    \label{fig:density_all}
\end{figure}

To further clarify the context of the study, we provide some metadata about the studied corridor.
The density of pedestrians in the recording area is shown in \fref{fig:density_all}. We compute it by counting the number of pedestrians in the recording area during 1~min time windows and dividing it by the area of the recording area ($40$~m $\times$ $7$~m $= 280$~m$^2$). The density of pedestrians in the recording area varies between 0 and 0.06 pedestrians/m$^2$, with an average value of 0.04 pedestrians/m$^2$, which can be considered as a low density.

\begin{figure}[htb]
    \centering
    \includegraphics[width=0.6\textwidth]{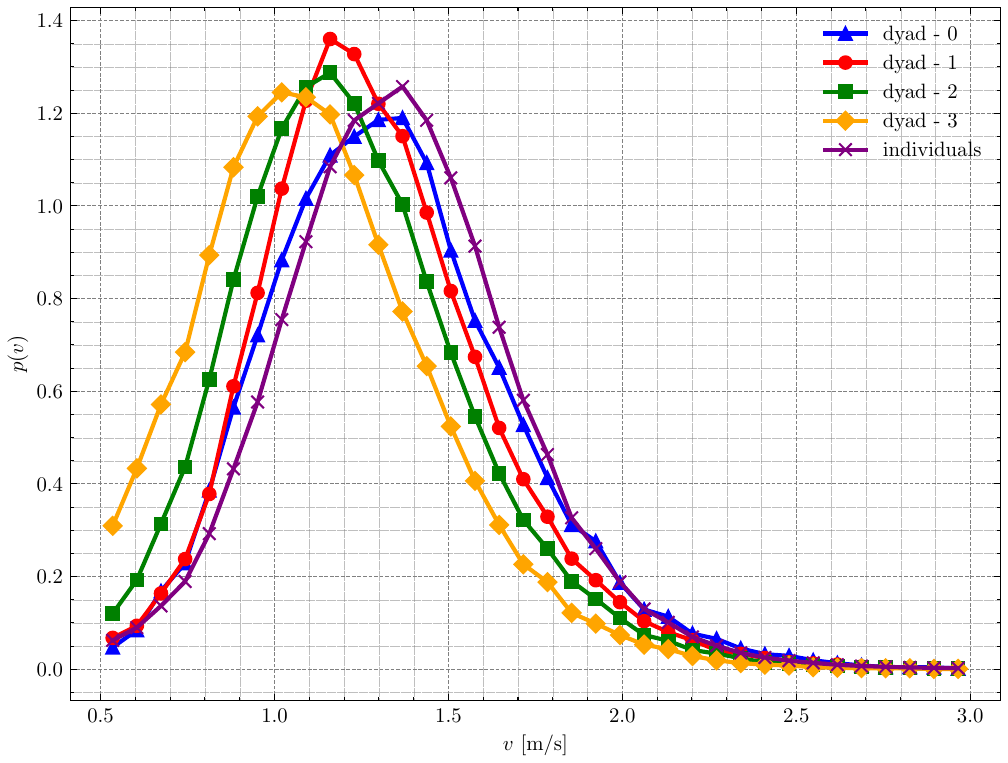}
    \caption{The probability density function of the velocity of individuals and dyads.}
    \label{fig:velocity_pdf}
\end{figure}

The probability density function of the velocity of individuals and dyads in the recording area is shown in \fref{fig:velocity_pdf}. The dyads are categorised based on their level of interaction, as defined in~\sref{sec:data set}. We observe the effect of interaction previously reported in ~\cite{yucel2018modeling}, namely that strongly interacting dyads have a lower average velocity compared to weakly interacting dyads. In addition, we observe that the velocity of non-interacting dyads is comparable to that of individuals.

\subsection{Data preparation}
\label{sec:data preparation}

The raw trajectories in the DIAMOR data set are sampled at a non-uniform rate, with a frequency varying between 20 and 50~Hz~\cite{Diamor_dataset} due to holes in the data caused by occlusions or tracking errors. To ensure that the analysis is not affected by this, we resample the trajectories at a constant rate $f_s$ of 33~Hz with cubic spline interpolation~\cite{knorr2016influence}.

As mentioned in~\sref{sec:introduction}, the focus of this study lies on understanding the dynamics of typical individual-dyad collision avoidance in public settings.
%Specifically, we aim to investigate collision avoidance scenarios among walking pedestrians. 
To achieve this, we choose to keep only the trajectories that align with typical walking speeds in public spaces and disregard those that do not. For establishing the common velocity range of urban pedestrians, we refer to literature on human locomotion and, basing our reasoning on the findings of~\cite{zanlungo2014potential}, we consider the trajectories with an average velocity falling within the range of $[0.5, 3]$~m/sec to belong to typical urban walking motion and the rest to be associated with other states (e.g.\ standing/running or tracking artifacts).

Following this, to address the impact of sensing noise and natural swaying resulting from human gait, we applied filtering to the trajectories. Commonly, low-pass filtering is employed for this purpose~\cite{olivier2013collision, fajen2013guiding, rio2018local}, and in this study we opted for a Savitzky-Golay filter~\cite{savitzky1964smoothing}, which is known to be well-suited for smoothing noisy data. Specifically, we adjusted the polynomial order used to fit the samples to 2 and the length of filter window to 3~s. This decision was guided by the observation that typical gait cycles last between 1 and 2 seconds~\cite{kirtley1985influence,hediyeh2014pedestrian}.

In what follows, we will use the notation $\vb{p}(t)$ to denote the position of a pedestrian at time $t$, and $\vb{v}(t)$ to denote the velocity of the pedestrian at time $t$. We will also use the notation $\vb{p}_i(t)$ and $\vb{v}_i(t)$ to denote the position and velocity of the individual $i$ at time $t$, and $\vb{p}_d(t)$ and $\vb{v}_d(t)$ to denote the position and velocity of the dyad $d$ at time $t$\footnote{Note that when this notation is used, we consider the dyad as a single entity and use the average position and velocity of the dyad members. In other situations, we will consider the dyad members separately to avoid biasing the results with an artificial smoothing caused by averaging.}.

A \textit{trajectory} $T$ is defined as the sequence of positions $\vb{p}(t_k)$ and velocities $\vb{v}(t_k)$ of a pedestrian, where $t_k$ is the time at which the positions are recorded, with $k \in [0, N-1]$ and $N$ being the number of samples.

The velocity $\vb{v}(t_k)$ is derived from the positions using a simple forward Euler difference, i.e.\

\begin{equation}
    \vb{v}(t_k) =
    \begin{cases}
        \frac{\vb{p}(t_{k+1}) - \vb{p}(t_k)}{t_{k+1} - t_k} & \text{if } k < N-1 \\
        \vb{v}(t_{k-1})                                     & \text{if } k = N-1
    \end{cases}
    \label{eq:velocity}
\end{equation}

\begin{equation}
    T = \left[(\vb{p}(t_0), \vb{v}(t_0)), (\vb{p}(t_1), \vb{v}(t_1)), \ldots, (\vb{p}(t_{N-1}), \vb{v}(t_{N-1}))\right]
\end{equation}

\subsection{Intended direction of motion}
\label{sec:method_desired_direction}

Our primary assumption in assessing the trajectory deviation is that pedestrians aim to minimise the distance traveled and will therefore select the straightest path to reach their destination whenever possible. This assumption, notably introduced in~\cite{hoogendoorn2004pedestrian} and adopted by numerous other studies~\cite{bhojwani2022impact, olivier2012minimal}, posits that at the tactical level, where pedestrians make decisions about their desired area and route, they do so by minimizing a cost function. This function takes into account factors such as distance traveled, trajectory comfort, or anticipated encounters with other pedestrians~\cite{hoogendoorn2004pedestrian}. In the context of a straight corridor, such as the one examined in our study, a straight line is reasonably expected to be the optimal route to cross the corridor\footnote{Assuming that the pedestrian is not aiming to exit the corridor through a side passage.}$^,$\footnote{It is worth noting that if the corridor is sufficiently wide, the optimal path may still be straight, but not perfectly aligned with the corridor axis, as pedestrians may cross it diagonally.}.

To compute the intended straight line trajectory, we first need to identify the \textit{intended direction of motion} of the pedestrian. We define it as the line going through $\vb{p}(t_0)$ and guided by $\vb{v_0}$, which is the average velocity vector over a 0.5~s window starting at $t_0$. At the sampling frequency of 33~Hz, this corresponds to $N_e = \lfloor33 \times 0.5\rfloor = 16$ samples and\footnote{Since the velocity vectors are computed using a forward Euler difference, the velocity time at $t_k$ is a vector pointing from $\vb{p}(t_k)$ to $\vb{p}(t_{k+1})$. Therefore, the average velocity vector $\vb{v_0}$ is a vector pointing from $\vb{p}(t_0)$ to $\vb{p}(t_{N_e})$.}
\begin{equation}
    \vb{v_0} = \frac{1}{N_e} \sum_{k=0}^{N_e-1} \vb{v}(t_k).
\end{equation}

We believe that $0.5$~s is a reasonable window size for this purpose, as, since it is approximately the time it takes for a pedestrian to take one step, it is large enough to capture the general direction of motion, while being small enough to avoid capturing the effects of the collision avoidance behavior.

The intended direction of motion $L_0$ is therefore formally defined as
\begin{equation}
    L_0 = \{\vb{p}(t_0) + \lambda \vb{v_0} \mid \lambda \in \mathbb{R}\}.
    \label{eq:desired_direction_of_motion}
\end{equation}

We argue that this line better represents the intended motion of the pedestrian than taking the line going through $\vb{p}(t_0)$ and $\vb{p}(t_{N-1})$, as the latter would be influenced by the trajectory deviations that we aim to quantify, particularly if the pedestrian does not return to its original intended path after the deviation.

In addition, the \textit{straight line trajectory} $T_0$ is defined as the trajectory that the pedestrian would follow, should she maintain her desired motion direction walking at a constant speed. The points of the straight line trajectory are all on the line $L_0$ and verify
\begin{equation}
    \left\{\begin{array}{@{}l@{}}
        \vb{\tilde{p}}(t_k) = \vb{p}(t_0) + \vb{v_0}t_k \\
        \vb{\tilde{v}}(t_k) = \vb{v_0}
    \end{array}\right. \quad \forall k \in [0, N-1]
    \label{eq:straight_line_traj}
\end{equation}

We emphasise that the straight line trajectory, composed of a sequence of $N$ discrete positions and velocities, differs from the desired direction of motion, which represents a line with an infinite number of points.

\subsection{Situations of interest}

Once again, drawing from the terminology introduced in~\cite{hoogendoorn2004pedestrian}, at the operational level -where pedestrians execute their selected route- they deviate from the straight line trajectory, possibly influenced by such factors as gait characteristics~\cite{uetake1992can} and the presence of other pedestrians or obstacles. In this study, since we aim to quantify and compare trajectory deviations of individuals and dyads, magnifying on the impact of groups' social interaction (and the amount of attentional resources available for navigation planning), we examine two types of situations: (1) encounters, where a dyad and an individual are on a frontal collision or close-to-collision course, and (2) undisturbed segments, where neither the individual nor the dyad encounters any other pedestrian in a reasonably large area around them, allowing them to move freely without needing to perform avoidance behaviour. In the following discussion, we begin by providing detailed insights into the undisturbed case, since it serves as a baseline representing the deviation in the absence of specific collision avoidance maneuvers.

\subsubsection{Undisturbed situations}
\label{sec:method_undisturbed}

As explained in \sref{sec:method_desired_direction}, we argue that although pedestrians aim to minimise the traveled distance and will therefore select the straightest path to reach their destination when possible, it is unrealistic to anticipate them to walk on a perfectly straight line, even when there are no other pedestrians present. This meandering~\cite{uetake1992can} can be explained by the natural swaying resulting from the human gait and the impact of factors such as the cognitive load~\cite{ho2019using} and was modeled using a Langevin-like model in~\cite{corbetta2017fluctuations}.
%fluctuations from the preferred path have even been 

In this respect, we wish to obtain a baseline of straightness for both individuals and dyads, when they are not forced to perform collision avoidance behavior. To that end, we define undisturbed segments as the portions of a trajectory where no other pedestrian is located at less than 4~m away~\cite{cinelli2008locomotor,kitazawa2010pedestrian}, the same window size adopted in the computation of trajectory deviation during encounters.

However, although the encounters defined in~\sref{sec:method_encounters} are by definition spatially bounded\footnote{Usually these portions measure 3 to 4~m, depending on the lateral distance and speed of the two parties.}, undisturbed segments can be arbitrarily long\footnote{In less densely populated environments, longer undisturbed segments can be observed.}. To enable comparability of segments from the two scenarios, we must select undisturbed segments of similar lengths to those in the spatially constrained encounter cases. To that end, we extract undisturbed segments of 4~m, ensuring there are no overlaps between them\footnote{In addition, we require that the direction of motion at the beginning and end of the segment are aligned with the horizontal axis. This is ensured by verifying that the absolute value of the angles of the velocity vector in 0.5~s windows at the beginning and end of the segment, wrapped in the range $[-\pi, \pi]$ are less than $\frac{\pi}{8}$ or greater than $\frac{7\pi}{8}$ in more than 90\% of all $N_e$ time steps in these windows. This is to ensure that the pedestrian is not turning at the beginning or end of the segment.}.

According to the above, one individual or one dyad may have multiple undisturbed segments (in particular if they are observed for a long time and if there are few other pedestrians around). Since these segments might not be independent, given that the pedestrians' behavior in one segment may be influenced by the behavior in the previous one, we will consider the average deviation over all undisturbed segments of an individual or dyad as a single data point in the analysis. In \tref{tab:undisturbed_encounter_count}-(a), we show the number of individuals and dyads in undisturbed situations and in \tref{tab:undisturbed_encounter_count}-(b), we show a breakdown of the number of undisturbed dyads according to the intensity of interaction of the dyad.

\begin{table}[!htb]

    \caption{
        (a) Number of individuals and dyads in undisturbed situations and number of encounters.
        (b) Breakdown of the number of undisturbed dyads and encounters according to the intensity of interaction of the dyad.
    }
    \label{tab:undisturbed_encounter_count}
    \begin{center}

        \begin{tabular}{cc}
            (a)
             &
            (b)
            \\
            \begin{tabular}[t]{ccc}
                \toprule
                \multicolumn{2}{c}{Undisturbed} &
                Encounter                                     \\
                Individuals                     & Dyads &
                \\
                \midrule
                1966                            & 457   & 609
                \\
                \bottomrule
            \end{tabular}
             &
            \begin{tabular}[t]{lrr}
                \toprule
                Intensity of interaction & Undisturbed & Encounter \\
                \midrule
                0                        & 18          & 45        \\
                1                        & 60          & 88        \\
                2                        & 299         & 380       \\
                3                        & 80          & 96        \\
                \bottomrule
            \end{tabular}
        \end{tabular}
    \end{center}

\end{table}

\subsubsection{Encounters}
\label{sec:method_encounters}

We define encounters as situations where a dyad $d$ and an individual $i$ are moving towards each other and are on a collision or close-to-collision course. In such scenarios, it is likely that one or both parties will engage in collision avoidance behaviour to ensure a comfortable passage.

Although the superposition assumption, commonly employed in many models (e.g.\cite{helbing1995social}) suggests that the collective effects on a pedestrian from multiple neighbors can be linearly combined~\cite{rio2018local}, there is ongoing debate regarding whether neighborhood is determined by metric or topological distances (e.g.\ degree of neighborhood). In this work, we choose to use metric distance, since the density associated with our data set is not high enough to result in collective behaviour~\cite{ballerini2008interaction}. In that respect, we consider only those dyads and individuals, who approach each other \textit{sufficiently} close. Specifically, sufficiency is assessed by the condition $\exists t \mid d_{di}(t) \leq 4$m, where $t$ represents time and $d_{di}$ denotes the instantaneous distance between the dyad and the individual.

The choice of a 4~m threshold is grounded in prior research on collision avoidance. Cinelli and Patla found that the ``safety zone,'' which is the distance individuals allow a moving object to approach before initiating an avoidance behaviour, averages around 3.73~m~\cite{cinelli2008locomotor}. In addition, G{\'e}rin-Lajoie et al.\ showed that anticipatory locomotor phase starts with an initial path deviation which occurs about 4.5~m from the obstacle~\cite{gerin2005negotiation}. Moreover, Kitazawa et al.\ demonstrated that pedestrians focus their gaze most intensely on approaching individuals when they are, on average, approximately 3.97~m away, seldom directing their attention to pedestrians at greater distances~\cite{kitazawa2010pedestrian}.

Among dyad-individual pairs that approach sufficiently close, we solely consider those engaging in \textit{frontal encounters}, namely moving in opposite directions. There are two main reasons for this selective approach. Firstly, given our focus on collision avoidance inside an environment with predominantly bi-directional flow, we contend that encounters involving pedestrians from opposite flows are more pertinent than those within the same flow. Secondly, since we aim to discern the implications of social interaction level (and in turn speculate on the allocation of attentional demands), only those instances where involved parties can visually examine each other (particularly in which the individual can judge groups relation and assess dyad's level of interaction) are considered~\cite{knorr2016influence,van2009exploiting}. In contrast, non-frontal encounters are omitted, since collision avoidance is less prominent within low-density bi-directional flow settings, and individuals are not likely to react to dyads' characteristics due to limited observation capabilities.

% dyad's level of interaction and will therefore be likely to (e.g., interpersonal distance, speed, etc.) rather than social factors.

To address this, we compute the predominant relative motion direction of $d$ and $i$ at the beginning of the encounter by calculating the cosine of the angle between their velocity vectors,
\begin{equation}
    c_{di}(t) =
    \frac
    {
        \vb{v}_d(t) \cdot \vb{v_i}(t)
    }{
        ||\vb{v}_d(t)||||\vb{v}_i(t)||
    }.
\end{equation}

We classify an encounter as frontal if, during a 0.5~second window starting at $t_0$, the cosine of the angle between the velocity vectors is smaller than $-\cos(\frac{\pi}{8})$ for at least $90\%$ of all $N_e$ time steps (indicating an angle range of $[\frac{7\pi}{8},\frac{9\pi}{8}]$).
This condition is formally expressed as

\begin{equation}
    \frac{1}{N_e} \sum_{k=0}^{N_e-1} \mathbbm{1}_{\{c_{di}(t_k) < -\cos(\frac{\pi}{8})\}} \geq 0.9,
    \label{eq:relative_motion}
\end{equation}
where $\mathbbm{1}$ is the indicator function that is equal to 1 if the condition inside the brackets is true and 0 otherwise.

For further ensuring anticipatory locomotor adjustments during frontal encounters, we calculate extrapolated straight line trajectories (see \eref{eq:straight_line_traj}) at the initial instant of the encounter and require the closest approach distance on such paths to be less than 2~m\footnote{Note that despite starting the encounter at a distance of 4~m, it is possible that the dyad and the individual have sufficient lateral distance to comfortably clear each other, rendering such encounters irrelevant for the scope of this work.}. Specifically, we derive the approach distance using the average velocity of the dyad and the individual over a 0.5~s window (the same time window used for determining the relative motion direction, see also \eref{eq:relative_motion}), employing a method akin to computing the impact parameter for examining the interaction of charged particles.

% remediate to situations where the trajectories during the encounter might not contain enough points, for instance if the trajectories are cut prematurely due to tracking issues. We aim to include 
Finally, we ensure that both the dyad and the individual's behaviour is captured even after they have laterally passed each other. This conditioning is motivated by the findings reported in~\cite{corbetta2018physics}, where the authors noted that, in cases of very close encounters, lateral distance kept increasing even after ensuring avoidance. Hence, we necessitate that both the individual and the dyad clear each other entirely, with adequately long trajectories preceding and following the clearance. To be precise, we specify that they must maintain a distance of at least 3~m apart at the beginning and end of the encounter, demonstrating an initial approach succeeded by subsequent distancing.

After applying the conditions described above, the number of encounters which will be subject to an analysis in the upcoming sections, turn out to be as illustrated in \tref{tab:undisturbed_encounter_count}-(b).

\subsection{Measures of deviation}
\label{sec:method_measures}

In this section, we provide the definitions of a set of measures for quantifying the deviation of an actual trajectory from an intended straight line trajectory (or, equivalently, its dissimilarity to such path) together with a discussion on their specifications. While most of the measures are compiled from existing literature on collision avoidance and trajectory clustering~\cite{hu2023spatio, besse2016review}, we also introduced several original measures to capture various aspects of deviation. Additionally, while some measures from the literature are directly applied to our data, others are adapted to suit our specific case.

When computing the deviation of a dyad, we consider the deviation of both members separately. This ensures that the deviation of the dyad is not artificially reduced by averaging the positions of the two members.

\subsubsection{Position based measures}

\begin{figure}[htb]
    \centering
    \includegraphics[width=0.8\textwidth]{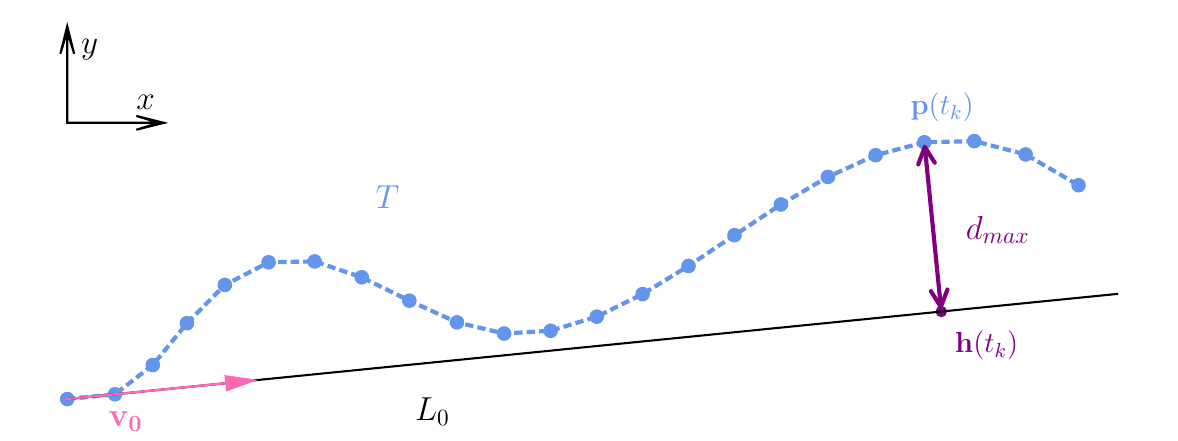}
    \caption{Illustration of the maximum lateral deviation  $d_{max}$. The intended direction of motion $L_0$ is shown with a solid line. The maximum lateral deviation $d_{max}$ is the maximum distance between a point on $T$ and $L_0$.}
    \label{fig:maximum_lateral_deviation}
\end{figure}

In collision avoidance literature, it is common to assess avoidance behaviour as deviation from a straight line. In studies with handcrafted experimental settings, such as frontal encounters in a corridor~\cite{jia2019experimental,huber2014adjustments,daamen2012interaction},
such straight line is considered directly as the environment axis and deviation is measured along the direction orthogonal to that (typically denoted as $y$-axis). In our case, rather than using the environment (i.e.\ corridor) axis, we derive an \textit{intended direction of motion} for each individual and group (see \eref{eq:desired_direction_of_motion} and \eref{eq:straight_line_traj} and $L_0$, for instance, in~\fref{fig:maximum_lateral_deviation}). We compare an observed trajectory to a straight line along that intended direction of motion in various ways as explained below.

\paragraph{Euclidean deviation $\delta_{E}$}

The Euclidean distance $\delta_{E}$~\cite{tao2021comparative} is probably one of the most intuitive and straightforward measures of deviation. It is computed as the average distance between the points of the observed trajectory $T$ and the points on the straight line trajectory $T_0$, i.e.\

\begin{equation}
    \delta_{E} = \frac{1}{N} \sum_{k=0}^{N-1} ||\vb{p}(t_k) - \vb{\tilde{p}}(t_k)||.
\end{equation}

\begin{figure}
    \centering
    \includegraphics[width=0.8\textwidth]{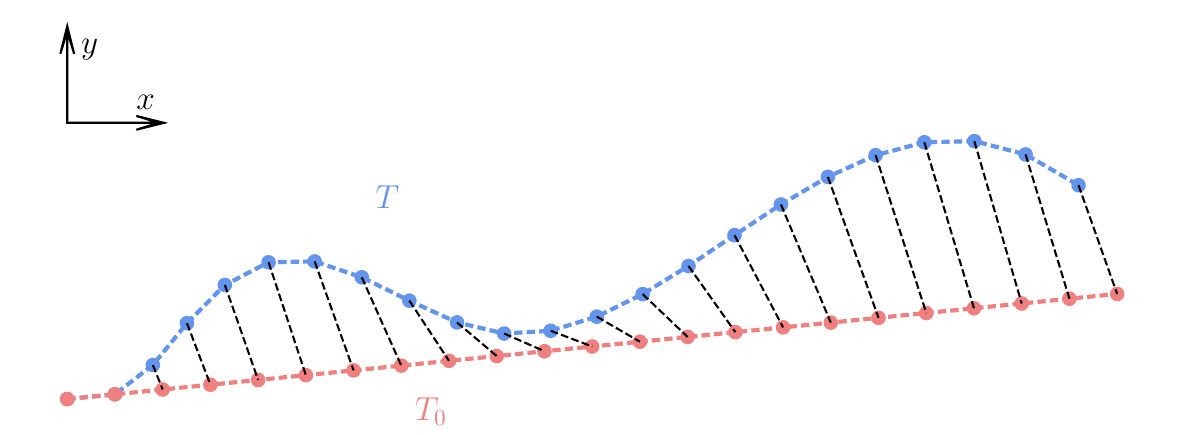}
    \caption{Illustration of the Euclidean deviation $\delta_{E}$, which  is the average of the distances between the observed trajectory of the pedestrian $T$ and the straight line trajectory $T_0$.}
    \label{fig:euclidean_distance}
\end{figure}

In \fref{fig:euclidean_distance}, we illustrate the Euclidean deviation $\delta_{E}$ for some hypothetical $T$ and $T_0$.

\paragraph{Lockstep maximum deviation $\delta_{max}$}

The lockstep maximum deviation $\delta_{max}$ is defined as the maximum distance between simultaneous pairs of points of the trajectory and the straight line trajectory. While the Euclidean deviation $\delta_{E}$ provides an average measure of deviation across the entire trajectory, the lockstep maximum deviation $\delta_{max}$ captures the maximum deviation, at a certain time, between the observed trajectory $T$ of the pedestrian and her straight line trajectory $T_0$.

Formally,

\begin{equation}
    \delta_{max} = \max_{k \in [0, N-1]} ||\vb{p}(t_k) - \vb{\tilde{p}}(t_k)||.
\end{equation}

Note that this measure is sometimes called the lockstep Euclidean distance in the literature~\cite{tao2021comparative}.

\begin{figure}[htb]
    \centering
    \includegraphics[width=0.8\textwidth]{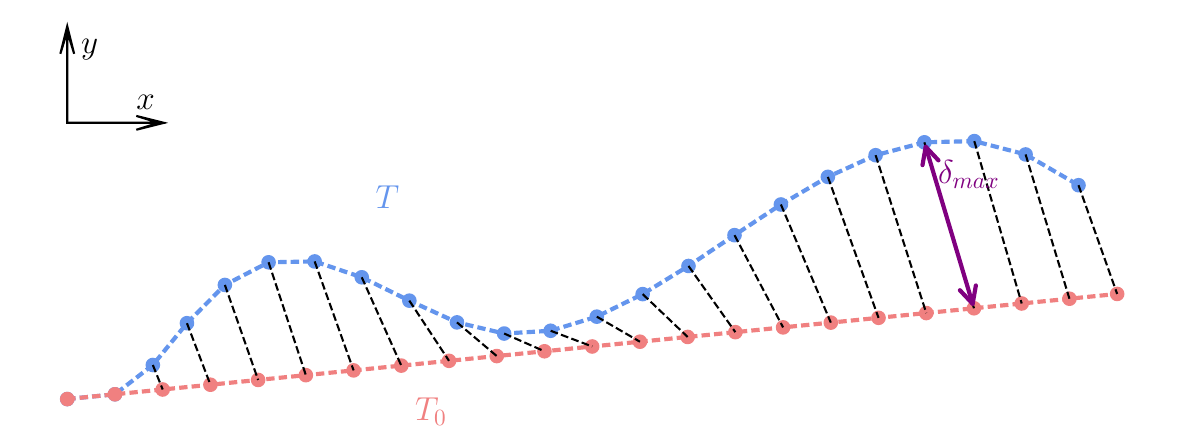}
    \caption{Illustration of the lockstep maximum deviation $\delta_{max}$, which is the maximum distance between simultaneous pairs of points of the trajectory $T$ and the straight line trajectory $T_0$.}
    \label{fig:lockstep_maximum_deviation}
\end{figure}

In \fref{fig:lockstep_maximum_deviation}, we show an example for the computation of lockstep maximum deviation $\delta_{max}$.

\paragraph{Discrete Fréchet deviation $\delta_{F}$}

We derive the discrete Fréchet deviation from the standard discrete Fréchet distance between two trajectories. Originally, Fréchet distance is a measure of similarity between two continuous curves and is often colloquially referred to as the ``dog leash'' distance, as it can be interpreted as the length of the shortest leash that would allow a dog to walk along one trajectory, while its owner walks along the other trajectory (without backtracking, but allowing to stop).
The discrete Fréchet distance is its discrete version and assesses the dissimilarity of two discrete curves~\cite{guo2017efficient}\footnote{In the figures, we illustrate $T$ and $T_0$ as piecewise continuous for making them easier to understand. However, we use only the end points of the line segments in evaluating the measures.}. To that end, an optimal mapping between the points of the two discrete curves is identified as the one minimizing the maximum distance between the matched points. Given such mapping, discrete Fréchet deviation $\delta_{F}$ is the maximum of the distances between each pair of matched points along $T$ and $T_0$.

\begin{figure}[htb]
    \centering
    \includegraphics[width=0.8\textwidth]{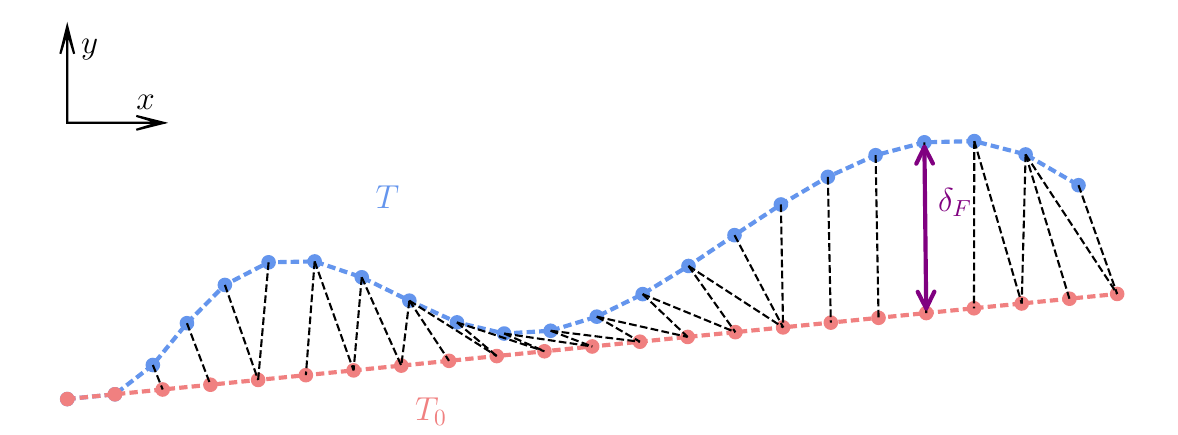}
    \caption{Illustration of the discrete Fréchet deviation $\delta_{F}$, which is the maximum distance in the optimal mapping (shown in black dashed lines) between the observed trajectory of the pedestrian $T$ and the straight line trajectory $T_0$.}
    \label{fig:frechet_deviation}
\end{figure}

In \fref{fig:frechet_deviation}, we show the discrete Fréchet distance $\delta_{F}$ for some hypothetical $T$ and $T_0$ and illustrate the two points in space that correspond to the $\delta_{F}$ of such $T$ and $T_0$.

\paragraph{Maximum lateral deviation $d_{max}$}

The maximum lateral deviation $d_{max}$ is defined as the maximum distance between a point on the trajectory of the pedestrian and the intended direction of motion $L_0$ (see \fref{fig:maximum_lateral_deviation}).

Formally,
\begin{equation}
    d_{max} = \max_{k \in [0, N-1]} |d_k|
\end{equation}
where $d_k$ is the signed distance between the point $\vb{p}(t_k)$ and its projection $\vb{h}(t_k)$ on $L_0$ computed as
\begin{equation}
    d_k =
    \left(
    \left(\vb{p}(t_k) - \vb{p}(t_0)\right) \times \frac{\vb{v_0}}{||\vb{v_0||}}
    \right)_z
    \label{eq:d_k}
\end{equation}

where $\times$ denotes the cross product\footnote{The cross product is defined in 3D space, while we are working in 2D space. However, we can consider the cross product of two 2D vectors $\vb{a} = (a_x, a_y)$ and $\vb{b} = (b_x, b_y)$ as $(\vb{a} \times \vb{b})_z = a_xb_y - a_yb_x$, i.e.\ the $z$-component of the 3D cross product of the two 3D vectors $(a_x, a_y, 0)$ and $(b_x, b_y, 0)$. It can also be interpreted as the determinant of the matrix $\begin{pmatrix} a_x & b_x \\ a_y & b_y \end{pmatrix}$.}.

The maximum lateral deviation $d_{max}$ and variables involved in computing it are illustrated in \fref{fig:maximum_lateral_deviation}.

We draw the reader's attention to the fact that the maximum lateral deviation is not the same as the lockstep maximum deviation $\delta_{max}$, which is the maximum distance between simultaneous pairs of points of the observed trajectory and the straight line trajectory. Although it is slightly more fine grained, as it does not require the discretization of the intended direction of motion, it also does not take into consideration velocity (magnitude) information.

\paragraph{Integral of lateral deviation $\Delta$}

The integral of lateral deviation is defined as the integral of the distance between a pedestrian's observed trajectory $T$ and the intended direction of motion $L_0$. Using the definitions of $d_k$ and $\vb{h}(t_k)$ given above, it can be approximated with the trapezoidal rule as
\begin{equation}
    \Delta =
    \sum_{k=0}^{N-2}
    ||
    \vb{h}(t_{k+1})
    -
    \vb{h}(t_{k})
    ||
    \frac{d_k+d_{k+1}}{2}.
\end{equation}
%where $h(\vb{p}(t_k))$ is the orthogonal projection of the point $\vb{p}(t_k)$ on the intended direction of motion $L_0$.

One notable difference of the integral of lateral deviation with the maximum lateral deviation is that deviation on both sides of the intended direction of motion will tend to cancel each other out (since $d_k$ is signed, see also \eref{eq:d_k}). This measure is therefore more sensitive to the overall deviation of the trajectory of the pedestrian from the intended direction of motion (over the entire course), rather than to the maximum deviation (at a single point in space).

\begin{figure}[htb]
    \centering
    \includegraphics[width=0.8\textwidth]{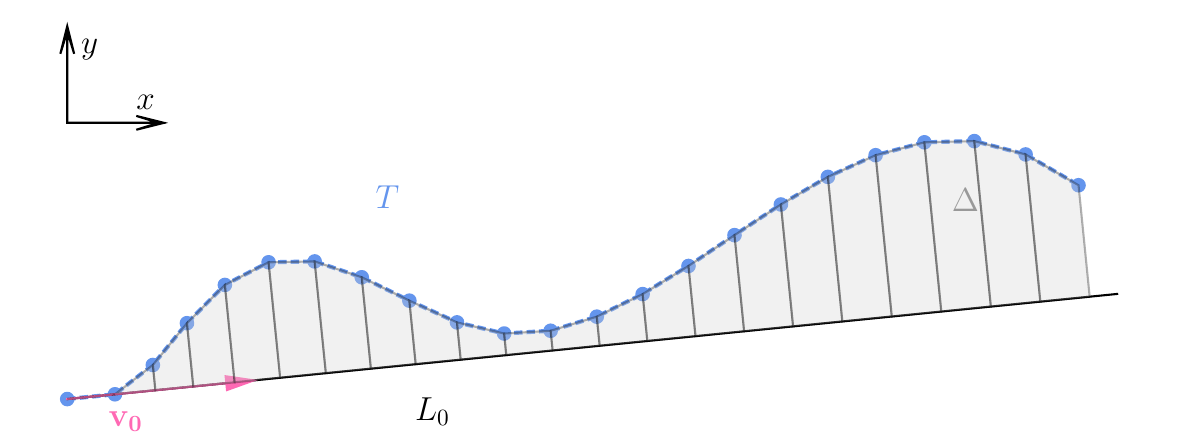}
    \caption{Illustration of the integral of lateral deviation $\Delta$. The intended direction of motion $L_0$ is shown with a solid line. The integral of lateral deviation $\Delta$ is the sum of the areas of the shaded region.}
    \label{fig:integral_lateral_deviation}
\end{figure}

In \fref{fig:integral_lateral_deviation}, we show an example for the computation of integral of lateral deviation $\Delta$, which corresponds to area of the shaded region.

\paragraph{Dynamic time warping deviation $\delta_{DTW}$}

The dynamic time warping (DTW) deviation is computed using the time warped distance between a pedestrian's observed trajectory $T$ and her straight line trajectory $T_0$. Notably, it allows for non-linear alignment of points~\cite{su2020survey}, where each point on $T$ is matched to a point on $T_0$, ensuring that the mapping is monotonically increasing in time (i.e.\ not going back in time). Additionally, the first (resp. last) point of one trajectory has to be matched to the first (resp. last) point of the other one.
The cost of such mapping is computed as the sum of the distances between the matched points, and the DTW deviation $\delta_{DTW}$ is simply the cost of the optimal mapping and can efficiently be computed using dynamic programming.

\begin{figure}[htb]
    \centering
    \includegraphics[width=0.8\textwidth]{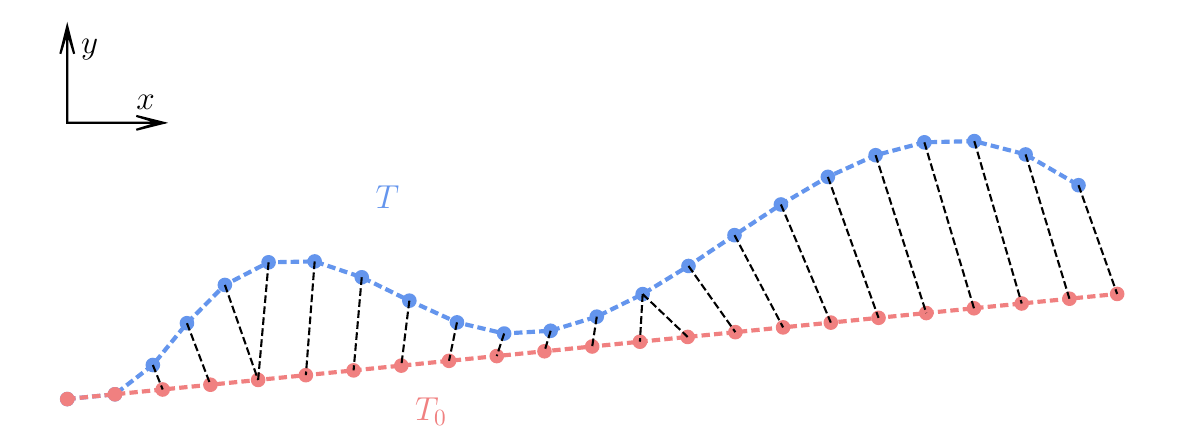}
    \caption{Illustration of the dynamic time warping deviation $\delta_{DTW}$. The optimal mapping between the trajectory of the pedestrian and the straight line trajectory is shown with black dashed lines. The dynamic time warping deviation $\delta_{DTW}$ is the sum of the distances between the matched points.}
    \label{fig:dtw_deviation}
\end{figure}

%We define the dynamic time warping deviation as the dynamic time warping distance between the trajectory of the pedestrian $T$ and the straight line trajectory $T_0$.
In \fref{fig:dtw_deviation}, we show a hypothetical optimal mapping between a pedestrian's trajectory $T$ and her straight line trajectory $T_0$. For this example, $\delta_{DTW}$ is the sum of the lengths of the dashed lines.

\paragraph{Longest common subsequence deviation $\delta_{LCSS}$}

Originally, longest common subsequence is defined for assessing the similarity of two sequences of symbols (e.g.\ text strings). Specifically, it is computed as the length of the longest subsequence that is common to these sequences, where a subsequence can be derived from a sequence by deleting some elements without changing the order of the remaining ones.

In the context of trajectories, the longest common subsequence~\cite{vlachos2002discovering} concerning two trajectories $T, T^\prime$ denoted as $l_{LCSS}(T, T^\prime)$ is defined as the length of the longest subsequence of points that are considered \textit{close enough}, i.e.\ the distance between the points is less than a given threshold $\epsilon$.

The choice of threshold is crucial. Namely, a threshold that is too small would make the measure overly sensitive to any residual noise that is not filtered out during preprocessing, while a threshold that is too large would result in considering all points as close enough, rendering $l_{LCSS}(T, T^\prime))$ equal to the length of the trajectories, which would not be informative. Given that deviations of pedestrians are usually of the order of a few tens of centimeters, we consider that two points are close enough, if the distance between them is less than 5~cm. This value was empirically selected to be small enough to capture the deviations of pedestrians, while being large enough to remain robust to noise (a position accuracy of 34~mm was reported for the tracking system used in this work in~\cite{glas2014automatic}).

Finally, we turn this measure of similarity into a distance with the normalization proposed in~\cite{tao2021comparative}, and apply it on a pedestrian's observed trajectory $T$ and her corresponding straight line trajectory $T_0$,
\begin{equation}
    \delta_{LCSS}(T, T_0)
    = 1 - \frac{l_{LCSS}(T, T_0)}{N}.
\end{equation}

\begin{figure}[htb]
    \centering
    \includegraphics[width=0.8\textwidth]{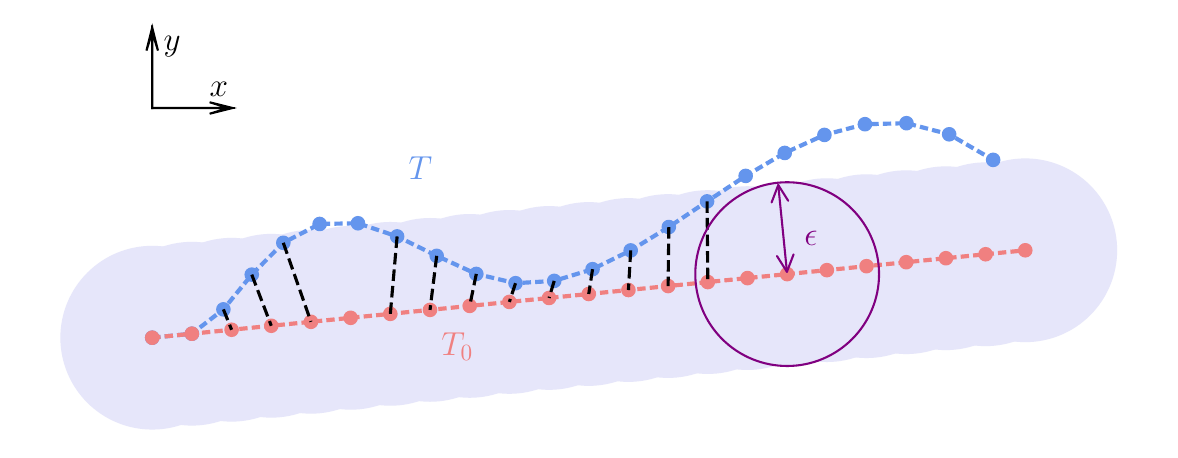}
    \caption{Illustration of the longest common subsequence deviation $\delta_{LCSS}$. The circles of radius $\epsilon$ represent the threshold on vicinity. Points connected with black dashed lines are considered {\textit close enough}.}
    \label{fig:lcss_deviation}
\end{figure}

In \fref{fig:lcss_deviation}, we illustrate the computation of the longest common subsequence deviation. The circles of radius $\epsilon$ represent the threshold under which two points are considered close enough. Points matched when computing the longest common subsequence are shown with dashed lines.

\paragraph{Levenshtein deviation $\delta_{Lev}$}

Levenshtein distance~\cite{chen2005robust} $d_{Lev}$ (also referred to as edit distance on real sequence, or simply as edit distance, even though the latter actually refers to a family of distance metrics) is a measure of similarity between two trajectories. It is defined as the minimum number of edit operations that transforms one trajectory into the other, with possible edit operations being insertions or deletions of points. Similar to the longest common subsequence distance, we consider that two points can be mapped, provided that the distance between them is less than $\epsilon = 5$~cm.

Levenshtein distance and LCSS are closely related and in some situations complementary (when no points is mapped to more than one point, $N = l_{LCSS} + d_{Lev}$, as the number of deletion/insertion is the number of points that are not matched by the longest common subsequence).

The Levenshtein deviation $\delta_{Lev}$ is defined as the Levenshtein distance between a pedestrian's observed trajectory $T_0$ and her straight line trajectory, i.e.\ $\delta_{Lev} = d_{Lev}(T, T_0)$.

\paragraph{Deviation index $\tilde{\tau}$}
\label{sec:straigtness_index}

% https://books.google.co.jp/books/about/Circular_Statistics_in_Biology.html?id=RJlYwAEACAAJ&redir_esc=y

The \textit{straightness index}~\cite{batschelet1981circular} of a discrete trajectory is defined as the ratio of the net displacement to the gross displacement (thus, also referred to as the net-to-gross displacement ratio)~\cite{codling2008random}, where the net displacement $D$, is simply
\begin{equation}
    D =\left\|\vb{p}(t_{N-1}) - \vb{p}(t_0)\right\|,
    \label{eq:deviation_index_D}
\end{equation}
and the gross displacement $L$ is the sum of the distances between consecutive positions:
\begin{equation}
    L=\sum_{k=1}^{N-1}\left\|\vb{p}(t_{k}) - \vb{p}(t_{k-1})\right\|.
    \label{eq:deviation_index_L}
\end{equation}

The straightness index is then defined as
\begin{equation}
    \tau = \frac{D}{L}.
\end{equation}

From \eref{eq:deviation_index_D} and \eref{eq:deviation_index_L}, one can see that $\tau$ takes values between 0 and 1. In particular, for an infinitely tortuous trajectory ($D$ fixed, $L \to \infty$) or a closed curve ($D=0$, $L$ fixed), $\tau$ assumes the value of 0, whereas for a perfectly straight trajectory (i.e.\ $D=L$), it leads to the value of 1. In order to be consistent with other measures of deviation, i.e.\ to have a lower value for straighter trajectories, we define the \textit{deviation index} $\tilde{\tau}$ as $1-\tau$.

\begin{figure}[htb]
    \centering
    \includegraphics[width=0.8\textwidth]{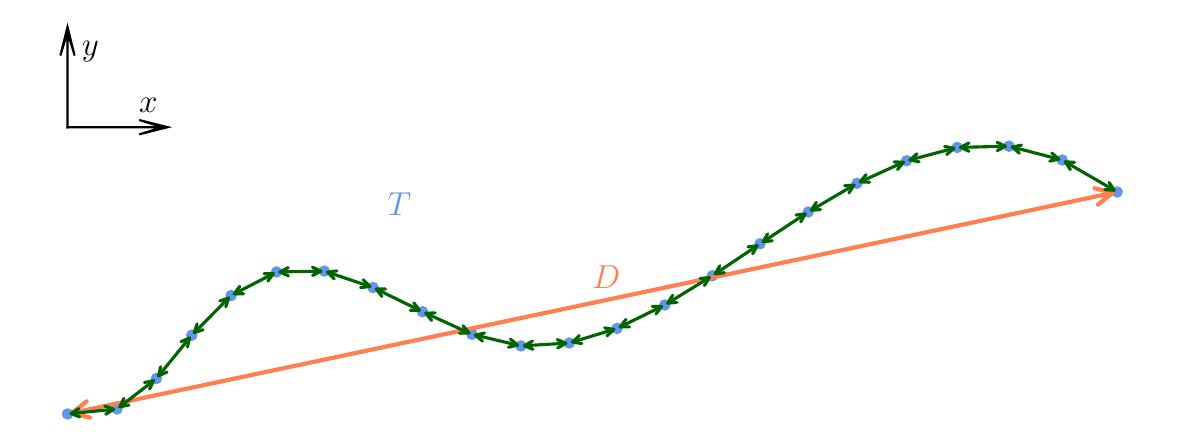}
    \caption{Illustration of  deviation index $\tilde{\tau}$. The net displacement $D$ is shown in orange and the gross displacement $L$ is the sum of the distances between consecutive positions show in green. The deviation index is $1- \frac{D}{L}$.}
    \label{fig:straightness_index}
\end{figure}

\fref{fig:straightness_index} illustrates the computation of the deviation index. The net displacement $D$ is shown in orange and all the distances between consecutive positions are shown in green. The gross displacement $L$ is the sum of these distances.

\subsubsection{Orientation based measures}

In addition to the position based measures, we introduce a set of measures that quantify the deviation of the trajectory of a pedestrian from a straight line trajectory based on the orientation of the velocity vectors, i.e.\ the direction of motion.

\paragraph{Maximum cumulative turning angle $\theta_{max}$}

To deviate from an intended trajectory, pedestrians naturally have to turn. We can therefore quantify the deviation of the trajectory by looking at the amount of turning performed. The cumulative turning angle until time $t_k$ is defined as the sum of the turning angles between consecutive velocity vectors until time $t_k$\footnote{Note that is not the same as computing the angle between $\vb{v}(t_0)$ and $\vb{v}(t_k)$, since it is theoretically possible to obtain an angle larger than $2\pi$ if the pedestrian makes a full turn. Nonetheless, this situation should not occur in practice.}.

Formally, it is defined as
\begin{equation}
    \theta_k = \sum_{j=0}^{k-1} d\theta_j,
\end{equation}

where $d\theta_j$ is the signed angle between the velocity vectors $\vb{v}(t_j)$ and $\vb{v}(t_{j+1})$, $d\theta_j = \angle (\vb{v}(t_j), \vb{v}(t_{j+1}))$.

\begin{figure}[htb]
    \centering
    \begin{subfigure}[t]{0.5\textwidth}
        \centering
        \includegraphics[width=\textwidth]{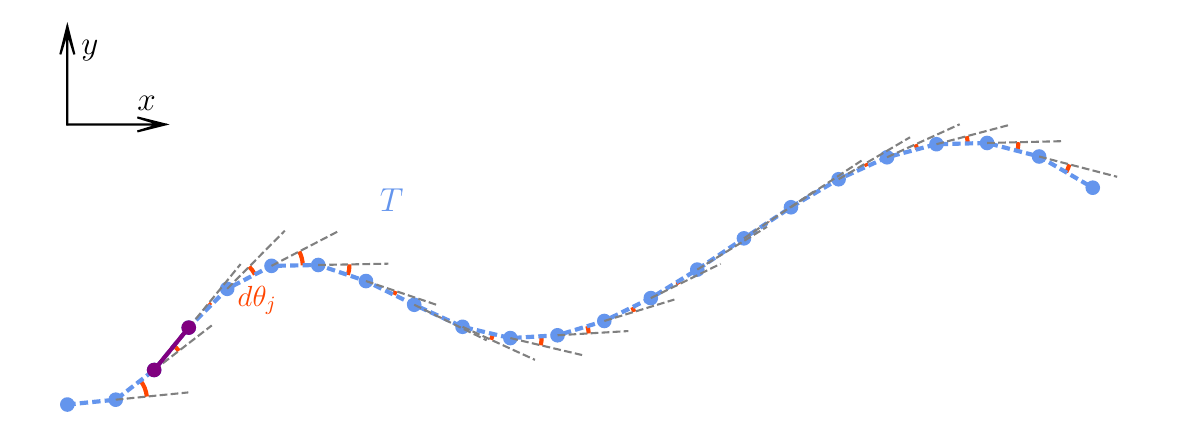}
        \caption{}
        \label{fig:maximum_cumulative_turning_angle_trajectory}
    \end{subfigure}
    \begin{subfigure}[t]{0.45\textwidth}
        \centering
        \includegraphics[width=\textwidth]{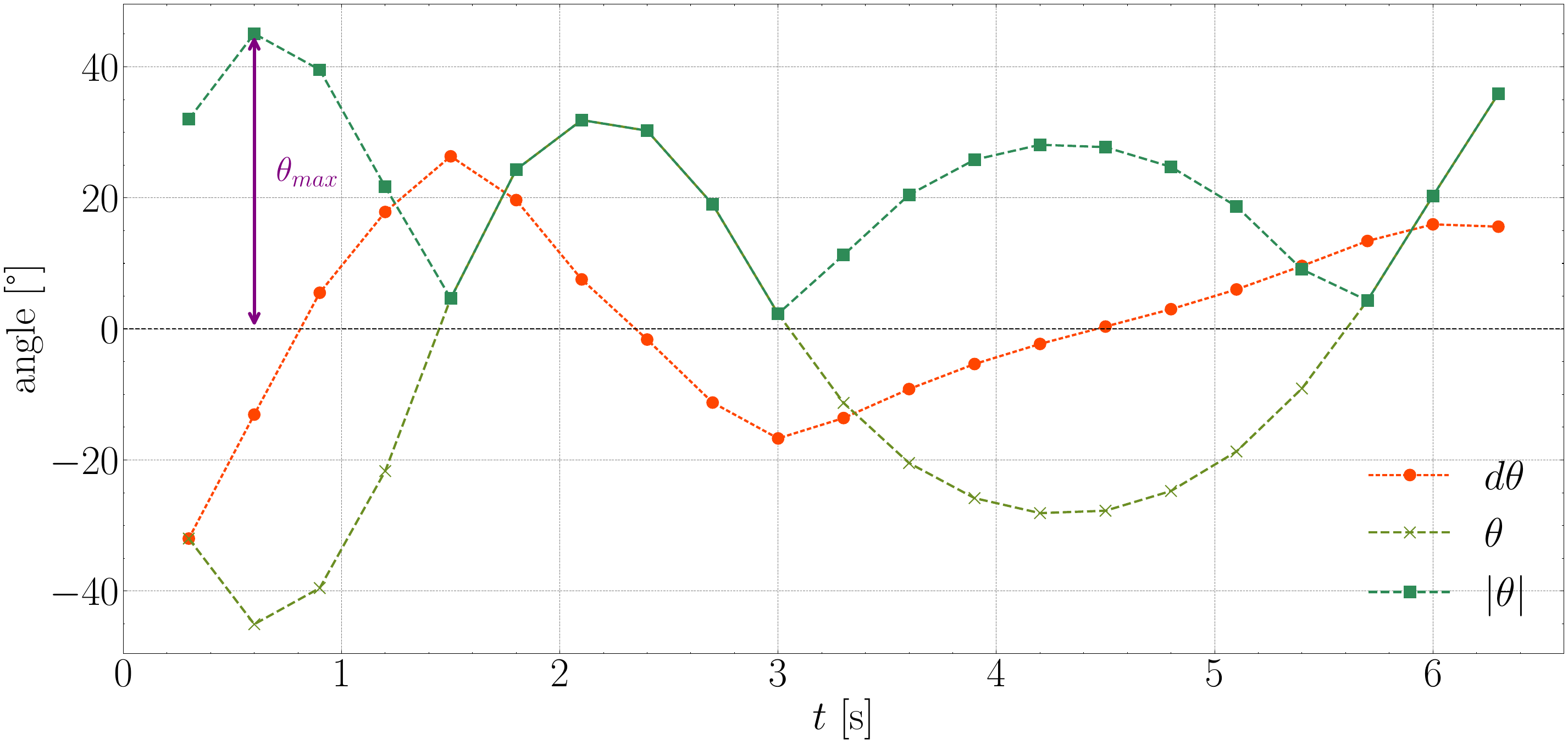}
        \caption{}
        \label{fig:maximum_cumulative_turning_angle_breakthrough}
    \end{subfigure}
    \caption{Illustration of the variables used in computing maximum cumulative turning angle. (a) The turning angles $d\theta_j$ are shown in red. The segment of the trajectory for which the cumulative turning angle is maximum is drawn in purple. (b) Corresponding values of turning angles $d\theta_j$, cumulative turning angles $\theta_k$, their absolute values $ |\theta_k|$ and the maximum cumulative turning angle $\theta_{max}$.
        %The integral of the cumulative turning angles $\Theta$ is the area of the shaded area.
    }
    \label{fig:maximum_cumulative_turning_angle}
\end{figure}

The turning angles being signed, the cumulative turning angle can be positive or negative, depending on the direction of the turning (see~\fref{fig:maximum_cumulative_turning_angle}). We are interested in the maximum cumulative turning angle $\theta_{max}$, which is the maximum of the absolute value of the cumulative turning angles over the trajectory.
\begin{equation}
    \theta_{max} = \max_{k \in [0, N-1]} |\theta_k|
\end{equation}

In \fref{fig:maximum_cumulative_turning_angle}, we illustrate the computation of the maximum cumulative turning angle. The left part of the figure shows the turning angles $d\theta_j$ and the segment of the trajectory where the cumulative turning angle is maximum. In the right part of the figure, we show a graph of the turning angles $d\theta_j$, the cumulative turning angles $\theta_k$, $ |\theta_k|$ and the maximum cumulative turning angle $\theta_{max}$.

We note that, in contrast with the position based measures, the maximum cumulative turning angle is an early indicator of the deviation of the trajectory. Typically, this maximum value is not reached at the same position, for instance, as the maximum lateral deviation. Specifically, by the time the pedestrian has begun turning back towards their intended direction of motion, the cumulative turning angle starts to decrease.

\paragraph{Average cumulative turning angle $\Theta$}

In the same way that we defined the Euclidean deviation as the average of the deviations, we can define the average cumulative turning angle $\Theta$ as the average of the cumulative turning angles over the trajectory. Formally,
\begin{equation}
    \Theta = \frac{1}{N} \sum_{k=0}^{N-1} |\theta_k|.
\end{equation}

\paragraph{Sinuosity $S$}

Sinuosity was introduced in~\cite{bovet1988spatial} as a way to quantify the randomness of an animal's path (especially in the case of foraging patterns, for instance when studying how ants locate and collect food). It is a measure proportional to the standard deviation of the distribution of the turning angles for a given trajectory and is formally defined as
\begin{equation}
    S = 1.18 \frac{\sigma_q} {\sqrt{q}},
    \label{eq:sinuosity}
\end{equation}
where $q$ is the step size of the trajectory. According to the definition of~\cite{bovet1988spatial}, the step size has to be constant across the trajectory. Thus, $S$ requires the trajectory $T$ to be re-discretised into another trajectory $\tilde{T}$ with equal spacing between all the points. $\sigma_q$ is the (circular) standard deviation of the turning angles $\theta$. The value $1.18$ has been numerically computed by the authors of~\cite{bovet1988spatial} using simulations of correlated random walks.

This measure is based on the assumption that the turning angles are drawn from a normal distribution wrapped on a circle (with some correlation between consecutive draws, to account for the forward tendency of the locomotion).

\begin{figure}[htb]
    \centering
    \includegraphics[width=0.8\textwidth]{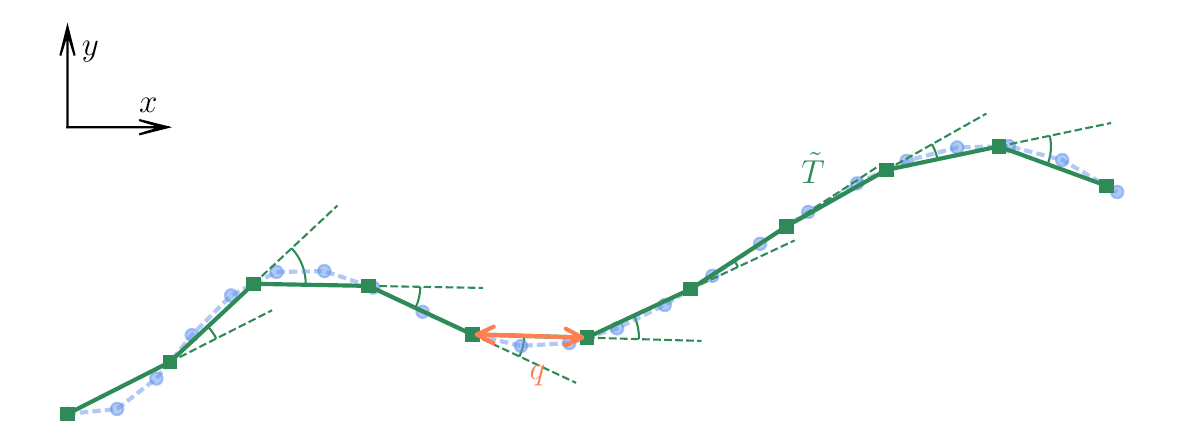}
    \caption{Illustration of  sinuosity $S$. The trajectory is re-discretised to obtain a trajectory $\tilde{T}$ with equal spacing $q$ between all the points. The turning angles are then computed.}
    \label{fig:sinuosity}
\end{figure}

In \fref{fig:sinuosity}, we illustrate the re-discretization of the trajectory and the computation of the turning angles.

\paragraph{Energy curvature $E_{\kappa}$}

In two dimensions, curvature is a measure of how much a curve deviates from a straight line. It describes the rate of change of the tangent angle of the curve and can be computed as
\begin{equation}
    \kappa(t_k) = \frac{||\vb{v}(t_k) \times \vb{a}(t_k)||}{||\vb{v}(t_k)||^3},
\end{equation}
where $\vb{v}(t_k)$ and $\vb{a}(t_k)$ are the velocity and acceleration vectors of the trajectory at times $t_k$ and $\vb{a}(t_k)$ is derived in a similar way to $\vb{v}(t_k)$ as in \eref{eq:velocity}.

Another way to understand curvature is to consider the circle that best approximates the curve at each point, i.e.\ the osculating circle. The curvature at a point is then defined as the reciprocal of the radius of the osculating circle.

\begin{figure}[htb]
    \centering
    \includegraphics[width=0.8\textwidth]{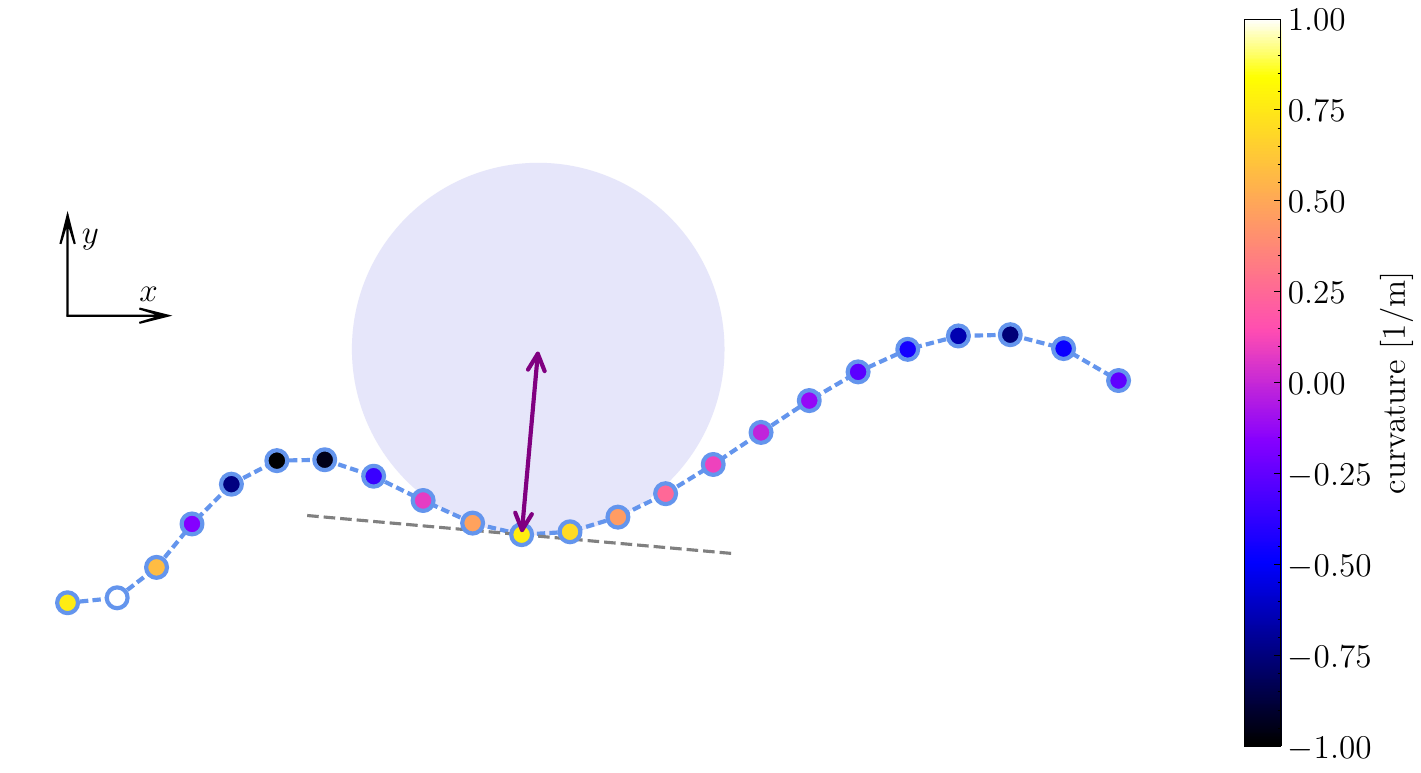}
    \caption{Illustration of  curvature $\kappa$. The color of a point on the trajectory correspond to the signed curvature at that point. The osculating circle at a point of the trajectory is shown (the curvature at that point is the reciprocal of the radius of the circle).}
    \label{fig:curvature}
\end{figure}

In \fref{fig:curvature}, we illustrate the computation of curvature. The color of the point on the trajectory corresponds to the signed curvature at that point. We also show the osculating circle at a point of the trajectory.

The energy curvature is defined as the integral of the square of the curvature over the trajectory. The integration is performed using the trapezoidal rule, and the energy curvature is given by

\begin{equation}
    E_{\kappa} = \sum_{k=0}^{N-2} (t_{k+1} - t_k) \frac{\kappa(t_k)^2 + \kappa(t_{k+1})^2}{2}.
\end{equation}

In path planning, minimising the energy curvature is often considered as a way to ensure smooth trajectories~\cite{bruckstein1990minimal}.

\subsubsection{Mixed measures}

Some works have proposed measures that combine position and orientation information to quantify the deviation of a trajectory. In this section, we present two such measures:  turn intensity and  suddenness of turn.

\paragraph{Turn intensity $i$}

The turn intensity was introduced in~\cite{murakami2021mutual}. In the authors' experiment, participants were moving and crossing each other along a straight elongated path ($x$-axis) and they consider instants at which the motion along the orthogonal axis ($y$-axis) changes direction, which they refer to as ``turning'' instants. A ``step'' is defined as the motion between two consecutive turning instants. The ``step length'' is defined as the $y$ component of a step and the ``step angle'' is defined as the absolute value of the angle deviation of a step from the horizontal axis. The turn intensity is then defined as the product of the step length and step angle.

In our case, since pedestrians might not be moving along the environment axis (i.e.\ $x$ axis, as detailed in \sref{sec:method_desired_direction}), we adapt the definitions of the turning instants and steps to consider the deviation from the intended direction of motion. In particular, we consider the signed angle $\psi_k$ between the initial velocity vector $\vb{v_0}$ and the velocity vector $\vb{v}(t_k)$ at each time step $t_k$, $\psi_k = \angle (\vb{v_0}, \vb{v}(t_k))$. The turning instants $t_s$ are then defined as the instants at which $\psi_k$ changes sign.

We then consider the $s$-th step angle $\omega_s$ to be the absolute value of the angle deviation of a step from the intended direction of motion and the step length $\lambda_s$ to be the orthogonal distance between the position of the pedestrian at the end of the step and the intended direction of motion.

\begin{equation}
    \omega_s = \Big| \angle (\vb{v_0}, (\vb{p_{s+1}} - \vb{p_s})) \Big |,
\end{equation}

\begin{equation}
    \lambda_s = \frac{||(\vb{p_{s+1}} - \vb{p_s}) \times \vb{v_0}||}{||\vb{v_0}||}.
\end{equation}

The turn intensity is then defined as average value of the product of the step lengths and step angles.

\begin{equation}
    i = \frac{\sum_{s=0}^{N_S-1} \omega_s \lambda_s}{N_S},
\end{equation}

where $N_S$ is the number of steps.

\begin{figure}[htb]
    \centering
    \includegraphics[width=0.8\textwidth]{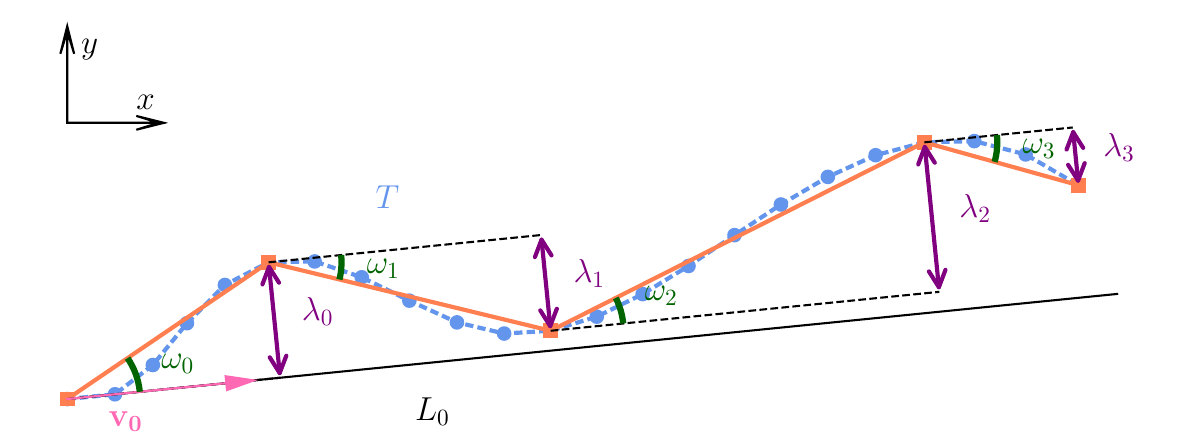}
    \caption{Illustration of the variables used in computing turn intensity $i$. The steps are shown in orange, step angles $\omega$  are shown  in green and step lengths $\lambda$  are shown  in purple.}
    \label{fig:turn_intensity}
\end{figure}

In \fref{fig:turn_intensity}, we illustrate
the variables used in computing turn intensity $i$.

\paragraph{Suddenness of turn $\sigma$}

The suddenness of turn was introduced in~\cite{murakami2021mutual}. It is defined as the absolute value of the product of the angle between the velocity vector at a given time $t$ and the horizontal axis (since the participants were moving along the $x$-axis in the experiments of~\cite{murakami2021mutual}) and the change in speed, i.e.\ the difference between the speed at time $t$ and the speed at time $t-dt$,  $dt$ being the sampling period.

Similar to the previous measures, we adapt this definition to consider the intended direction of motion rather than the environment (i.e.\ $x$) axis. In particular, we use the angles $\psi_k$ defined in the previous section and the suddenness of turn $\sigma$ is defined as

\begin{equation}
    \sigma = \frac{1}{N-2} \sum_{k=1}^{N-2} \left| \psi_k (||\vb{v}(t_k)|| - ||\vb{v}(t_{k-1})|| )\right|.
\end{equation}

\subsection{Toy trajectories}
\label{sec:toy_trajectories}

\begin{figure}[htb]
    \centering
    \includegraphics[width=0.6\textwidth]{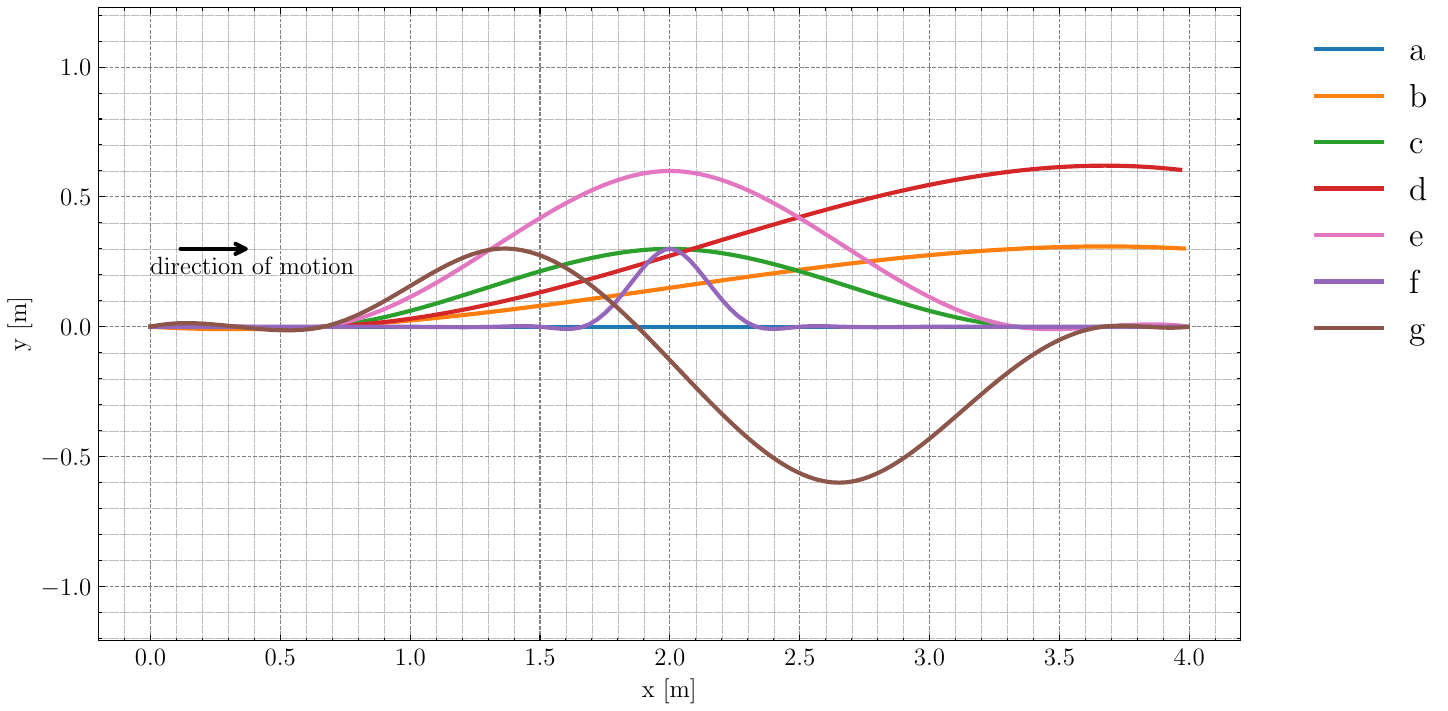}
    \caption{Toy trajectories, representing different types and levels of deviation.}
    \label{fig:toy_trajectories}
\end{figure}

To demonstrate how various deviation measures evaluate different trajectory shapes, we examine a set of 7 artificially generated toy trajectories (see \fref{fig:toy_trajectories}).
These trajectories, designed to be comparable with the actual undisturbed and encounter paths studied in this work, were generated by defining handcrafted waypoints and interpolating between them using a cubic spline, such that a constant velocity of 1.2~m/s is maintained at a sampling frequency of 33~Hz.

\begin{description}
    \item[a. Straight line:] perfectly straight line.
    \item[b. Small deviation without recovery:] the trajectory starts straight, then deviates to the left (up to 30~cm) and does not recover (i.e.\ the trajectory does not return to the straight line).
    \item[c. Small deviation with recovery:] the trajectory starts straight, then deviates to the left (up to 30~cm) and recovers (i.e.\ the trajectory returns to the straight line).
    \item[d. Big deviation without recovery:] the trajectory starts straight, then deviates to the left (up to 60~cm) and does not recover.
    \item[e. Big deviation with recovery:] the trajectory starts straight, then deviates to the left (up to 60~cm) and recovers.
    \item[f. Fast deviation with recovery:] the trajectory starts straight, then deviates to the left (up to 30~cm) and recovers, but the deviation is performed over a shorter distance than the previous cases.
    \item[g. Deviation on both sides:] the trajectory starts straight, then deviates to the left (up to 30~cm), then to the right (up to 60~cm) and recovers.
\end{description}

We calculate the value of deviation measures outlined in~\sref{sec:method_measures} for each of these toy trajectories and present the results in~\tref{tab:toy_measures}. As anticipated, we observe that all deviation measures yield a 0 or very small value for \textbf{a}, identifying it as the one with least deviation\footnote{While this information may seem trivial, it serves as a sort of sanity check}. To facilitate the comparison of deviation values, in each column of~\tref{tab:toy_measures} we underline the second smallest value and highlight the highest value in bold. Additionally, the rank of each value in a column is provided in parentheses on its right.

Just by observing the underlined and bold values, significant discrepancies between different deviation measures become apparent. For example, ``fast deviation with recovery''  \textbf{f} is identified to have very small deviation by measures such as $\delta_{F}$ and $d_{max}$, but it is identified as having the largest deviation by measures like $S$, $E_{\kappa}$, and $\sigma$. Upon closer examination of the ranks and comparison across different columns, we notice that trajectories other than \textbf{a} can be evaluated quite differently by various deviation measures.

In particular, three trajectories (\textbf{d}, \textbf{f}, and \textbf{g}) are identified as having the largest deviation by our deviation measures. Since distance-based measures are primarily influenced by the distance from the straight-line trajectory and the number of points significantly deviating from it, they consequently rank ``big deviation without recovery'' \textbf{d} and ``deviation on both sides'' \textbf{g} as the most deviated.
Furthermore, suddenness of  turn $\sigma$ is highest for ``fast deviation with recovery'' \textbf{f}. In addition, since angular measures are mainly affected by the number of turning points and the magnitude of the turning angles, they rank ``deviation on both sides'' \textbf{g} as having the largest deviation. Moreover, deviation index $\tilde{\tau}$ also ranks it as the most deviated due to its highest gross displacement among all toy trajectories.

By examining the second smallest (underlined) value in each column, we notice that position-based measures (with the exception of $\delta_{E}$, albeit with a small difference between second and third places) rank ``fast deviation with recovery'' \textbf{f} as the second least deviated. Since these measures are primarily influenced by the distance to the straight line trajectory and the number of points deviating, they evaluate \textbf{f} to be relatively straight, as it deviates over the shortest distance.
In contrast, position based measure $\tilde{\tau}$, orientation based measures such as $\theta_{max}$, $E_{\kappa}$, $S$, and mixed measures $i$ and $\sigma$ rank ``small deviation without recovery'' \textbf{b} as the second least deviated. This is consistent with the fact that \textbf{b} has the smallest gross displacement and undergoes a slower deviation.

We highlight the fact that ``fast deviation with recovery'' \textbf{f} is ranked as both the most deviated and the least deviated (ignoring the straight line) by different measures. This shows that the measures capture different aspects of the deviation, and no single measure is expected to fully capture the complexity of deviations of real (human) trajectories. In the specific scenarios investigated in this study, we nevertheless anticipate the deviations not to be excessively abrupt (i.e.\ with low acceleration and jerk) due to the density and geometry of the environment.

\begin{table}[!htb]
    \caption{Measures of deviation for the toy trajectories. The maximum value for each measure is highlighted in bold and the second smallest value is underlined (since the straight line trajectory is the smallest value for all the measures). The ranking of the trajectories (from the straightest to the most deviated) is also indicated in parenthesis.}
    \label{tab:toy_measures}
    \centering
    \begin{adjustbox}{angle=90}
        \scalebox{0.5}{
            \begin{tabular}{lrrrrrrrrrrrrrrr}
                \toprule
                Measures & $\delta_{E}$                          & $\delta_{max}$                        & $\delta_{F}$                          & $d_{max}$                             & $\Delta$                              & $\delta_{DTW}$                        & $\delta_{LCSS}$                       & $\delta_{Lev}$                    & $\tilde{\tau}$                        & $\theta_{max}$                        & $\Theta$                              & $S$                                   & $E_{\kappa}$                          & $i$                               & $\sigma$                              \\
                \midrule
                a        & $7.61 \times 10^{-17}$ (1)            & $4.55 \times 10^{-16}$ (1)            & $4.55 \times 10^{-16}$ (1)            & $0$ (1)                               & $0$ (1)                               & $7.61 \times 10^{-17}$ (1)            & $8.93 \times 10^{-3}$ (1)             & $0$ (1)                           & $0$ (1)                               & $0$ (1)                               & $0$ (1)                               & $0$ (1)                               & $0$ (1)                               & $0$ (1)                           & $0$ (1)                               \\
                b        & $1.65 \times 10^{-1}$ (4)             & $3.37 \times 10^{-1}$ (4)             & $3.36 \times 10^{-1}$ (4)             & $3.36 \times 10^{-1}$ (4)             & $6.50 \times 10^{-1}$ (5)             & $1.65 \times 10^{-1}$ (4)             & $7.14 \times 10^{-1}$ (4)             & $7.90 \times 10$ (4)              & $\underline{1.76 \times 10^{-3}}$ (2) & $\underline{1.97 \times 10^{-1}}$ (2) & $\underline{1.32 \times 10^{-1}}$ (2) & $\underline{1.32 \times 10^{-1}}$ (2) & $\underline{4.21 \times 10^{-5}}$ (2) & $\underline{6.58 \times 10}$ (2)  & $\underline{7.31 \times 10^{-2}}$ (2) \\
                c        & $\underline{1.35 \times 10^{-1}}$ (2) & $\underline{3.18 \times 10^{-1}}$ (2) & $3.16 \times 10^{-1}$ (3)             & $3.16 \times 10^{-1}$ (3)             & $4.85 \times 10^{-1}$ (4)             & $1.24 \times 10^{-1}$ (3)             & $5.75 \times 10^{-1}$ (3)             & $6.60 \times 10$ (3)              & $1.89 \times 10^{-2}$ (4)             & $3.77 \times 10^{-1}$ (4)             & $1.74 \times 10^{-1}$ (5)             & $5.18 \times 10^{-1}$ (4)             & $6.48 \times 10^{-4}$ (4)             & $1.20 \times 10^{2}$ (3)          & $2.22$ (4)                            \\
                d        & $2.98 \times 10^{-1}$ (5)             & $6.38 \times 10^{-1}$ (5)             & $\mathbf{6.34 \times 10^{-1}}$ (7)    & $\mathbf{6.34 \times 10^{-1}}$ (7)    & $\mathbf{1.17}$ (7)                   & $\mathbf{2.98 \times 10^{-1}}$ (7)    & $7.35 \times 10^{-1}$ (5)             & $8.20 \times 10$ (5)              & $7.10 \times 10^{-3}$ (3)             & $2.62 \times 10^{-1}$ (3)             & $1.44 \times 10^{-1}$ (3)             & $2.56 \times 10^{-1}$ (3)             & $1.62 \times 10^{-4}$ (3)             & $1.81 \times 10^{2}$ (5)          & $5.83 \times 10^{-1}$ (3)             \\
                e        & $3.10 \times 10^{-1}$ (6)             & $\mathbf{6.45 \times 10^{-1}}$ (7)    & $6.28 \times 10^{-1}$ (6)             & $6.28 \times 10^{-1}$ (6)             & $9.41 \times 10^{-1}$ (6)             & $2.55 \times 10^{-1}$ (6)             & $7.48 \times 10^{-1}$ (6)             & $9.50 \times 10$ (6)              & $6.81 \times 10^{-2}$ (6)             & $6.93 \times 10^{-1}$ (5)             & $3.28 \times 10^{-1}$ (6)             & $9.41 \times 10^{-1}$ (5)             & $2.36 \times 10^{-3}$ (5)             & $4.60 \times 10^{2}$ (6)          & $1.40 \times 10$ (5)                  \\
                f        & $1.37 \times 10^{-1}$ (3)             & $3.23 \times 10^{-1}$ (3)             & $\underline{3.00 \times 10^{-1}}$ (2) & $\underline{3.00 \times 10^{-1}}$ (2) & $\underline{1.00 \times 10^{-1}}$ (2) & $\underline{4.76 \times 10^{-2}}$ (2) & $\underline{1.78 \times 10^{-1}}$ (2) & $\underline{2.60 \times 10}$ (2)  & $5.91 \times 10^{-2}$ (5)             & $9.11 \times 10^{-1}$ (6)             & $1.64 \times 10^{-1}$ (4)             & $\mathbf{2.84}$ (7)                   & $\mathbf{2.71 \times 10^{-2}}$ (7)    & $1.70 \times 10^{2}$ (4)          & $\mathbf{6.09 \times 10}$ (7)         \\
                g        & $\mathbf{3.39 \times 10^{-1}}$ (7)    & $6.39 \times 10^{-1}$ (6)             & $5.43 \times 10^{-1}$ (5)             & $5.43 \times 10^{-1}$ (5)             & $2.04 \times 10^{-1}$ (3)             & $2.39 \times 10^{-1}$ (5)             & $\mathbf{7.70 \times 10^{-1}}$ (7)    & $\mathbf{1.04 \times 10^{2}}$ (7) & $\mathbf{1.27 \times 10^{-1}}$ (7)    & $\mathbf{9.97 \times 10^{-1}}$ (7)    & $\mathbf{4.40 \times 10^{-1}}$ (7)    & $1.57$ (6)                            & $7.32 \times 10^{-3}$ (6)             & $\mathbf{7.98 \times 10^{2}}$ (7) & $4.20 \times 10$ (6)                  \\
                \bottomrule
            \end{tabular}
        }
    \end{adjustbox}
\end{table}

\subsection{Impact parameter}
\label{sec:impact_parameter}

In this section, we briefly introduce the concept of the impact parameter, which we previously used in~\cite{gregorj2023social}.

\begin{figure}
    \centering
    \includegraphics[width=0.6\textwidth]{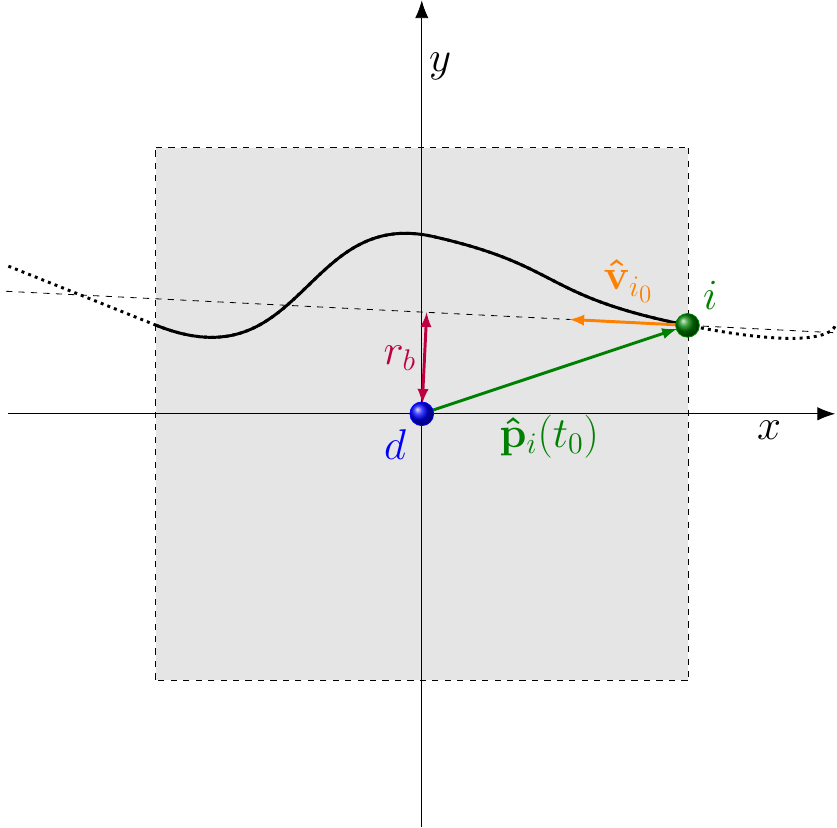}
    \caption{Illustration of the impact parameter. The individual's trajectory is transformed into a reference frame that moves with the dyad. The impact parameter $r_b$ is the distance from the dyad to the line guided by the individual's velocity vectors at the beginning of the encounter.}
    \label{fig:impact_parameter}
\end{figure}

In the scattering of particles in physics, the impact parameter is the distance between the path of an incoming particle and the target particle. To apply this concept to the study of pedestrian trajectories, we treat the deviation of an individual from a dyad as a scattering event. We start by transforming the trajectories of the dyad $d$ and the individual $i$ into a reference frame that moves with the dyad. Specifically, at each time instant the positions of the dyad and the individual are translated so that the dyad's center of mass is positioned at the origin, and their velocities are rotated such that the dyad's velocity is aligned with the positive x-axis\footnote{In physical scattering theory there is no need to build a special frame, since only relative positions and velocities are used. Nevertheless in our computation, since we use a box (see \fref{fig:impact_parameter}) which is not rotation invariant, the definition of frame became important. The reason we use a box is that our environment has walls and preferred directions.}. In this reference frame, the dyad remains fixed at the origin, while the individual moves towards it, analogous to a particle approaching a target in a scattering event.

We denote the position of the individual in the reference frame as $\vb{\hat{p}}_i$ and its velocity as $\vb{\hat{v}}_i$.

In this reference frame, the impact parameter $r_b$ is computed as the distance from the dyad (positioned at the origin) to the line guided by the individual's velocity vectors at the beginning of the encounter (as in \sref{sec:method_desired_direction}, we average the velocity vectors over $N_e$ time instants to alleviate the impact of orientation noise).It is a measure of how close the individual and the dyad would have passed each other if there were no collision avoidance. We illustrate the computation of the impact parameter in~\fref{fig:impact_parameter}.

Let $\vb{\hat{v}}_{i_0}$ be the average velocity vector of the individual over the first $N_e$ time instants of the encounter,

\begin{equation}
    \vb{\hat{v}}_{i_0} = \frac{1}{N_e} \sum_{k=0}^{N_e-1} \vb{\hat{v}}_i(t_k),
\end{equation}

where $N_e$ is the same as in the computation of the desired direction of motion (see \sref{sec:method_desired_direction}).

We can compute $r_b$ as

\begin{equation}
    r_b = \frac{||\vb{\hat{v}}_{i_0} \times \vb{\hat{p}}_i(t_0)||}{||\vb{\hat{v}}_{i_0}||}.
\end{equation}

As detailed in~\cite{gregorj2023social}, we scale the impact parameter by the width of the dyad (i.e.\ the average distance between the members of dyads with that level of interaction) to obtain a dimensionless measure $\bar{r}_b$ that better captures the relative distance between the individual and the member of the dyad. In particular, a value of $\bar{r}_b$ smaller than $0.5$ indicates that the individual would have passed through the dyad.

\section{Results}

\subsection{Undisturbed situations}

We begin by comparing the deviation of individuals and dyads in undisturbed situations, which we argue to constitute a baseline for the amount of deviation that can be expected in the absence of any encounter.
As noted in \sref{sec:method_undisturbed}, undisturbed situations occur when individuals or dyads are walking without any other pedestrian within a distance of at least 4~m along or perpendicular to their direction of motion.

In \tref{tab:undisturbed_euclidean_distance} $\sim$ \tref{tab:undisturbed_suddenness_turn}, we present the average deviation for dyads categorised by levels of interaction ranging from 0 to 3, all dyads, and individuals, with respect to each measure introduced in \sref{sec:method_measures}.

For convenient comparison with individuals, we also display the ratio of the average deviation of the dyads to the average deviation of the individuals. Additionally, we present the Kruskal-Wallis test $p$-value for the difference of means between dyads with various level of interaction, and the Welch T-test $p$-value for the difference of means between individuals and all dyads~\cite{kruskal1952use,welch1947generalization}.

The initial noteworthy finding is that, across the vast majority of measures, deviation is significantly higher for dyads (averaged across all levels of interaction) compared to individuals. However, to accurately interpret this observation, it is essential to delve into the allocation of attentional resources during social interaction, achieved by contrasting the deviations of dyads with varying levels of interaction.
Examining the breakdown of normalised deviation values according to the level of interaction of the dyad, we observe a trend where the level of interaction correlates with an increase in deviation. Interestingly, the deviations of non-interacting dyads appear to be smaller to those of individuals, while interacting dyads which exhibit comparable or higher deviations.

\subsection{Encounters}

We now turn to the encounter situations, which correspond to the situations in which dyads and individuals pass each other frontally at a distance less than 4~m~\cite{gerin2005negotiation,cinelli2008locomotor,kitazawa2010pedestrian}.
We first contrast deviations during encounters with undisturbed situations. Subsequently, we conduct a closer examination of encounters, comparing deviations of individuals and dyads during these cases. Finally, we provide insights into the effect of dyads' level of interaction on both their own deviation and the deviation of the individuals involved.

\subsubsection{Deviations during encounters}
\label{sec:deviations_during_encounters}

In \tref{tab:encounter_euclidean_distance} $\sim$  \tref{tab:encounter_suddenness_turn} we show the amount of deviation of individuals and dyads for each measure introduced in \sref{sec:method_measures}.
It is crucial to emphasise that when reporting deviations related to encounter situations, we categorise not only the values of dyads but also those of individuals, with respect to the level of social interaction of the dyad involved in the encounter.

We notice that individuals have a tendency to deviate more when encountering dyads with high or medium interaction levels (i.e.\ levels 3, 2) compared to when they encounter non- or weakly interacting ones (i.e.\ levels 0, 1). In particular, the deviation when encountering a strongly interacting dyad (level 3) is almost always\footnote{Except for $\Delta$, $\delta_{LCSS}$ an $\sigma$.} larger than for other levels of interaction. However, upon subjecting the data to a Kruskal-Wallis test, we see that the differences in deviation for the individual when encountering dyads with varying levels of interaction are not statistically significant for most metrics (all except energy curvature $E_{\kappa}$ and sinuosity $S$).
Conversely, the variation in dyad deviation is statistically significant but does not exhibit a clear pattern, with dyads of interaction levels 0 and 3 deviating more than those with interaction levels of 1 and 2.
Finally, averaging across all interaction levels, individuals deviate more than dyads in a statistically significant manner for all measures.

\subsubsection{Comparison of deviations during encounters and undisturbed situations}
\label{sec:comparison_of_deviations}

The results from \tref{tab:encounter_euclidean_distance} $\sim$ \tref{tab:encounter_suddenness_turn} can be better understood when contrasted with undisturbed situations. To facilitate this comparison, \tref{tab:ratio_euclidean_distance} $\sim$ \tref{tab:ratio_suddenness_turn} provide the ratio of average deviations during encounters to the corresponding values observed in undisturbed scenarios.

One notable observation regarding individuals is the consistent and steady increase in deviation during encounters across the majority of measures (see ratios of individuals in \tref{tab:ratio_euclidean_distance} $\sim$ \tref{tab:ratio_suddenness_turn}). Specifically, the ratios are consistently greater than 1 for all measures except suddenness of turn $\sigma$ and energy curvature $E_{\kappa}$, which we believe to be caused by the noise introduced by further differentiation when deriving the accelerations.

A similar increase is noted for dyads with 0 interaction levels, exhibiting generally large ration ($>1.5$), whereas higher interaction levels show a ratio close to 1, suggesting minimal change. Furthermore, we observe a decrease in the ratios with increasing interaction level, with statistically significant differences between various interaction levels. Additionally, a statistically significant difference is evident between the ratios of individuals and dyads after averaging across all interaction levels (see T-test $p$-values in the bottom line of \tref{tab:ratio_euclidean_distance} $\sim$ \tref{tab:ratio_suddenness_turn})

These findings are further clarified by \tref{tab:p_values_encounters_undisturbed_euclidean_distance} $\sim$ \tref{tab:p_values_encounters_undisturbed_suddenness_turn}, which provide $p$-values for the difference between undisturbed situations and encounters for both individuals and dyads across all interaction levels. Upon averaging over interaction levels, both individuals and dyads exhibit a statistically significant difference between undisturbed and encounter scenarios for all measures. However, while this disparity holds true for most measures for individuals even when considering the interaction level of the encountered dyad, it appears that only dyads with 0 interaction levels demonstrate a distinct behavior in encounters compared to undisturbed situations.

Finally, we show in \tref{tab:p_values_encounters_euclidean_distance} $\sim$  \fref{tab:p_values_encounters_suddenness_turn} the $p$-values for the difference of means between dyads with various level of interaction and individuals encountering these dyads. We see that the differences in the deviation between the dyads and the individuals are often significant for  interaction (levels 1, 2 and 3), but not for the non-interacting dyads.

The results presented thus far do not take into account the initial risk of collision, as we only set a threshold of 4~m to qualify an encounter. Among these encounters, some might have been more critical than others, with individual and dyad even facing a risk of collision. In the following section, we investigate the impact of the initial risk of collision on the deviation of individuals and dyads, by considering the impact parameter.

\subsection{Effect of the impact parameter}

As detailed in \sref{sec:impact_parameter}, the impact parameter is a measure of how close the individual would have passed the dyad if there were no collision avoidance behavior involved. In particular, $\bar{r}_b$ is a dimensionless measure that indicates the initial risk of collision, since it measures the distance between the individual and the dyad relative to the dyad's width\footnote{The width of the dyad is the average distance between the members of dyads with that level of interaction.}.

We chose to bin the values of $\bar{r}_b$ into 4 bins of equal size, which we interpret as follows: a value of $\bar{r}_b$ smaller than 1 indicates that the individual is on track to pass through or collide with the dyad. Between 1 and 2, the individual is close to the dyad, but not on track to pass through it. Between 2 and 3, the individual is further away from the dyad and may not need to deviate significantly to pass comfortably. Finally, a value of $\bar{r}_b$ larger than 3 indicates that the individual is far from the dyad and does not need to deviate.

\begin{figure}[htb]
    \centering
    \begin{subfigure}[t]{\textwidth}
        \centering
        \includegraphics[width=\textwidth]{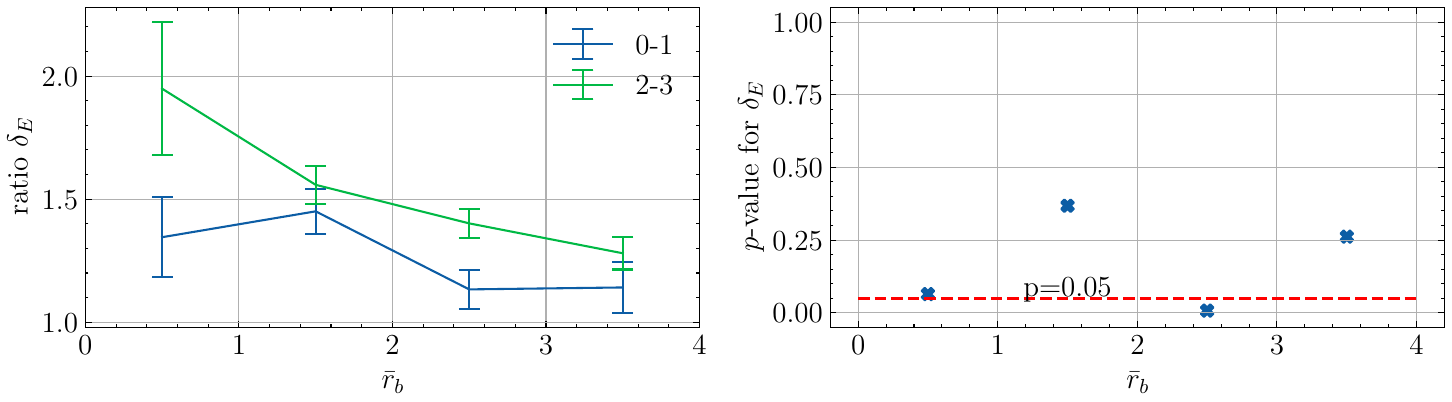}
        \caption{Individual}
        \label{fig:ratios_euclidean_distance_individuals_wrt_impact_parameter_01_23}
    \end{subfigure}
    \begin{subfigure}[t]{\textwidth}
        \centering
        \includegraphics[width=\textwidth]{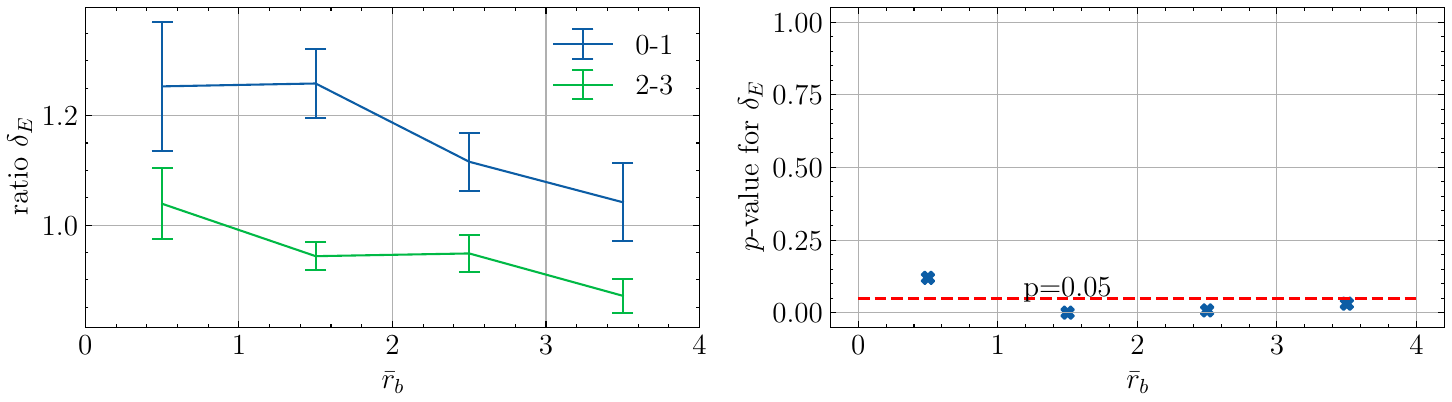}
        \caption{Dyad}
        \label{fig:ratios_euclidean_distance_groups_wrt_impact_parameter_01_23}
    \end{subfigure}
    \caption{Ratio of the value of the Euclidean distance $\delta_{E}$ in encounters to the undisturbed value for binned normalized impact parameter $\bar{r}_b$. The ratio are shown separately for encounters involving dyads with a low (0-1, in blue) and high (2-3, in green) level of interaction. The error bars represent the standard error of the mean. The $p$-values for the difference of means between 0-1 and 2-3 are also shown. The red dashed line represents the threshold $p=0.05$.}
    \label{fig:ratios_euclidean_distance_wrt_impact_parameter_01_23}
\end{figure}

\begin{figure}[htb]
    \centering
    \begin{subfigure}[t]{\textwidth}
        \centering
        \includegraphics[width=\textwidth]{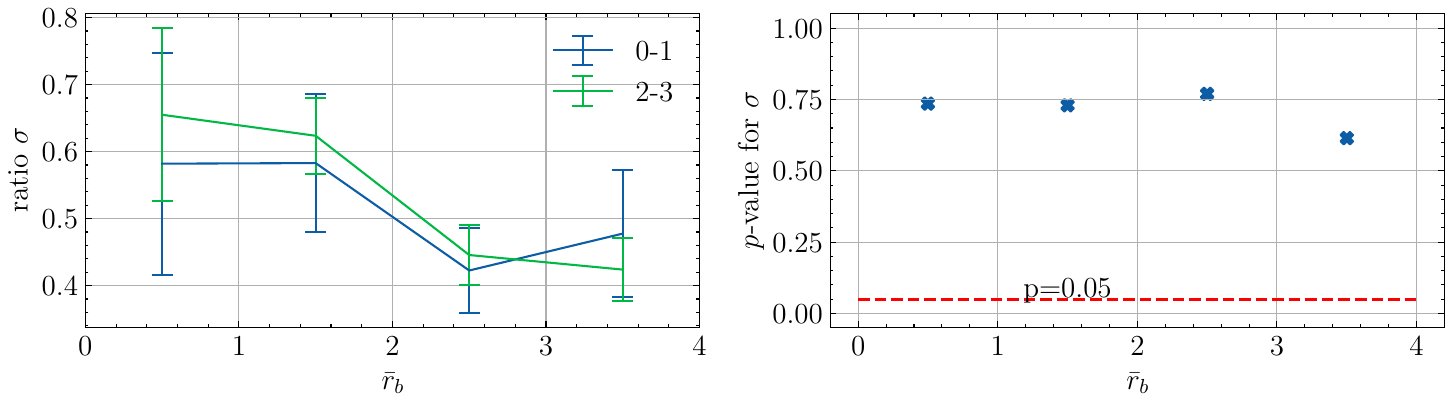}
        \caption{Individual}
        \label{fig:ratios_suddenness_turn_individuals_wrt_impact_parameter_01_23}
    \end{subfigure}
    \begin{subfigure}[t]{\textwidth}
        \centering
        \includegraphics[width=\textwidth]{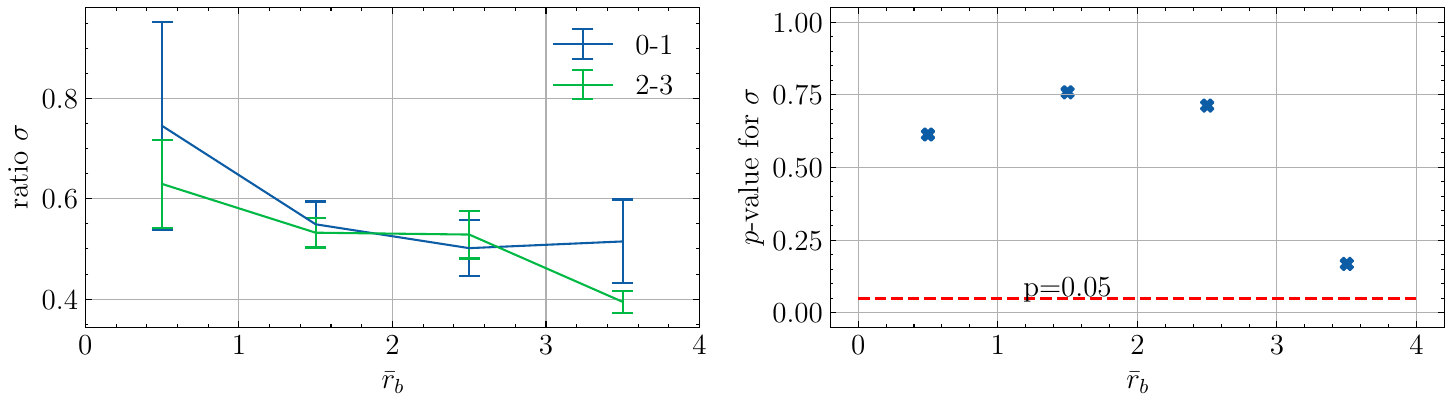}
        \caption{Dyad}
        \label{fig:ratios_suddenness_turn_groups_wrt_impact_parameter_01_23}
    \end{subfigure}
    \caption{Ratio of the value of the suddenness of turn $\sigma$ in encounters to the undisturbed value for binned normalized impact parameter $\bar{r}_b$. The ratio are shown separately for encounters involving dyads with a low (0-1, in blue) and high (2-3, in green) level of interaction. The error bars represent the standard error of the mean. The $p$-values for the difference of means between 0-1 and 2-3 are also shown. The red dashed line represents the threshold $p=0.05$.}
    \label{fig:ratios_suddenness_turn_wrt_impact_parameter_01_23}
\end{figure}

In \fref{fig:ratios_euclidean_distance_wrt_impact_parameter_01_23} $\sim$ \fref{fig:ratios_suddenness_turn_wrt_impact_parameter_01_23}, we illustrate the ratio of the average deviation during encounters to the average deviation in undisturbed situations for individuals and dyads with respect to the normalised impact parameter $\bar{r}_b$ for each measure. We also provide the T-test $p$-values for the difference in the ratio between dyads with low interaction levels (0 and 1) and dyads with high interaction levels (2 and 3) for each bin of $\bar{r}_b$.

Because the binning necessarily reduces the amount of data, in particular for interaction with already limited data (e.g.\ levels 0 and 3), we chose to separate encounters into two categories based on the level of interaction of the dyad: one with interaction levels 0 and 1, and another with levels 2 and 3. This helps balancing the number of data points in each and to get comparable sizes of samples for both classes. We argue that this categorisation is not unreasonable, as it allows to contrast low and high interaction levels.

For individuals, we observe that the ratio is generally\footnote{Except for the suddenness of turn $\sigma$.} higher when encountering dyads with higher interaction levels (2 and 3) compared to those with lower interaction levels (0 and 1). This is consistent with the results presented in \sref{sec:comparison_of_deviations}, where we observed that the deviation of individuals was higher when encountering dyads with higher interaction levels.
We observe that the associated T-test $p$-values generally follow a similar pattern for the position-based measures\footnote{Except for $\delta_{LCSS}$ and $\delta_{Lev}$.}, where the difference in the ratio between the two classes of dyads is statistically significant (or close to being significant) in the first and third bins of $\bar{r}_b$ (i.e.\ when the individual is on track to pass through the dyad or when the individual is relatively far from the dyad).

For the dyads, in accordance with the results of \sref{sec:comparison_of_deviations}, we observe that the ratio of the lower interaction levels (0 and 1) is systematically higher than that of the higher interaction levels (2 and 3) for all measures and all values of $\bar{r}_b$.
For all measures except $\sigma$, the difference in the ratio between the two classes of dyads is statistically significant in the second bin of $\bar{r}_b$ (i.e.\ when the individual is close to the dyad but not on track to pass through it).

\subsection{Correlation between measures}
\label{sec:method_correlation}

In~\sref{sec:toy_trajectories}, we demonstrated that different measures of deviation may capture distinct aspects of deviation. Some measures may be more sensitive to the magnitude of deviation, while others may be more sensitive to its abruptness. In this section, our objective is to assess the consistency of deviation measures to some extent.

\begin{figure}[htb]
    \centering
    \includegraphics[width=0.6\textwidth]{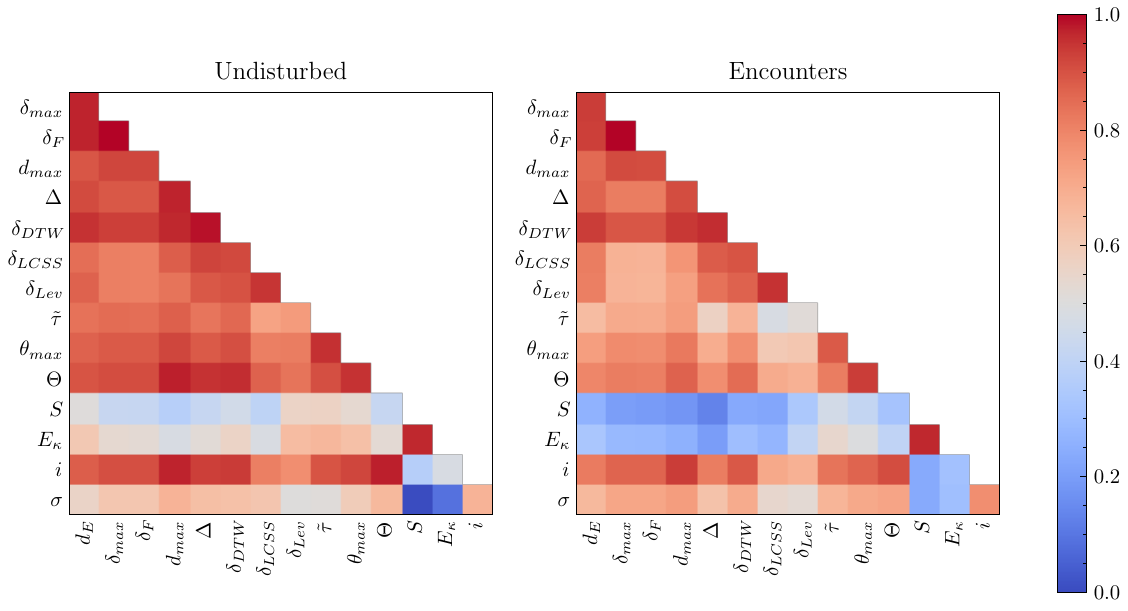}
    \caption{Spearman correlation between the measures of deviation.}
    \label{fig:correlation_measures}
\end{figure}

To that end, we illustrate Spearman correlation between the measures of deviation for all trajectories (dyads and individuals) in both undisturbed and encounter situations in \fref{fig:correlation_measures}. It is evident from this figure that the deviation measures generally exhibit correlation, with coefficients exceeding 0.5 for most measures.

Notably, sinuosity, curvature, and suddenness of turn display lower correlations with other measures. We attribute this observation to the fact that their computation involves some form of differentiation, such as deriving turning angles from velocity vectors for $S$, and calculating $\sigma$ and $E_{\kappa}$ using velocity and acceleration values.

We also note that the correlation between the measures is generally larger for the undisturbed situations than for the encounters. We believe this result to be partly due to the fact that the number of encounters is smaller than the number of undisturbed situations, which might make the correlation less reliable.

\begin{figure}[htb]
    \centering
    \includegraphics[width=0.8\textwidth]{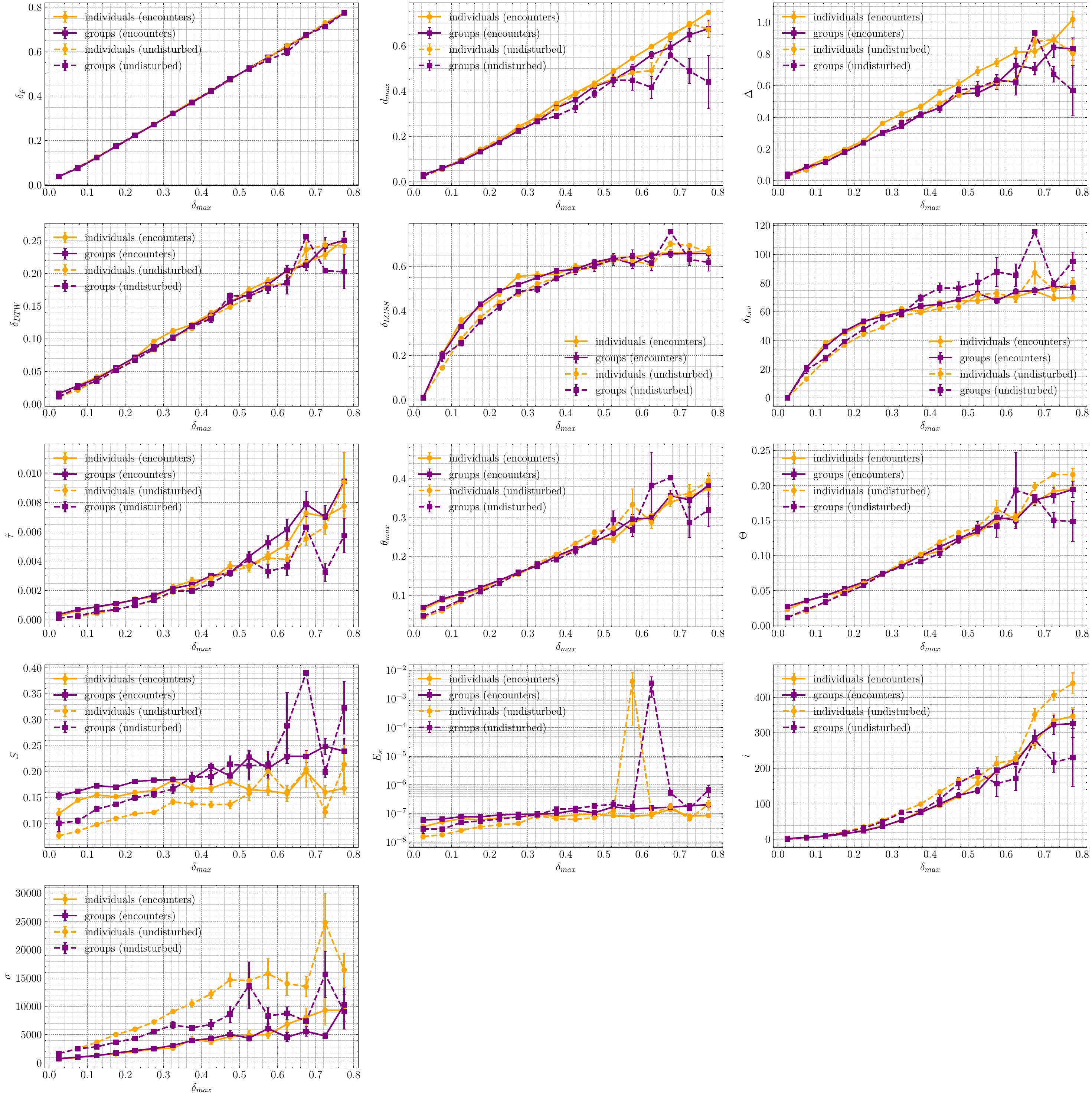}
    \caption{Average value of the different measures of deviation as a function of the lockstep maximum deviation. Values are binned in 16 bins of equal size. The scale of the y-axis is logarithmic for $E_{\kappa}$.}
    \label{fig:correlation_binned}
\end{figure}

To further investigate the correlation between the measures, we visualise the dependence of the different measures against the lockstep maximum deviation in \fref{fig:correlation_binned}. The deviation $\delta_{max}$ is binned in 16 bins of equal size, and the average value of the different measures is computed for each bin. We can see that the measures are all increasing with the lockstep maximum deviation, which is consistent with the idea that the measures are correlated. We see that although most measures seem to be linearly correlated with the lockstep maximum deviation, the rate of increase for $\delta_{LCSS}$ and $\delta_{Lev}$ slows down for large values of $\delta_{max}$. The main reason is that these two measures are dimensionless values (see \sref{sec:method_measures}) which are impacted by the number of points in the trajectory that are far away from the undisturbed trajectory rather than the actual amplitude of the deviation.

\section{Discussion}

We interpret the presented results considering two kinds of effects: the \textit{dynamic stability} of the considered entity (individual or dyad) and the amount of \textit{awareness} regarding the environment. We argue that the amount of deviation observed is a result of the interplay between these two factors.

The deviation of the non- and weakly interacting dyads is seen to be significantly smaller than the deviation of the individuals, which implies that dyads keep a more straight trajectory than individuals. This suggests that dyads maintain a more consistent trajectory compared to individuals, potentially due to their inclination to remain physically close, thereby constraining deviations from the intended path. Using a physical analogy, the inertia of the system composed of the two members of the dyad can be expected to be larger than the inertia of the individual, which would make the dyad more stable.

In addition, awareness about the changes in the environment is argued to be directly related to the level of interaction of the dyad. Namely, it is reasonable to assume that non interacting dyads in undisturbed situations are more attentive to their surroundings, since they do not need to allocate attention towards their social interaction. Consequently, in undisturbed scenarios, these dyads are more aware of the availability of economical straight paths and are more likely to adopt them compared to interacting dyads, which may deviate due to their internal -social interaction- dynamics.

During encounters, the fact that all dyads deviate significantly less than the individuals suggests that they contribute less to the avoidance process. In addition, the deviation of non-interacting dyads is higher than that of interacting dyads, which suggests that the level of interaction affects the dyad's ability to focus on the environment. This is further supported by the fact that the deviation of individuals increases significantly during encounters, regardless of the level of interaction of the dyad. This indicates that individuals anticipate (or react to) the diminished involvement of dyads in collision avoidance and adjust their deviations accordingly\footnote{Despite their simplicity, force based models such as the traditional social force model~\cite{helbing1995social} might predict such results, since the repulsive force perceived by the individual would be the sum of the repulsive forces perceived by the two members of the dyad}.

All the used measures show an increase in individuals' deviation when encountering dyads with higher interaction levels, although no measure suggests statistical significance.
Nevertheless, our prior investigation focusing on relative dynamics~\cite{gregorj2023social} revealed an interaction-level dependent statistical difference in encounters, particularly when such encounters could lead to a collision without deviation. By combining these findings with those of the current study, which show that in encounters between an interacting dyad and an individual, the latter bears the burden of collision avoidance, we can deduce that the individual's behavior is influenced by the dyad's level of interaction.

The impact parameter analysis provides further insights into the dynamics of the encounters. The observed patterns in the ratios of deviation during encounters to undisturbed situations for individuals and dyads with respect to the impact parameter suggest that the initial risk of collision influences the deviation of both individuals and dyads.

Concerning the individual, the difference between high and low interaction levels is most pronounced when the individual is on track to pass close to the dyad, with the deviation of the high interaction levels being significantly higher than that of the low interaction levels. We believe that is an effect of the intrusion phenomena that we observed in~\cite{gregorj2023social}, where the individual is more likely to pass through the dyad when the dyad is less reactive, essentially non-deviating and maintaining a straight trajectory. In situations where the encounter is expected to be close but not on a collision course, the deviation of the individual is less dependent on the dyad's level of interaction, which might be due to the fact that the individual has to deviate regardless of the dyad's involvement in social interaction. For less critical situations, the deviation of the individuals is again more dependent on the dyads' level of interaction.
We believe that this is due to the fact that in such situations the individuals mostly do not need to avoid low interacting dyads, since they walk on straight paths, while they more often need to avoid high interacting ones, due to their wandering behaviour.
%, which we attribute to the fact that, since there is less urgency to avoid a collision, the individual can adapt its deviation to the dyad's behavior, in particular since strongly interacting dyads tend to have less straight trajectories.} 
Finally, for high values of the impact parameter, the deviation of the individual does not depend on the dyad's level of interaction, since there is not much risk of collision and the individual essentially doesn't need to deviate.

For the dyad, the difference between high and low interaction levels is most pronounced when the individual and dyad are on track to pass close to each other, with the deviation of the low interaction levels being significantly higher than that of the high interaction levels. We hypothesise that this is where the effect of the dyad's awareness of the environment is most pronounced, as the dyad with low interaction levels is more likely to be focused on the environment and react to the individual's presence. For very low values of the impact parameter, there is no difference between the deviation of the dyads with different interaction levels, which might arise from the fact that the situation is so critical that the dyad has to react regardless of its involvement in social interaction by either opening up or deviating. At higher values of the impact parameter, the deviation of the dyads with different interaction levels is also similar, which might be due to the fact that the situation is less critical and the dyad can afford to be less reactive and the stability takes over.

% However, the results are not sufficiently definitive to presume a "Theory of Mind" effect (i.e., individuals' awareness of the dyad's inner state), as individuals may simply be dynamically reacting to the dyad's limited response.

While previous research has suggested that utilizing angular variables~\cite{fajen2003behavioral,murakami2021mutual} or velocity adjustments~\cite{huber2014adjustments,yamamoto2019body-rotation} to assess deviations from the intended path could offer a promising approach, our study faced limitations in this regard. As discussed in \sref{sec:method_correlation}, the trajectory data we utilised may not adequately support differentiation, resulting in certain measures (such as $\sigma$, $E_{\kappa}$, and $S$) being less effective in capturing deviations compared to others.
While controlled experiments can employ various wearable sensors to accurately register position or acceleration~\cite{tetsushi2010person, kaji2016estimating}, implementing such methods in real-world settings poses challenges. Environmental sensors may be sparse or susceptible to measurement noise and clutter, making it difficult to derive accurate values in naturalistic environments. Although higher-quality tracking data can be obtained in laboratory settings using wearable sensors, it remains uncertain whether the collected data accurately reflects the behaviour of human subjects in naturalistic settings, as discussed in~\sref{sec:introduction}~\cite{mobbs2021promises}. Moreover, with more detailed data, such as limb orientations, it could become feasible to investigate how pedestrians manage deviations. This could involve analyzing adjustments in step lengths or cephalocaudal reorientations~\cite{gerin2005negotiation, buhler2022coordinating}.

Evidence from ethology suggests that interactions among collectively moving non-human animals, such as bird flocks, are primarily influenced by topological distance rather than metric distance~\cite{ballerini2008interaction}. While there is ongoing debate regarding whether neighborhood in human motion is determined in a topological or metric manner~\cite{rio2018local}, it is important to consider that the conclusions drawn from ethological studies may not directly apply to our scenario. This is because collective behaviour, which is characterised by emergent complex patterns, typically arises in situations involving a large number of individuals, whereas our scenario involves low density.

% There remains a lot to be explored about the extent to which the addition of a cognitive load impacts on pedestrian collision avoidance. An evident extension to this work would be investigation of collision avoidance of triads. As mentioned above, they constitute the other building block of crowds and an understanding of triad dynamics may make it possible to modularly ...

%first decide goal position, then direct gaze towards the goal (because manual response is faster than saccadic response)~\cite{nummenmaa2009ll}, then follow head-to-tail motion~\cite{buhler2022coordinating} 

Research on collision avoidance often examines metrics such as the time or distance to the predicted point of collision~\cite{zanlungo2011social,karamouzas2014universal}. For instance, Olivier et al. define the Minimal Predicted Distance (MDP) as the anticipated closest distance between participants at a given time, assuming no adjustments are made to their velocity vectors~\cite{olivier2012minimal}. Similarly, Bhojwani et al. introduce the Theoretical Point of Collision (TPC), representing the point where a collision with an approaching virtual pedestrian would occur if no locomotor adjustments are made~\cite{bhojwani2022impact}. These metrics, akin to $d_{max}$, primarily focus on the point of collision rather than the deviation from the intended path. They also consider the involved parties in the encounter in a more integrated manner compared to our isolated approach. In future research, it would be beneficial to explore how deviation metrics evolve over time and their relationship with these collision distance measures.

Another aspect of pedestrian dynamics that should be investigated is the behaviour of overtaking or following others, whether they are individuals or groups.  Given the typically faster and more flexible nature of individuals compared to groups, it is reasonable to expect that individuals would engage in overtaking more frequently than groups. Moreover, individuals may tend to overtake groups more frequently than other individuals, again due to their greater speed and maneuverability. Interestingly, the decision of individuals to overtake groups may not necessarily depend on the social relation or interaction within the group. Since individuals approaching from behind may have limited visibility of the group's social features or intentions, their overtaking behaviour may primarily be driven by considerations such as speed and convenience rather than social dynamics.

\section{Conclusion}

Over the last few decades, numerous pedestrian collision avoidance models have emerged, offering insights into various aspects of pedestrian behavior~\cite{seyfried2006basics,karamouzas2014universal}. However, despite this extensive body of work, most microscopic models predominantly focus on one-on-one collision avoidance scenarios, neglecting the dynamics involving groups~\cite{nicolas2023social}. This oversight is notable considering that groups represent a substantial portion of crowds in real-world settings~\cite{moussaid2010walking,schultz2014group}. Thus, understanding the mechanisms underlying group-individual collision avoidance has remained a gap in pedestrian dynamics research.

Addressing this gap, our study aims to contribute to the understanding of how groups and individuals navigate and avoid collisions in crowded environments, shedding light on this understudied aspect of pedestrian behavior. In particular, by analyzing ecological data and considering various deviation measures, this study presents insights into the stability and deviation dynamics of dyads and  individuals.

In particular, concerning individuals, we demonstrate a heightened inclination to avoid interacting dyads compared to non-interacting ones. This suggests that individuals not only anticipate the future paths of others but also assess their contribution to collision avoidance. It is probable that upon detecting a lack of involvement, individuals adjust their behavior based on their internal model of human navigation.

Regarding dyads, our findings suggest that, in undisturbed situations, non-interacting ones, exhibit small deviations from their intended trajectories, indicating a high level of efficiency and stability. Conversely, interacting dyads show larger deviations,  possibly attributable to being more consumed with their internal dynamics. Additionally, interacting dyads are less impacted by encounters compared to individuals, while non-interacting dyads display a similar pattern to individuals.

Based on these observations, we believe that being part of a group affects the behavior of a pedestrian in two, possibly contrasting, ways. A dynamical effect, i.e.\ the tendency to move together, appears to increase the stability and economy of motion. On the other hand, a social effect diverts attention from the environment, resulting in a less economical movement, in a way similar to how the usage of smart phones affects pedestrian behavior~\cite{murakami2021mutual}.

We believe that this study not only provides valuable insights into the complex dynamics of group-individual collision avoidance, but it also has tangible implications and has the potential to open up  new avenues of technical  improvement.
%, e.g.\ for improving the mobility and safety of  public space users  crowd simulations, robot navigation, and understanding human behavior in various contexts. 
In particular, it holds practical significance in various real-world scenarios, such as urban planning and crowd management. Namely, with such insights, we can inform the development of more effective pedestrian flow management strategies, improve the design of urban spaces to leverage safety and mobility efficiency, and enhance the planning of events or gatherings to reduce congestion and potential hazards. Additionally, our results can help the development of intelligent systems for autonomous vehicles, robots, or ubiquitous technology, providing safer and more seamless interactions with human pedestrians in diverse smart environments.

\begin{table}[!htb]
    \caption{Average and standard deviation of the Euclidean distance $\delta_{E}$ (in~m), for dyads and individuals in undisturbed situations. The Kruskal-Wallis $p$-value for the difference of means between dyads with various level of interaction and the Welch T-test $p$-value for the difference of means between all individuals against all dyads are also shown.}
    \label{tab:undisturbed_euclidean_distance}
    \begin{center}
\begin{tabular}{lccc}
            \toprule
            Intensity of interaction & $\delta_{E}$                                & Normalized   & $p$-value                                       \\
            \midrule
            0                        & $6.74 \times 10^{-2}\pm4.54 \times 10^{-2}$ & $0.80$       & \multirow{4}{*}{$\mathbf{3.21 \times 10^{-4}}$} \\
            1                        & $8.71 \times 10^{-2}\pm5.71 \times 10^{-2}$ & $1.03$       &                                                 \\
            2                        & $9.93 \times 10^{-2}\pm8.97 \times 10^{-2}$ & $1.18$       &                                                 \\
            3                        & $1.22 \times 10^{-1}\pm7.54 \times 10^{-2}$ & $1.44$       &                                                 \\
            \midrule
            All dyads                & $1.00 \times 10^{-1}\pm8.31 \times 10^{-2}$ & $\num{1.19}$ & \multirow{2}{*}{$\mathbf{2.18 \times 10^{-4}}$} \\
            Individuals              & $8.44 \times 10^{-2}\pm7.86 \times 10^{-2}$ & $\num{1.00}$ &                                                 \\
            \bottomrule
        \end{tabular}
    \end{center}
\end{table}
\begin{table}[!htb]
    \caption{Average and standard deviation of the lockstep maximum deviation $\delta_{max}$ (in~m), for dyads and individuals in undisturbed situations. The Kruskal-Wallis $p$-value for the difference of means between dyads with various level of interaction and the Welch T-test $p$-value for the difference of means between all individuals against all dyads are also shown.}
    \label{tab:undisturbed_simultaneous_frechet_deviation}
    \begin{center}
        \begin{tabular}{lccc}
            \toprule
            Intensity of interaction & $\delta_{max}$                              & Normalized   & $p$-value                                       \\
            \midrule
            0                        & $1.89 \times 10^{-1}\pm1.32 \times 10^{-1}$ & $0.77$       & \multirow{4}{*}{$\mathbf{1.01 \times 10^{-3}}$} \\
            1                        & $2.49 \times 10^{-1}\pm1.66 \times 10^{-1}$ & $1.01$       &                                                 \\
            2                        & $2.69 \times 10^{-1}\pm2.07 \times 10^{-1}$ & $1.09$       &                                                 \\
            3                        & $3.21 \times 10^{-1}\pm1.83 \times 10^{-1}$ & $1.31$       &                                                 \\
            \midrule
            All dyads                & $2.72 \times 10^{-1}\pm1.97 \times 10^{-1}$ & $\num{1.11}$ & \multirow{2}{*}{$\mathbf{9.37 \times 10^{-3}}$} \\
            Individuals              & $2.46 \times 10^{-1}\pm1.92 \times 10^{-1}$ & $\num{1.00}$ &                                                 \\
            \bottomrule
        \end{tabular}
    \end{center}
\end{table}
\begin{table}[!htb]
    \caption{Average and standard deviation of the Fréchet deviation $\delta_{F}$ (in~m), for dyads and individuals in undisturbed situations. The Kruskal-Wallis $p$-value for the difference of means between dyads with various level of interaction and the Welch T-test $p$-value for the difference of means between all individuals against all dyads are also shown.}
    \label{tab:undisturbed_frechet_deviation}
    \begin{center}
        \begin{tabular}{lccc}
            \toprule
            Intensity of interaction & $\delta_{F}$                                & Normalized   & $p$-value                                       \\
            \midrule
            0                        & $1.89 \times 10^{-1}\pm1.32 \times 10^{-1}$ & $0.77$       & \multirow{4}{*}{$\mathbf{1.21 \times 10^{-3}}$} \\
            1                        & $2.48 \times 10^{-1}\pm1.66 \times 10^{-1}$ & $1.01$       &                                                 \\
            2                        & $2.65 \times 10^{-1}\pm1.98 \times 10^{-1}$ & $1.08$       &                                                 \\
            3                        & $3.17 \times 10^{-1}\pm1.80 \times 10^{-1}$ & $1.29$       &                                                 \\
            \midrule
            All dyads                & $2.69 \times 10^{-1}\pm1.91 \times 10^{-1}$ & $\num{1.10}$ & \multirow{2}{*}{$\mathbf{1.60 \times 10^{-2}}$} \\
            Individuals              & $2.45 \times 10^{-1}\pm1.92 \times 10^{-1}$ & $\num{1.00}$ &                                                 \\
            \bottomrule
        \end{tabular}
    \end{center}
\end{table}
\begin{table}[!htb]
    \caption{Average and standard deviation of the maximum lateral deviation $d_{max}$ (in~m), for dyads and individuals in undisturbed situations. The Kruskal-Wallis $p$-value for the difference of means between dyads with various level of interaction and the Welch T-test $p$-value for the difference of means between all individuals against all dyads are also shown.}
    \label{tab:undisturbed_max_lateral_deviation}
    \begin{center}
        \begin{tabular}{lccc}
            \toprule
            Intensity of interaction & $d_{max}$                                   & Normalized   & $p$-value                                       \\
            \midrule
            0                        & $1.52 \times 10^{-1}\pm1.30 \times 10^{-1}$ & $0.75$       & \multirow{4}{*}{$\mathbf{2.27 \times 10^{-2}}$} \\
            1                        & $1.99 \times 10^{-1}\pm1.50 \times 10^{-1}$ & $0.98$       &                                                 \\
            2                        & $2.02 \times 10^{-1}\pm1.36 \times 10^{-1}$ & $1.00$       &                                                 \\
            3                        & $2.37 \times 10^{-1}\pm1.50 \times 10^{-1}$ & $1.17$       &                                                 \\
            \midrule
            All dyads                & $2.06 \times 10^{-1}\pm1.41 \times 10^{-1}$ & $\num{1.02}$ & \multirow{2}{*}{$6.76 \times 10^{-1}$}          \\
            Individuals              & $2.03 \times 10^{-1}\pm1.49 \times 10^{-1}$ & $\num{1.00}$ &                                                 \\
            \bottomrule
        \end{tabular}
    \end{center}
\end{table}
\begin{table}[!htb]
    \caption{Average and standard deviation of the integral of lateral deviation $\Delta$ (in~m$^2$), for dyads and individuals in undisturbed situations. The Kruskal-Wallis $p$-value for the difference of means between dyads with various level of interaction and the Welch T-test $p$-value for the difference of means between all individuals against all dyads are also shown.}
    \label{tab:undisturbed_integral_deviation}
    \begin{center}
        \begin{tabular}{lccc}
            \toprule
            Intensity of interaction & $\Delta$                                    & Normalized   & $p$-value                                       \\
            \midrule
            0                        & $2.07 \times 10^{-1}\pm1.76 \times 10^{-1}$ & $0.79$       & \multirow{4}{*}{$\mathbf{1.39 \times 10^{-2}}$} \\
            1                        & $2.62 \times 10^{-1}\pm1.95 \times 10^{-1}$ & $0.99$       &                                                 \\
            2                        & $2.81 \times 10^{-1}\pm2.02 \times 10^{-1}$ & $1.07$       &                                                 \\
            3                        & $3.37 \times 10^{-1}\pm2.26 \times 10^{-1}$ & $1.28$       &                                                 \\
            \midrule
            All dyads                & $2.85 \times 10^{-1}\pm2.07 \times 10^{-1}$ & $\num{1.08}$ & \multirow{2}{*}{$\mathbf{4.22 \times 10^{-2}}$} \\
            Individuals              & $2.63 \times 10^{-1}\pm2.05 \times 10^{-1}$ & $\num{1.00}$ &                                                 \\
            \bottomrule
        \end{tabular}
    \end{center}
\end{table}
\begin{table}[!htb]
    \caption{Average and standard deviation of the dynamic time warping deviation $\delta_{DTW}$ (in~m), for dyads and individuals in undisturbed situations. The Kruskal-Wallis $p$-value for the difference of means between dyads with various level of interaction and the Welch T-test $p$-value for the difference of means between all individuals against all dyads are also shown.}
    \label{tab:undisturbed_dynamic_time_warping_deviation}
    \begin{center}
        \begin{tabular}{lccc}
            \toprule
            Intensity of interaction & $\delta_{DTW}$                              & Normalized   & $p$-value                                       \\
            \midrule
            0                        & $5.88 \times 10^{-2}\pm4.70 \times 10^{-2}$ & $0.78$       & \multirow{4}{*}{$\mathbf{6.02 \times 10^{-3}}$} \\
            1                        & $7.58 \times 10^{-2}\pm5.55 \times 10^{-2}$ & $1.00$       &                                                 \\
            2                        & $8.23 \times 10^{-2}\pm6.35 \times 10^{-2}$ & $1.09$       &                                                 \\
            3                        & $9.71 \times 10^{-2}\pm6.17 \times 10^{-2}$ & $1.28$       &                                                 \\
            \midrule
            All dyads                & $8.31 \times 10^{-2}\pm6.21 \times 10^{-2}$ & $\num{1.10}$ & \multirow{2}{*}{$\mathbf{2.64 \times 10^{-2}}$} \\
            Individuals              & $7.57 \times 10^{-2}\pm7.36 \times 10^{-2}$ & $\num{1.00}$ &                                                 \\
            \bottomrule
        \end{tabular}
    \end{center}
\end{table}
\begin{table}[!htb]
    \caption{Average and standard deviation of the longest common subsequence deviation $\delta_{LCSS}$ (dimensionless), for dyads and individuals in undisturbed situations. The Kruskal-Wallis $p$-value for the difference of means between dyads with various level of interaction and the Welch T-test $p$-value for the difference of means between all individuals against all dyads are also shown.}
    \label{tab:undisturbed_lcss_deviation}
    \begin{center}
        \begin{tabular}{lccc}
            \toprule
            Intensity of interaction & $\delta_{LCSS}$                             & Normalized   & $p$-value                                       \\
            \midrule
            0                        & $3.08 \times 10^{-1}\pm2.10 \times 10^{-1}$ & $0.77$       & \multirow{4}{*}{$\mathbf{3.63 \times 10^{-2}}$} \\
            1                        & $4.01 \times 10^{-1}\pm1.63 \times 10^{-1}$ & $1.01$       &                                                 \\
            2                        & $4.16 \times 10^{-1}\pm1.70 \times 10^{-1}$ & $1.05$       &                                                 \\
            3                        & $4.53 \times 10^{-1}\pm1.65 \times 10^{-1}$ & $1.14$       &                                                 \\
            \midrule
            All dyads                & $4.17 \times 10^{-1}\pm1.72 \times 10^{-1}$ & $\num{1.05}$ & \multirow{2}{*}{$\mathbf{4.17 \times 10^{-2}}$} \\
            Individuals              & $3.98 \times 10^{-1}\pm1.77 \times 10^{-1}$ & $\num{1.00}$ &                                                 \\
            \bottomrule
        \end{tabular}
    \end{center}
\end{table}
\begin{table}[!htb]
    \caption{Average and standard deviation of the Levenshtein deviation $\delta_{Lev}$ (dimensionless), for dyads and individuals in undisturbed situations. The Kruskal-Wallis $p$-value for the difference of means between dyads with various level of interaction and the Welch T-test $p$-value for the difference of means between all individuals against all dyads are also shown.}
    \label{tab:undisturbed_edit_distance_deviation}
    \begin{center}
        \begin{tabular}{lccc}
            \toprule
            Intensity of interaction & $\delta_{Lev}$                    & Normalized   & $p$-value                                       \\
            \midrule
            0                        & $3.31 \times 10\pm2.38 \times 10$ & $0.80$       & \multirow{4}{*}{$\mathbf{4.81 \times 10^{-4}}$} \\
            1                        & $4.33 \times 10\pm1.99 \times 10$ & $1.04$       &                                                 \\
            2                        & $5.03 \times 10\pm2.65 \times 10$ & $1.21$       &                                                 \\
            3                        & $5.84 \times 10\pm2.76 \times 10$ & $1.41$       &                                                 \\
            \midrule
            All dyads                & $5.01 \times 10\pm2.64 \times 10$ & $\num{1.21}$ & \multirow{2}{*}{$\mathbf{< 10^{-4}}$}           \\
            Individuals              & $4.14 \times 10\pm2.19 \times 10$ & $\num{1.00}$ &                                                 \\
            \bottomrule
        \end{tabular}
    \end{center}
\end{table}
\begin{table}[!htb]
    \caption{Average and standard deviation of the deviation index $\tilde{\tau}$ (dimensionless), for dyads and individuals in undisturbed situations. The Kruskal-Wallis $p$-value for the difference of means between dyads with various level of interaction and the Welch T-test $p$-value for the difference of means between all individuals against all dyads are also shown.}
    \label{tab:undisturbed_straightness_index}
    \begin{center}
        \begin{tabular}{lccc}
            \toprule
            Intensity of interaction & $\tilde{\tau}$                              & Normalized   & $p$-value                                       \\
            \midrule
            0                        & $8.40 \times 10^{-4}\pm8.84 \times 10^{-4}$ & $0.58$       & \multirow{4}{*}{$\mathbf{5.46 \times 10^{-3}}$} \\
            1                        & $1.30 \times 10^{-3}\pm1.59 \times 10^{-3}$ & $0.89$       &                                                 \\
            2                        & $1.73 \times 10^{-3}\pm5.65 \times 10^{-3}$ & $1.19$       &                                                 \\
            3                        & $1.77 \times 10^{-3}\pm1.63 \times 10^{-3}$ & $1.21$       &                                                 \\
            \midrule
            All dyads                & $1.65 \times 10^{-3}\pm4.66 \times 10^{-3}$ & $\num{1.13}$ & \multirow{2}{*}{$4.19 \times 10^{-1}$}          \\
            Individuals              & $1.46 \times 10^{-3}\pm3.85 \times 10^{-3}$ & $\num{1.00}$ &                                                 \\
            \bottomrule
        \end{tabular}
    \end{center}
\end{table}
\begin{table}[!htb]
    \caption{Average and standard deviation of the maximum cumulative turning angle $\theta_{max}$ (in~rad), for dyads and individuals in undisturbed situations. The Kruskal-Wallis $p$-value for the difference of means between dyads with various level of interaction and the Welch T-test $p$-value for the difference of means between all individuals against all dyads are also shown.}
    \label{tab:undisturbed_max_cumulative_turning_angle}
    \begin{center}
        \begin{tabular}{lccc}
            \toprule
            Intensity of interaction & $\theta_{max}$                              & Normalized   & $p$-value                                       \\
            \midrule
            0                        & $1.18 \times 10^{-1}\pm6.98 \times 10^{-2}$ & $0.81$       & \multirow{4}{*}{$\mathbf{3.92 \times 10^{-3}}$} \\
            1                        & $1.37 \times 10^{-1}\pm6.87 \times 10^{-2}$ & $0.95$       &                                                 \\
            2                        & $1.56 \times 10^{-1}\pm1.92 \times 10^{-1}$ & $1.08$       &                                                 \\
            3                        & $1.83 \times 10^{-1}\pm1.38 \times 10^{-1}$ & $1.27$       &                                                 \\
            \midrule
            All dyads                & $1.57 \times 10^{-1}\pm1.69 \times 10^{-1}$ & $\num{1.08}$ & \multirow{2}{*}{$1.44 \times 10^{-1}$}          \\
            Individuals              & $1.45 \times 10^{-1}\pm1.24 \times 10^{-1}$ & $\num{1.00}$ &                                                 \\
            \bottomrule
        \end{tabular}
    \end{center}
\end{table}
\begin{table}[!htb]
    \caption{Average and standard deviation of the integral cumulative turning angle $\Theta$ (in~rad), for dyads and individuals in undisturbed situations. The Kruskal-Wallis $p$-value for the difference of means between dyads with various level of interaction and the Welch T-test $p$-value for the difference of means between all individuals against all dyads are also shown.}
    \label{tab:undisturbed_integral_cumulative_turning_angle}
    \begin{center}
        \begin{tabular}{lccc}
            \toprule
            Intensity of interaction & $\Theta$                                    & Normalized   & $p$-value                                       \\
            \midrule
            0                        & $5.12 \times 10^{-2}\pm3.93 \times 10^{-2}$ & $0.77$       & \multirow{4}{*}{$\mathbf{9.48 \times 10^{-3}}$} \\
            1                        & $6.52 \times 10^{-2}\pm4.58 \times 10^{-2}$ & $0.97$       &                                                 \\
            2                        & $6.91 \times 10^{-2}\pm6.78 \times 10^{-2}$ & $1.03$       &                                                 \\
            3                        & $8.53 \times 10^{-2}\pm8.10 \times 10^{-2}$ & $1.27$       &                                                 \\
            \midrule
            All dyads                & $7.07 \times 10^{-2}\pm6.74 \times 10^{-2}$ & $\num{1.06}$ & \multirow{2}{*}{$2.58 \times 10^{-1}$}          \\
            Individuals              & $6.69 \times 10^{-2}\pm5.37 \times 10^{-2}$ & $\num{1.00}$ &                                                 \\
            \bottomrule
        \end{tabular}
    \end{center}
\end{table}
\begin{table}[!htb]
    \caption{Average and standard deviation of the sinuosity $S$ (in rad/m$^{\frac{1}{2}}$), for dyads and individuals in undisturbed situations. The Kruskal-Wallis $p$-value for the difference of means between dyads with various level of interaction and the Welch T-test $p$-value for the difference of means between all individuals against all dyads are also shown.}
    \label{tab:undisturbed_sinuosity}
    \begin{center}
        \begin{tabular}{lccc}
            \toprule
            Intensity of interaction & $S$                                         & Normalized   & $p$-value                                       \\
            \midrule
            0                        & $1.41 \times 10^{-1}\pm4.63 \times 10^{-2}$ & $1.15$       & \multirow{4}{*}{$\mathbf{1.03 \times 10^{-2}}$} \\
            1                        & $1.40 \times 10^{-1}\pm3.75 \times 10^{-2}$ & $1.15$       &                                                 \\
            2                        & $1.73 \times 10^{-1}\pm2.42 \times 10^{-1}$ & $1.42$       &                                                 \\
            3                        & $1.83 \times 10^{-1}\pm1.02 \times 10^{-1}$ & $1.50$       &                                                 \\
            \midrule
            All dyads                & $1.69 \times 10^{-1}\pm2.01 \times 10^{-1}$ & $\num{1.39}$ & \multirow{2}{*}{$\mathbf{< 10^{-4}}$}           \\
            Individuals              & $1.22 \times 10^{-1}\pm1.04 \times 10^{-1}$ & $\num{1.00}$ &                                                 \\
            \bottomrule
        \end{tabular}
    \end{center}
\end{table}
\begin{table}[!htb]
    \caption{Average and standard deviation of the energy curvature $E_{\kappa}$ (in s/m$^2$), for dyads and individuals in undisturbed situations. The Kruskal-Wallis $p$-value for the difference of means between dyads with various level of interaction and the Welch T-test $p$-value for the difference of means between all individuals against all dyads are also shown.}
    \label{tab:undisturbed_energy_curvature}
    \begin{center}
        \begin{tabular}{lccc}
            \toprule
            Intensity of interaction & $E_{\kappa}$                                & Normalized            & $p$-value                                       \\
            \midrule
            0                        & $5.79 \times 10^{-8}\pm4.68 \times 10^{-8}$ & $5.85 \times 10^{-5}$ & \multirow{4}{*}{$\mathbf{2.53 \times 10^{-3}}$} \\
            1                        & $5.64 \times 10^{-8}\pm4.55 \times 10^{-8}$ & $5.71 \times 10^{-5}$ &                                                 \\
            2                        & $1.99 \times 10^{-3}\pm3.43 \times 10^{-2}$ & $2.01$                &                                                 \\
            3                        & $4.48 \times 10^{-4}\pm2.95 \times 10^{-3}$ & $0.45$                &                                                 \\
            \midrule
            All dyads                & $1.38 \times 10^{-3}\pm2.78 \times 10^{-2}$ & $\num{1.40}$          & \multirow{2}{*}{$7.80 \times 10^{-1}$}          \\
            Individuals              & $9.88 \times 10^{-4}\pm2.34 \times 10^{-2}$ & $\num{1.00}$          &                                                 \\
            \bottomrule
        \end{tabular}
    \end{center}
\end{table}
\begin{table}[!htb]
    \caption{Average and standard deviation of the turn intensity $i$ (in~deg$\times$cm), for dyads and individuals in undisturbed situations. The Kruskal-Wallis $p$-value for the difference of means between dyads with various level of interaction and the Welch T-test $p$-value for the difference of means between all individuals against all dyads are also shown.}
    \label{tab:undisturbed_turn_intensity}
    \begin{center}
        \begin{tabular}{lccc}
            \toprule
            Intensity of interaction & $i$                                   & Normalized   & $p$-value                                       \\
            \midrule
            0                        & $3.38 \times 10\pm5.04 \times 10$     & $0.59$       & \multirow{4}{*}{$\mathbf{6.51 \times 10^{-3}}$} \\
            1                        & $5.59 \times 10\pm1.06 \times 10^{2}$ & $0.98$       &                                                 \\
            2                        & $5.50 \times 10\pm8.71 \times 10$     & $0.96$       &                                                 \\
            3                        & $7.42 \times 10\pm9.49 \times 10$     & $1.30$       &                                                 \\
            \midrule
            All dyads                & $5.77 \times 10\pm9.04 \times 10$     & $\num{1.01}$ & \multirow{2}{*}{$9.29 \times 10^{-1}$}          \\
            Individuals              & $5.72 \times 10\pm1.05 \times 10^{2}$ & $\num{1.00}$ &                                                 \\
            \bottomrule
        \end{tabular}
    \end{center}
\end{table}
\begin{table}[!htb]
    \caption{Average and standard deviation of the suddenness of turn $\sigma$ (in~deg$\times$cm/s), for dyads and individuals in undisturbed situations. The Kruskal-Wallis $p$-value for the difference of means between dyads with various level of interaction and the Welch T-test $p$-value for the difference of means between all individuals against all dyads are also shown.}
    \label{tab:undisturbed_suddenness_turn}
    \begin{center}
        \begin{tabular}{lccc}
            \toprule
            Intensity of interaction & $\sigma$                                  & Normalized   & $p$-value                              \\
            \midrule
            0                        & $5.04 \times 10^{3}\pm6.02 \times 10^{3}$ & $0.75$       & \multirow{4}{*}{$1.38 \times 10^{-1}$} \\
            1                        & $5.10 \times 10^{3}\pm4.30 \times 10^{3}$ & $0.76$       &                                        \\
            2                        & $5.15 \times 10^{3}\pm4.30 \times 10^{3}$ & $0.76$       &                                        \\
            3                        & $5.29 \times 10^{3}\pm3.30 \times 10^{3}$ & $0.79$       &                                        \\
            \midrule
            All dyads                & $5.17 \times 10^{3}\pm4.22 \times 10^{3}$ & $\num{0.77}$ & \multirow{2}{*}{$\mathbf{< 10^{-4}}$}  \\
            Individuals              & $6.74 \times 10^{3}\pm6.39 \times 10^{3}$ & $\num{1.00}$ &                                        \\
            \bottomrule
        \end{tabular}
    \end{center}
\end{table}
\clearpage
\begin{table}[!htb]
    \caption{Average and standard deviation of the Euclidean distance $\delta_{E}$ (in~m), for dyads and individuals during encounters. Kruskal-Wallis $p$-values for the difference of means between dyads (and individuals encountering dyads) with various level of interaction and the Welch T-test $p$-value for the difference of means between all individuals against all dyads are also shown.}
    \label{tab:encounter_euclidean_distance}
    \begin{center}
        \begin{tabular}{lcccc}
            \toprule
                             & \multicolumn{2}{c}{Individual}              & \multicolumn{2}{c}{Dyad}                                                                                                               \\
            \midrule
            Intensity of     & $\delta_{E}$                                & Kruskal-Wallis                         & $\delta_{E}$                                & Kruskal-Wallis                                  \\
            interaction      &                                             & $p$-value                              &                                             & $p$-value                                       \\
            \midrule
            0                & $1.25 \times 10^{-1}\pm8.73 \times 10^{-2}$ & \multirow{4}{*}{$9.58 \times 10^{-1}$} & $1.02 \times 10^{-1}\pm5.90 \times 10^{-2}$ & \multirow{4}{*}{$\mathbf{2.61 \times 10^{-2}}$} \\
            1                & $1.16 \times 10^{-1}\pm7.02 \times 10^{-2}$ &                                        & $8.91 \times 10^{-2}\pm6.56 \times 10^{-2}$ &                                                 \\
            2                & $1.30 \times 10^{-1}\pm1.14 \times 10^{-1}$ &                                        & $9.65 \times 10^{-2}\pm7.05 \times 10^{-2}$ &                                                 \\
            3                & $1.31 \times 10^{-1}\pm1.00 \times 10^{-1}$ &                                        & $1.09 \times 10^{-1}\pm8.43 \times 10^{-2}$ &                                                 \\
            \midrule
            All              & $1.28 \times 10^{-1}\pm1.05 \times 10^{-1}$ &                                        & $9.78 \times 10^{-2}\pm7.17 \times 10^{-2}$ &                                                 \\
\bottomrule
        \end{tabular}
    \end{center}
\end{table}
\begin{table}[!htb]
    \caption{Average and standard deviation of the lockstep maximum deviation $\delta_{max}$ (in~m), for dyads and individuals during encounters. Kruskal-Wallis $p$-values for the difference of means between dyads (and individuals encountering dyads) with various level of interaction and the Welch T-test $p$-value for the difference of means between all individuals against all dyads are also shown.}
    \label{tab:encounter_simultaneous_frechet_deviation}
    \begin{center}
        \begin{tabular}{lcccc}
            \toprule
                             & \multicolumn{2}{c}{Individual}              & \multicolumn{2}{c}{Dyad}                                                                                                               \\
            \midrule
            Intensity of     & $\delta_{max}$                              & Kruskal-Wallis                         & $\delta_{max}$                              & Kruskal-Wallis                                  \\
            interaction      &                                             & $p$-value                              &                                             & $p$-value                                       \\
            \midrule
            0                & $3.53 \times 10^{-1}\pm2.60 \times 10^{-1}$ & \multirow{4}{*}{$9.55 \times 10^{-1}$} & $2.92 \times 10^{-1}\pm1.78 \times 10^{-1}$ & \multirow{4}{*}{$\mathbf{2.94 \times 10^{-2}}$} \\
            1                & $3.50 \times 10^{-1}\pm2.23 \times 10^{-1}$ &                                        & $2.42 \times 10^{-1}\pm1.87 \times 10^{-1}$ &                                                 \\
            2                & $3.76 \times 10^{-1}\pm3.21 \times 10^{-1}$ &                                        & $2.64 \times 10^{-1}\pm2.10 \times 10^{-1}$ &                                                 \\
            3                & $3.78 \times 10^{-1}\pm2.86 \times 10^{-1}$ &                                        & $2.90 \times 10^{-1}\pm2.30 \times 10^{-1}$ &                                                 \\
            \midrule
            All              & $3.71 \times 10^{-1}\pm2.99 \times 10^{-1}$ &                                        & $2.67 \times 10^{-1}\pm2.09 \times 10^{-1}$ &                                                 \\
\bottomrule
        \end{tabular}
    \end{center}
\end{table}
\begin{table}[!htb]
    \caption{Average and standard deviation of the Fréchet deviation $\delta_{F}$ (in~m), for dyads and individuals during encounters. Kruskal-Wallis $p$-values for the difference of means between dyads (and individuals encountering dyads) with various level of interaction and the Welch T-test $p$-value for the difference of means between all individuals against all dyads are also shown.}
    \label{tab:encounter_frechet_deviation}
    \begin{center}
        \begin{tabular}{lcccc}
            \toprule
                             & \multicolumn{2}{c}{Individual}              & \multicolumn{2}{c}{Dyad}                                                                                                               \\
            \midrule
            Intensity of     & $\delta_{F}$                                & Kruskal-Wallis                         & $\delta_{F}$                                & Kruskal-Wallis                                  \\
            interaction      &                                             & $p$-value                              &                                             & $p$-value                                       \\
            \midrule
            0                & $3.52 \times 10^{-1}\pm2.61 \times 10^{-1}$ & \multirow{4}{*}{$9.55 \times 10^{-1}$} & $2.92 \times 10^{-1}\pm1.79 \times 10^{-1}$ & \multirow{4}{*}{$\mathbf{2.62 \times 10^{-2}}$} \\
            1                & $3.50 \times 10^{-1}\pm2.23 \times 10^{-1}$ &                                        & $2.41 \times 10^{-1}\pm1.88 \times 10^{-1}$ &                                                 \\
            2                & $3.75 \times 10^{-1}\pm3.21 \times 10^{-1}$ &                                        & $2.62 \times 10^{-1}\pm2.11 \times 10^{-1}$ &                                                 \\
            3                & $3.78 \times 10^{-1}\pm2.87 \times 10^{-1}$ &                                        & $2.89 \times 10^{-1}\pm2.31 \times 10^{-1}$ &                                                 \\
            \midrule
            All              & $3.70 \times 10^{-1}\pm2.99 \times 10^{-1}$ &                                        & $2.66 \times 10^{-1}\pm2.10 \times 10^{-1}$ &                                                 \\
\bottomrule
        \end{tabular}
    \end{center}
\end{table}
\begin{table}[!htb]
    \caption{Average and standard deviation of the maximum lateral deviation $d_{max}$ (in~m), for dyads and individuals during encounters. Kruskal-Wallis $p$-values for the difference of means between dyads (and individuals encountering dyads) with various level of interaction and the Welch T-test $p$-value for the difference of means between all individuals against all dyads are also shown.}
    \label{tab:encounter_max_lateral_deviation}
    \begin{center}
        \begin{tabular}{lcccc}
            \toprule
                             & \multicolumn{2}{c}{Individual}              & \multicolumn{2}{c}{Dyad}                                                                                                               \\
            \midrule
            Intensity of     & $d_{max}$                                   & Kruskal-Wallis                         & $d_{max}$                                   & Kruskal-Wallis                                  \\
            interaction      &                                             & $p$-value                              &                                             & $p$-value                                       \\
            \midrule
            0                & $3.24 \times 10^{-1}\pm2.58 \times 10^{-1}$ & \multirow{4}{*}{$9.86 \times 10^{-1}$} & $2.62 \times 10^{-1}\pm1.77 \times 10^{-1}$ & \multirow{4}{*}{$\mathbf{1.20 \times 10^{-2}}$} \\
            1                & $3.18 \times 10^{-1}\pm2.11 \times 10^{-1}$ &                                        & $2.06 \times 10^{-1}\pm1.75 \times 10^{-1}$ &                                                 \\
            2                & $3.35 \times 10^{-1}\pm2.63 \times 10^{-1}$ &                                        & $2.11 \times 10^{-1}\pm1.70 \times 10^{-1}$ &                                                 \\
            3                & $3.41 \times 10^{-1}\pm2.89 \times 10^{-1}$ &                                        & $2.41 \times 10^{-1}\pm2.20 \times 10^{-1}$ &                                                 \\
            \midrule
            All              & $3.33 \times 10^{-1}\pm2.60 \times 10^{-1}$ &                                        & $2.19 \times 10^{-1}\pm1.81 \times 10^{-1}$ &                                                 \\
\bottomrule
        \end{tabular}
    \end{center}
\end{table}
\begin{table}[!htb]
    \caption{Average and standard deviation of the integral of lateral deviation $\Delta$ (in~m$^2$), for dyads and individuals during encounters. Kruskal-Wallis $p$-values for the difference of means between dyads (and individuals encountering dyads) with various level of interaction and the Welch T-test $p$-value for the difference of means between all individuals against all dyads are also shown.}
    \label{tab:encounter_integral_deviation}
    \begin{center}
        \begin{tabular}{lcccc}
            \toprule
                             & \multicolumn{2}{c}{Individual}              & \multicolumn{2}{c}{Dyad}                                                                                                      \\
            \midrule
            Intensity of     & $\Delta$                                    & Kruskal-Wallis                         & $\Delta$                                    & Kruskal-Wallis                         \\
            interaction      &                                             & $p$-value                              &                                             & $p$-value                              \\
            \midrule
            0                & $4.45 \times 10^{-1}\pm3.82 \times 10^{-1}$ & \multirow{4}{*}{$8.82 \times 10^{-1}$} & $3.26 \times 10^{-1}\pm2.41 \times 10^{-1}$ & \multirow{4}{*}{$8.92 \times 10^{-2}$} \\
            1                & $4.13 \times 10^{-1}\pm2.82 \times 10^{-1}$ &                                        & $2.72 \times 10^{-1}\pm2.37 \times 10^{-1}$ &                                        \\
            2                & $4.54 \times 10^{-1}\pm3.69 \times 10^{-1}$ &                                        & $2.66 \times 10^{-1}\pm2.25 \times 10^{-1}$ &                                        \\
            3                & $4.52 \times 10^{-1}\pm4.34 \times 10^{-1}$ &                                        & $3.11 \times 10^{-1}\pm3.14 \times 10^{-1}$ &                                        \\
            \midrule
            All              & $4.47 \times 10^{-1}\pm3.70 \times 10^{-1}$ &                                        & $2.78 \times 10^{-1}\pm2.45 \times 10^{-1}$ &                                        \\
\bottomrule
        \end{tabular}
    \end{center}
\end{table}
\begin{table}[!htb]
    \caption{Average and standard deviation of the dynamic time warping deviation $\delta_{DTW}$ (in~m), for dyads and individuals during encounters. Kruskal-Wallis $p$-values for the difference of means between dyads (and individuals encountering dyads) with various level of interaction and the Welch T-test $p$-value for the difference of means between all individuals against all dyads are also shown.}
    \label{tab:encounter_dynamic_time_warping_deviation}
    \begin{center}
        \begin{tabular}{lcccc}
            \toprule
                             & \multicolumn{2}{c}{Individual}              & \multicolumn{2}{c}{Dyad}                                                                                                      \\
            \midrule
            Intensity of     & $\delta_{DTW}$                              & Kruskal-Wallis                         & $\delta_{DTW}$                              & Kruskal-Wallis                         \\
            interaction      &                                             & $p$-value                              &                                             & $p$-value                              \\
            \midrule
            0                & $1.17 \times 10^{-1}\pm9.04 \times 10^{-2}$ & \multirow{4}{*}{$9.73 \times 10^{-1}$} & $9.43 \times 10^{-2}\pm6.23 \times 10^{-2}$ & \multirow{4}{*}{$5.21 \times 10^{-2}$} \\
            1                & $1.09 \times 10^{-1}\pm6.83 \times 10^{-2}$ &                                        & $7.93 \times 10^{-2}\pm6.49 \times 10^{-2}$ &                                        \\
            2                & $1.21 \times 10^{-1}\pm1.12 \times 10^{-1}$ &                                        & $8.25 \times 10^{-2}\pm6.81 \times 10^{-2}$ &                                        \\
            3                & $1.21 \times 10^{-1}\pm1.04 \times 10^{-1}$ &                                        & $9.47 \times 10^{-2}\pm8.50 \times 10^{-2}$ &                                        \\
            \midrule
            All              & $1.19 \times 10^{-1}\pm1.04 \times 10^{-1}$ &                                        & $8.48 \times 10^{-2}\pm7.04 \times 10^{-2}$ &                                        \\
\bottomrule
        \end{tabular}
    \end{center}
\end{table}
\begin{table}[!htb]
    \caption{Average and standard deviation of the longest common subsequence deviation $\delta_{LCSS}$ (dimensionless), for dyads and individuals during encounters. Kruskal-Wallis $p$-values for the difference of means between dyads (and individuals encountering dyads) with various level of interaction and the Welch T-test $p$-value for the difference of means between all individuals against all dyads are also shown.}
    \label{tab:encounter_lcss_deviation}
    \begin{center}
        \begin{tabular}{lcccc}
            \toprule
                             & \multicolumn{2}{c}{Individual}              & \multicolumn{2}{c}{Dyad}                                                                                                      \\
            \midrule
            Intensity of     & $\delta_{LCSS}$                             & Kruskal-Wallis                         & $\delta_{LCSS}$                             & Kruskal-Wallis                         \\
            interaction      &                                             & $p$-value                              &                                             & $p$-value                              \\
            \midrule
            0                & $5.02 \times 10^{-1}\pm1.95 \times 10^{-1}$ & \multirow{4}{*}{$7.63 \times 10^{-1}$} & $4.72 \times 10^{-1}\pm1.92 \times 10^{-1}$ & \multirow{4}{*}{$2.54 \times 10^{-1}$} \\
            1                & $4.86 \times 10^{-1}\pm1.90 \times 10^{-1}$ &                                        & $4.22 \times 10^{-1}\pm2.29 \times 10^{-1}$ &                                        \\
            2                & $5.03 \times 10^{-1}\pm1.97 \times 10^{-1}$ &                                        & $4.33 \times 10^{-1}\pm2.19 \times 10^{-1}$ &                                        \\
            3                & $4.97 \times 10^{-1}\pm2.01 \times 10^{-1}$ &                                        & $4.54 \times 10^{-1}\pm2.15 \times 10^{-1}$ &                                        \\
            \midrule
            All              & $5.00 \times 10^{-1}\pm1.96 \times 10^{-1}$ &                                        & $4.38 \times 10^{-1}\pm2.19 \times 10^{-1}$ &                                        \\
\bottomrule
        \end{tabular}
    \end{center}
\end{table}
\begin{table}[!htb]
    \caption{Average and standard deviation of the Levenshtein deviation $\delta_{Lev}$ (dimensionless), for dyads and individuals during encounters. Kruskal-Wallis $p$-values for the difference of means between dyads (and individuals encountering dyads) with various level of interaction and the Welch T-test $p$-value for the difference of means between all individuals against all dyads are also shown.}
    \label{tab:encounter_edit_distance_deviation}
    \begin{center}
        \begin{tabular}{lcccc}
            \toprule
                             & \multicolumn{2}{c}{Individual}           & \multicolumn{2}{c}{Dyad}                                                                                                     \\
            \midrule
            Intensity of     & $\delta_{Lev}$                           & Kruskal-Wallis                         & $\delta_{Lev}$                    & Kruskal-Wallis                                  \\
            interaction      &                                          & $p$-value                              &                                   & $p$-value                                       \\
            \midrule
            0                & $5.22 \times 10\pm2.19 \times 10$        & \multirow{4}{*}{$1.73 \times 10^{-1}$} & $4.90 \times 10\pm2.07 \times 10$ & \multirow{4}{*}{$\mathbf{1.16 \times 10^{-2}}$} \\
            1                & $5.05 \times 10\pm2.05 \times 10$        &                                        & $4.44 \times 10\pm2.47 \times 10$ &                                                 \\
            2                & $5.49 \times 10\pm2.29 \times 10$        &                                        & $4.78 \times 10\pm2.50 \times 10$ &                                                 \\
            3                & $5.76 \times 10\pm2.56 \times 10$        &                                        & $5.29 \times 10\pm2.60 \times 10$ &                                                 \\
            \midrule
            All              & $5.45 \times 10\pm2.31 \times 10$        &                                        & $4.82 \times 10\pm2.49 \times 10$ &                                                 \\
            T-test $p$-value & \multicolumn{4}{c}{$\mathbf{< 10^{-4}}$}                                                                                                                                \\
            \bottomrule
        \end{tabular}
    \end{center}
\end{table}
\begin{table}[!htb]
    \caption{Average and standard deviation of the deviation index $\tilde{\tau}$ (dimensionless), for dyads and individuals during encounters. Kruskal-Wallis $p$-values for the difference of means between dyads (and individuals encountering dyads) with various level of interaction and the Welch T-test $p$-value for the difference of means between all individuals against all dyads are also shown.}
    \label{tab:encounter_straightness_index}
    \begin{center}
        \begin{tabular}{lcccc}
            \toprule
                             & \multicolumn{2}{c}{Individual}              & \multicolumn{2}{c}{Dyad}                                                                                                               \\
            \midrule
            Intensity of     & $\tilde{\tau}$                              & Kruskal-Wallis                         & $\tilde{\tau}$                              & Kruskal-Wallis                                  \\
            interaction      &                                             & $p$-value                              &                                             & $p$-value                                       \\
            \midrule
            0                & $2.84 \times 10^{-3}\pm3.01 \times 10^{-3}$ & \multirow{4}{*}{$6.50 \times 10^{-1}$} & $2.52 \times 10^{-3}\pm2.92 \times 10^{-3}$ & \multirow{4}{*}{$\mathbf{1.82 \times 10^{-3}}$} \\
            1                & $2.92 \times 10^{-3}\pm3.65 \times 10^{-3}$ &                                        & $1.72 \times 10^{-3}\pm2.43 \times 10^{-3}$ &                                                 \\
            2                & $3.39 \times 10^{-3}\pm4.96 \times 10^{-3}$ &                                        & $2.24 \times 10^{-3}\pm5.60 \times 10^{-3}$ &                                                 \\
            3                & $3.73 \times 10^{-3}\pm4.59 \times 10^{-3}$ &                                        & $2.57 \times 10^{-3}\pm3.46 \times 10^{-3}$ &                                                 \\
            \midrule
            All              & $3.33 \times 10^{-3}\pm4.62 \times 10^{-3}$ &                                        & $2.24 \times 10^{-3}\pm4.79 \times 10^{-3}$ &                                                 \\
\bottomrule
        \end{tabular}
    \end{center}
\end{table}
\begin{table}[!htb]
    \caption{Average and standard deviation of the maximum cumulative turning angle $\theta_{max}$ (in~rad), for dyads and individuals during encounters. Kruskal-Wallis $p$-values for the difference of means between dyads (and individuals encountering dyads) with various level of interaction and the Welch T-test $p$-value for the difference of means between all individuals against all dyads are also shown.}
    \label{tab:encounter_max_cumulative_turning_angle}
    \begin{center}
        \begin{tabular}{lcccc}
            \toprule
                             & \multicolumn{2}{c}{Individual}              & \multicolumn{2}{c}{Dyad}                                                                                                               \\
            \midrule
            Intensity of     & $\theta_{max}$                              & Kruskal-Wallis                         & $\theta_{max}$                              & Kruskal-Wallis                                  \\
            interaction      &                                             & $p$-value                              &                                             & $p$-value                                       \\
            \midrule
            0                & $1.95 \times 10^{-1}\pm1.02 \times 10^{-1}$ & \multirow{4}{*}{$9.39 \times 10^{-1}$} & $1.76 \times 10^{-1}\pm8.65 \times 10^{-2}$ & \multirow{4}{*}{$\mathbf{1.10 \times 10^{-2}}$} \\
            1                & $1.92 \times 10^{-1}\pm1.07 \times 10^{-1}$ &                                        & $1.49 \times 10^{-1}\pm9.05 \times 10^{-2}$ &                                                 \\
            2                & $1.97 \times 10^{-1}\pm1.18 \times 10^{-1}$ &                                        & $1.63 \times 10^{-1}\pm1.22 \times 10^{-1}$ &                                                 \\
            3                & $2.09 \times 10^{-1}\pm1.37 \times 10^{-1}$ &                                        & $1.73 \times 10^{-1}\pm1.04 \times 10^{-1}$ &                                                 \\
            \midrule
            All              & $1.98 \times 10^{-1}\pm1.19 \times 10^{-1}$ &                                        & $1.63 \times 10^{-1}\pm1.13 \times 10^{-1}$ &                                                 \\
\bottomrule
        \end{tabular}
    \end{center}
\end{table}
\begin{table}[!htb]
    \caption{Average and standard deviation of the integral cumulative turning angle $\Theta$ (in~rad), for dyads and individuals during encounters. Kruskal-Wallis $p$-values for the difference of means between dyads (and individuals encountering dyads) with various level of interaction and the Welch T-test $p$-value for the difference of means between all individuals against all dyads are also shown.}
    \label{tab:encounter_integral_cumulative_turning_angle}
    \begin{center}
        \begin{tabular}{lcccc}
            \toprule
                             & \multicolumn{2}{c}{Individual}              & \multicolumn{2}{c}{Dyad}                                                                                                               \\
            \midrule
            Intensity of     & $\Theta$                                    & Kruskal-Wallis                         & $\Theta$                                    & Kruskal-Wallis                                  \\
            interaction      &                                             & $p$-value                              &                                             & $p$-value                                       \\
            \midrule
            0                & $9.81 \times 10^{-2}\pm6.41 \times 10^{-2}$ & \multirow{4}{*}{$9.42 \times 10^{-1}$} & $8.34 \times 10^{-2}\pm4.82 \times 10^{-2}$ & \multirow{4}{*}{$\mathbf{1.25 \times 10^{-2}}$} \\
            1                & $9.10 \times 10^{-2}\pm5.31 \times 10^{-2}$ &                                        & $7.08 \times 10^{-2}\pm5.25 \times 10^{-2}$ &                                                 \\
            2                & $9.63 \times 10^{-2}\pm6.69 \times 10^{-2}$ &                                        & $7.53 \times 10^{-2}\pm5.30 \times 10^{-2}$ &                                                 \\
            3                & $1.02 \times 10^{-1}\pm7.52 \times 10^{-2}$ &                                        & $8.33 \times 10^{-2}\pm5.77 \times 10^{-2}$ &                                                 \\
            \midrule
            All              & $9.66 \times 10^{-2}\pm6.64 \times 10^{-2}$ &                                        & $7.65 \times 10^{-2}\pm5.35 \times 10^{-2}$ &                                                 \\
\bottomrule
        \end{tabular}
    \end{center}
\end{table}
\begin{table}[!htb]
    \caption{Average and standard deviation of the sinuosity $S$ (in rad/m$^{\frac{1}{2}}$), for dyads and individuals during encounters. Kruskal-Wallis $p$-values for the difference of means between dyads (and individuals encountering dyads) with various level of interaction and the Welch T-test $p$-value for the difference of means between all individuals against all dyads are also shown.}
    \label{tab:encounter_sinuosity}
    \begin{center}
        \begin{tabular}{lcccc}
            \toprule
                             & \multicolumn{2}{c}{Individual}              & \multicolumn{2}{c}{Dyad}                                                                                                              \\
            \midrule
            Intensity of     & $S$                                         & Kruskal-Wallis                                  & $S$                                         & Kruskal-Wallis                        \\
            interaction      &                                             & $p$-value                                       &                                             & $p$-value                             \\
            \midrule
            0                & $1.56 \times 10^{-1}\pm5.93 \times 10^{-2}$ & \multirow{4}{*}{$\mathbf{9.19 \times 10^{-3}}$} & $1.73 \times 10^{-1}\pm6.24 \times 10^{-2}$ & \multirow{4}{*}{$\mathbf{< 10^{-4}}$} \\
            1                & $1.54 \times 10^{-1}\pm5.92 \times 10^{-2}$ &                                                 & $1.64 \times 10^{-1}\pm5.30 \times 10^{-2}$ &                                       \\
            2                & $1.65 \times 10^{-1}\pm6.67 \times 10^{-2}$ &                                                 & $1.95 \times 10^{-1}\pm2.40 \times 10^{-1}$ &                                       \\
            3                & $1.86 \times 10^{-1}\pm7.48 \times 10^{-2}$ &                                                 & $2.08 \times 10^{-1}\pm6.93 \times 10^{-2}$ &                                       \\
            \midrule
            All              & $1.66 \times 10^{-1}\pm6.72 \times 10^{-2}$ &                                                 & $1.91 \times 10^{-1}\pm1.94 \times 10^{-1}$ &                                       \\
\bottomrule
        \end{tabular}
    \end{center}
\end{table}
\begin{table}[!htb]
    \caption{Average and standard deviation of the energy curvature $E_{\kappa}$ (in s/m$^2$), for dyads and individuals during encounters. Kruskal-Wallis $p$-values for the difference of means between dyads (and individuals encountering dyads) with various level of interaction and the Welch T-test $p$-value for the difference of means between all individuals against all dyads are also shown.}
    \label{tab:encounter_energy_curvature}
    \begin{center}
        \begin{tabular}{lcccc}
            \toprule
                             & \multicolumn{2}{c}{Individual}              & \multicolumn{2}{c}{Dyad}                                                                                                              \\
            \midrule
            Intensity of     & $E_{\kappa}$                                & Kruskal-Wallis                                  & $E_{\kappa}$                                & Kruskal-Wallis                        \\
            interaction      &                                             & $p$-value                                       &                                             & $p$-value                             \\
            \midrule
            0                & $7.08 \times 10^{-8}\pm6.99 \times 10^{-8}$ & \multirow{4}{*}{$\mathbf{3.80 \times 10^{-3}}$} & $8.08 \times 10^{-8}\pm6.12 \times 10^{-8}$ & \multirow{4}{*}{$\mathbf{< 10^{-4}}$} \\
            1                & $6.59 \times 10^{-8}\pm6.36 \times 10^{-8}$ &                                                 & $6.96 \times 10^{-8}\pm5.68 \times 10^{-8}$ &                                       \\
            2                & $8.63 \times 10^{-8}\pm1.55 \times 10^{-7}$ &                                                 & $2.32 \times 10^{-4}\pm6.27 \times 10^{-3}$ &                                       \\
            3                & $1.11 \times 10^{-7}\pm1.18 \times 10^{-7}$ &                                                 & $1.30 \times 10^{-7}\pm1.12 \times 10^{-7}$ &                                       \\
            \midrule
            All              & $8.61 \times 10^{-8}\pm1.35 \times 10^{-7}$ &                                                 & $1.45 \times 10^{-4}\pm4.95 \times 10^{-3}$ &                                       \\
\bottomrule
        \end{tabular}
    \end{center}
\end{table}
\begin{table}[!htb]
    \caption{Average and standard deviation of the turn intensity $i$ (in~deg$\times$cm), for dyads and individuals during encounters. Kruskal-Wallis $p$-values for the difference of means between dyads (and individuals encountering dyads) with various level of interaction and the Welch T-test $p$-value for the difference of means between all individuals against all dyads are also shown.}
    \label{tab:encounter_turn_intensity}
    \begin{center}
        \begin{tabular}{lcccc}
            \toprule
                             & \multicolumn{2}{c}{Individual}            & \multicolumn{2}{c}{Dyad}                                                                                                         \\
            \midrule
            Intensity of     & $i$                                       & Kruskal-Wallis                         & $i$                                   & Kruskal-Wallis                                  \\
            interaction      &                                           & $p$-value                              &                                       & $p$-value                                       \\
            \midrule
            0                & $1.13 \times 10^{2}\pm1.97 \times 10^{2}$ & \multirow{4}{*}{$9.91 \times 10^{-1}$} & $7.20 \times 10\pm9.57 \times 10$     & \multirow{4}{*}{$\mathbf{1.69 \times 10^{-2}}$} \\
            1                & $9.34 \times 10\pm1.30 \times 10^{2}$     &                                        & $5.47 \times 10\pm1.11 \times 10^{2}$ &                                                 \\
            2                & $1.18 \times 10^{2}\pm2.09 \times 10^{2}$ &                                        & $5.40 \times 10\pm9.60 \times 10$     &                                                 \\
            3                & $1.22 \times 10^{2}\pm2.29 \times 10^{2}$ &                                        & $7.52 \times 10\pm1.71 \times 10^{2}$ &                                                 \\
            \midrule
            All              & $1.15 \times 10^{2}\pm2.03 \times 10^{2}$ &                                        & $5.88 \times 10\pm1.13 \times 10^{2}$ &                                                 \\
\bottomrule
        \end{tabular}
    \end{center}
\end{table}
\begin{table}[!htb]
    \caption{Average and standard deviation of the suddenness of turn $\sigma$ (in~deg$\times$cm/s), for dyads and individuals during encounters. Kruskal-Wallis $p$-values for the difference of means between dyads (and individuals encountering dyads) with various level of interaction and the Welch T-test $p$-value for the difference of means between all individuals against all dyads are also shown.}
    \label{tab:encounter_suddenness_turn}
    \begin{center}
        \begin{tabular}{lcccc}
            \toprule
                             & \multicolumn{2}{c}{Individual}                     & \multicolumn{2}{c}{Dyad}                                                                                                    \\
            \midrule
            Intensity of     & $\sigma$                                           & Kruskal-Wallis                         & $\sigma$                                  & Kruskal-Wallis                         \\
            interaction      &                                                    & $p$-value                              &                                           & $p$-value                              \\
            \midrule
            0                & $3.95 \times 10^{3}\pm6.09 \times 10^{3}$          & \multirow{4}{*}{$7.47 \times 10^{-1}$} & $3.15 \times 10^{3}\pm3.48 \times 10^{3}$ & \multirow{4}{*}{$5.19 \times 10^{-2}$} \\
            1                & $3.94 \times 10^{3}\pm7.39 \times 10^{3}$          &                                        & $2.74 \times 10^{3}\pm4.31 \times 10^{3}$ &                                        \\
            2                & $3.92 \times 10^{3}\pm5.94 \times 10^{3}$          &                                        & $2.89 \times 10^{3}\pm4.44 \times 10^{3}$ &                                        \\
            3                & $3.24 \times 10^{3}\pm5.13 \times 10^{3}$          &                                        & $2.18 \times 10^{3}\pm1.97 \times 10^{3}$ &                                        \\
            \midrule
            All              & $3.82 \times 10^{3}\pm6.07 \times 10^{3}$          &                                        & $2.77 \times 10^{3}\pm4.07 \times 10^{3}$ &                                        \\
\bottomrule
        \end{tabular}
    \end{center}
\end{table}
\clearpage
\begin{table}[!htb]
    \caption{Ratio of the average value for Euclidean distance $\delta_{E}$, for dyads and individuals during encounters to the undisturbed value. Kruskal-Wallis $p$-values for the difference of means between dyads (and individuals encountering dyads) with various level of interaction and the Welch T-test $p$-value for the difference of means between all individuals against all dyads are also shown.}
    \label{tab:ratio_euclidean_distance}
    \begin{center}
        \begin{tabular}{lcccc}
            \toprule
                             & \multicolumn{2}{c}{Individual}           & \multicolumn{2}{c}{Dyad}                                                                            \\
            \midrule
            Intensity of     & Ratio $\delta_{E}$                       & Kruskal-Wallis                         & Ratio $\delta_{E}$ & Kruskal-Wallis                        \\
            interaction      &                                          & $p$-value                              &                    & $p$-value                             \\
            \midrule
            0                & $1.49$                                   & \multirow{4}{*}{$9.58 \times 10^{-1}$} & $1.52$             & \multirow{4}{*}{$\mathbf{< 10^{-4}}$} \\
            1                & $1.38$                                   &                                        & $1.02$             &                                       \\
            2                & $1.54$                                   &                                        & $0.97$             &                                       \\
            3                & $1.55$                                   &                                        & $0.90$             &                                       \\
            \midrule
            T-test $p$-value & \multicolumn{4}{c}{$\mathbf{< 10^{-4}}$}                                                                                                       \\
            \bottomrule
        \end{tabular}
    \end{center}
\end{table}
\begin{table}[!htb]
    \caption{Ratio of the average value for lockstep maximum deviation $\delta_{max}$, for dyads and individuals during encounters to the undisturbed value. Kruskal-Wallis $p$-values for the difference of means between dyads (and individuals encountering dyads) with various level of interaction and the Welch T-test $p$-value for the difference of means between all individuals against all dyads are also shown.}
    \label{tab:ratio_simultaneous_frechet_deviation}
    \begin{center}
        \begin{tabular}{lcccc}
            \toprule
                             & \multicolumn{2}{c}{Individual}           & \multicolumn{2}{c}{Dyad}                                                                              \\
            \midrule
            Intensity of     & Ratio $\delta_{max}$                     & Kruskal-Wallis                         & Ratio $\delta_{max}$ & Kruskal-Wallis                        \\
            interaction      &                                          & $p$-value                              &                      & $p$-value                             \\
            \midrule
            0                & $1.44$                                   & \multirow{4}{*}{$9.55 \times 10^{-1}$} & $1.54$               & \multirow{4}{*}{$\mathbf{< 10^{-4}}$} \\
            1                & $1.43$                                   &                                        & $0.97$               &                                       \\
            2                & $1.53$                                   &                                        & $0.98$               &                                       \\
            3                & $1.54$                                   &                                        & $0.91$               &                                       \\
            \midrule
            T-test $p$-value & \multicolumn{4}{c}{$\mathbf{< 10^{-4}}$}                                                                                                         \\
            \bottomrule
        \end{tabular}
    \end{center}
\end{table}
\begin{table}[!htb]
    \caption{Ratio of the average value for Fréchet deviation $\delta_{F}$, for dyads and individuals during encounters to the undisturbed value. Kruskal-Wallis $p$-values for the difference of means between dyads (and individuals encountering dyads) with various level of interaction and the Welch T-test $p$-value for the difference of means between all individuals against all dyads are also shown.}
    \label{tab:ratio_frechet_deviation}
    \begin{center}
        %
    \end{center}
\end{table}
\begin{table}[!htb]
    \caption{Ratio of the average value for maximum lateral deviation $d_{max}$, for dyads and individuals during encounters to the undisturbed value. Kruskal-Wallis $p$-values for the difference of means between dyads (and individuals encountering dyads) with various level of interaction and the Welch T-test $p$-value for the difference of means between all individuals against all dyads are also shown.}
    \label{tab:ratio_max_lateral_deviation}
    \begin{center}
        %
    \end{center}
\end{table}
\begin{table}[!htb]
    \caption{Ratio of the average value for integral of lateral deviation $\Delta$, for dyads and individuals during encounters to the undisturbed value. Kruskal-Wallis $p$-values for the difference of means between dyads (and individuals encountering dyads) with various level of interaction and the Welch T-test $p$-value for the difference of means between all individuals against all dyads are also shown.}
    \label{tab:ratio_integral_deviation}
    \begin{center}
        %
    \end{center}
\end{table}
\begin{table}[!htb]
    \caption{Ratio of the average value for dynamic time warping deviation $\delta_{DTW}$, for dyads and individuals during encounters to the undisturbed value. Kruskal-Wallis $p$-values for the difference of means between dyads (and individuals encountering dyads) with various level of interaction and the Welch T-test $p$-value for the difference of means between all individuals against all dyads are also shown.}
    \label{tab:ratio_dynamic_time_warping_deviation}
    \begin{center}
        %
    \end{center}
\end{table}
\begin{table}[!htb]
    \caption{Ratio of the average value for longest common subsequence deviation $\delta_{LCSS}$, for dyads and individuals during encounters to the undisturbed value. Kruskal-Wallis $p$-values for the difference of means between dyads (and individuals encountering dyads) with various level of interaction and the Welch T-test $p$-value for the difference of means between all individuals against all dyads are also shown.}
    \label{tab:ratio_lcss_deviation}
    \begin{center}
        %
    \end{center}
\end{table}
\begin{table}[!htb]
    \caption{Ratio of the average value for Levenshtein deviation $\delta_{Lev}$, for dyads and individuals during encounters to the undisturbed value. Kruskal-Wallis $p$-values for the difference of means between dyads (and individuals encountering dyads) with various level of interaction and the Welch T-test $p$-value for the difference of means between all individuals against all dyads are also shown.}
    \label{tab:ratio_edit_distance_deviation}
    \begin{center}
        %
    \end{center}
\end{table}
\begin{table}[!htb]
    \caption{Ratio of the average value for deviation index $\tilde{\tau}$, for dyads and individuals during encounters to the undisturbed value. Kruskal-Wallis $p$-values for the difference of means between dyads (and individuals encountering dyads) with various level of interaction and the Welch T-test $p$-value for the difference of means between all individuals against all dyads are also shown.}
    \label{tab:ratio_straightness_index}
    \begin{center}
        %
    \end{center}
\end{table}
\clearpage

\begin{figure}[htb]
    \centering
    \begin{subfigure}[t]{\textwidth}
        \centering
        \includegraphics[width=\textwidth]{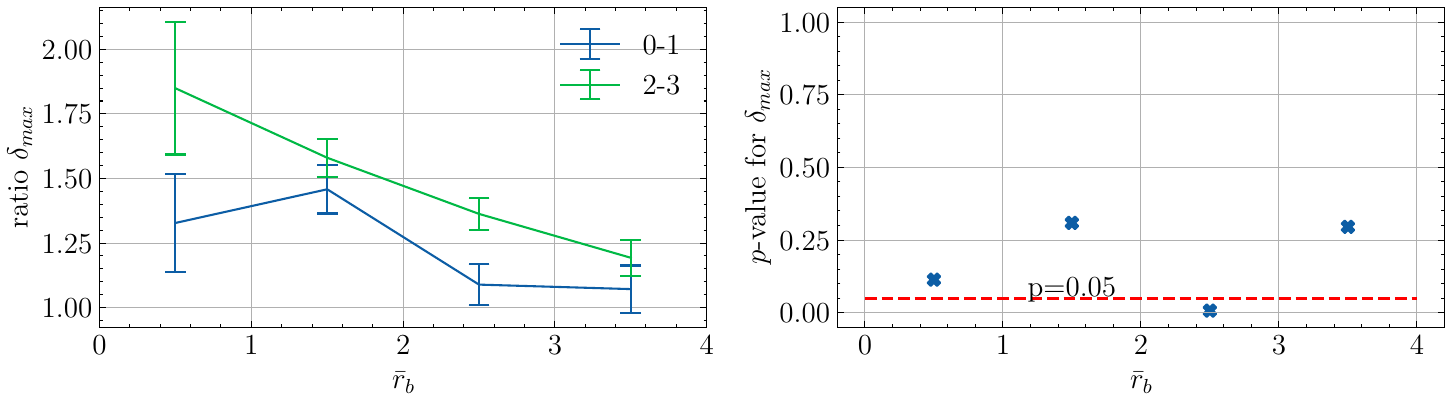}
        \caption{Individual}
        \label{fig:ratios_simultaneous_frechet_deviation_individuals_wrt_impact_parameter_01_23}
    \end{subfigure}
    \begin{subfigure}[t]{\textwidth}
        \centering
        \includegraphics[width=\textwidth]{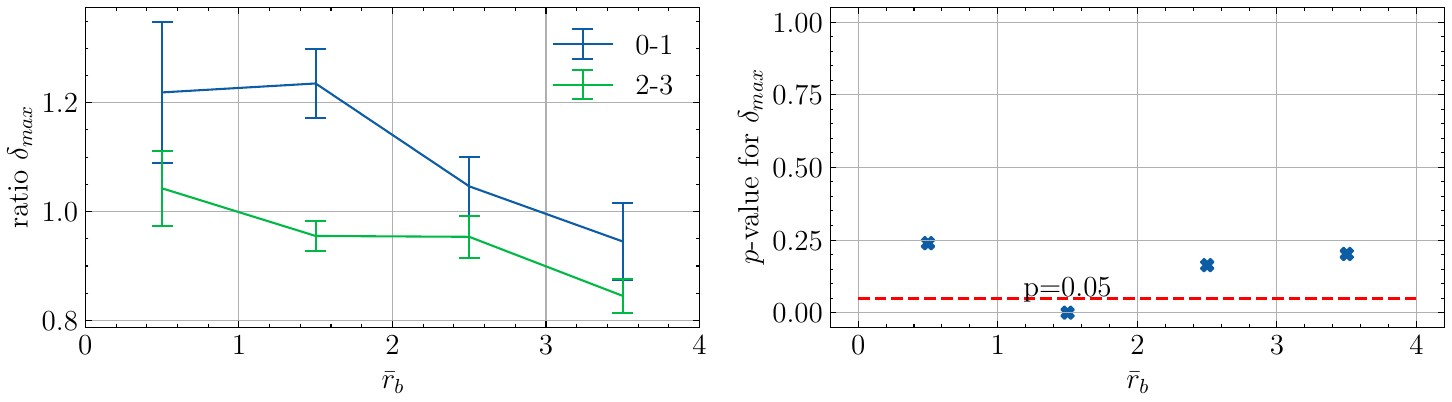}
        \caption{Dyad}
        \label{fig:ratios_simultaneous_frechet_deviation_groups_wrt_impact_parameter_01_23}
    \end{subfigure}
    \caption{Ratio of the value of the lockstep maximum deviation $\delta_{max}$ in encounters to the undisturbed value for binned normalized impact parameter $\bar{r}_b$. The ratio are shown separately for encounters involving dyads with a low (0-1, in blue) and high (2-3, in green) level of interaction. The error bars represent the standard error of the mean. The $p$-values for the difference of means between 0-1 and 2-3 are also shown. The red dashed line represents the threshold $p=0.05$.}
    \label{fig:ratios_simultaneous_frechet_deviation_wrt_impact_parameter_01_23}
\end{figure}

\begin{figure}[htb]
    \centering
    \begin{subfigure}[t]{\textwidth}
        \centering
        \includegraphics[width=\textwidth]{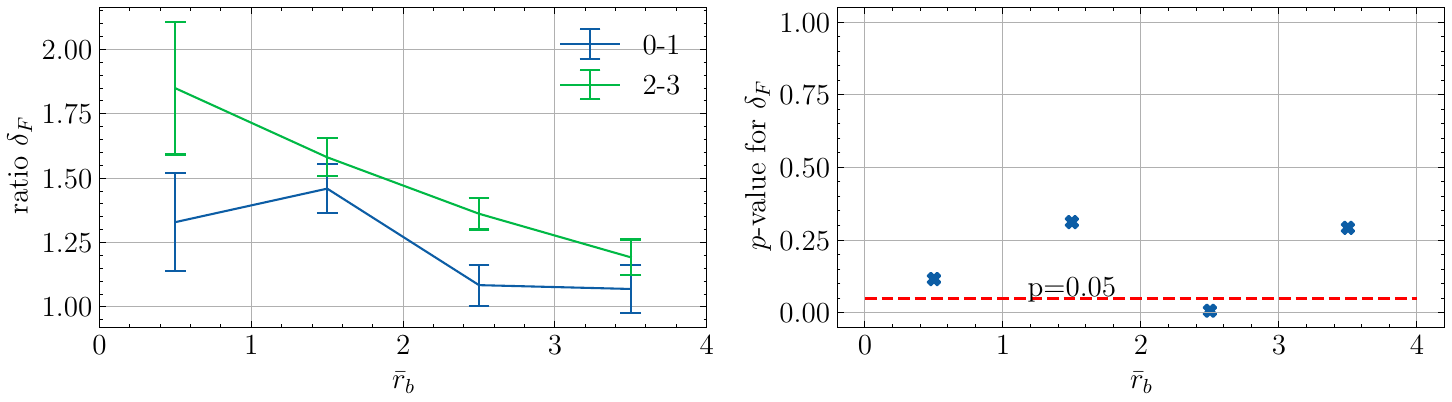}
        \caption{Individual}
        \label{fig:ratios_frechet_deviation_individuals_wrt_impact_parameter_01_23}
    \end{subfigure}
    \begin{subfigure}[t]{\textwidth}
        \centering
        \includegraphics[width=\textwidth]{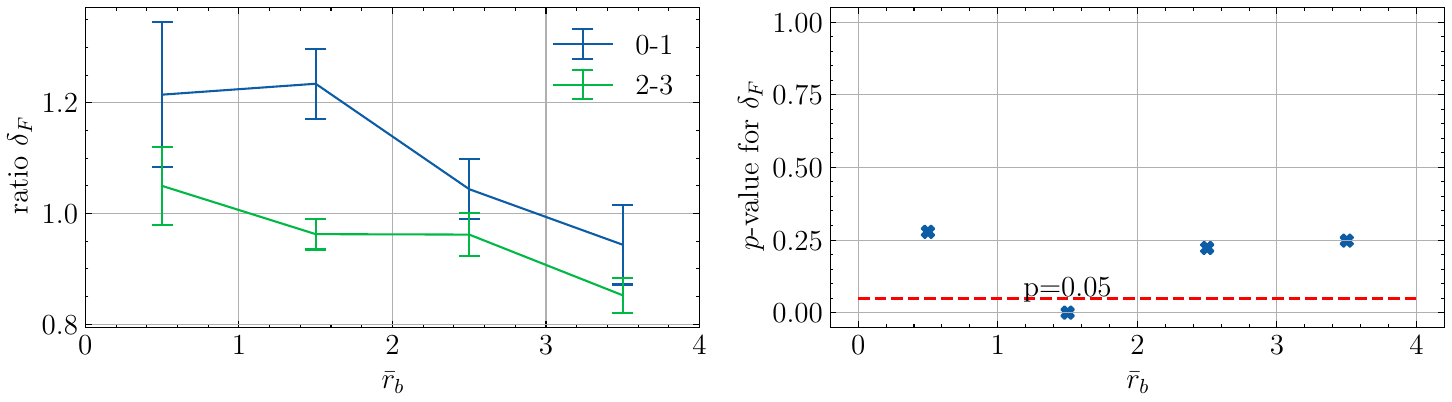}
        \caption{Dyad}
        \label{fig:ratios_frechet_deviation_groups_wrt_impact_parameter_01_23}
    \end{subfigure}
    \caption{Ratio of the value of the Fréchet deviation $\delta_{F}$ in encounters to the undisturbed value for binned normalized impact parameter $\bar{r}_b$. The ratio are shown separately for encounters involving dyads with a low (0-1, in blue) and high (2-3, in green) level of interaction. The error bars represent the standard error of the mean. The $p$-values for the difference of means between 0-1 and 2-3 are also shown. The red dashed line represents the threshold $p=0.05$.}
    \label{fig:ratios_frechet_deviation_wrt_impact_parameter_01_23}
\end{figure}

\begin{figure}[htb]
    \centering
    \begin{subfigure}[t]{\textwidth}
        \centering
        \includegraphics[width=\textwidth]{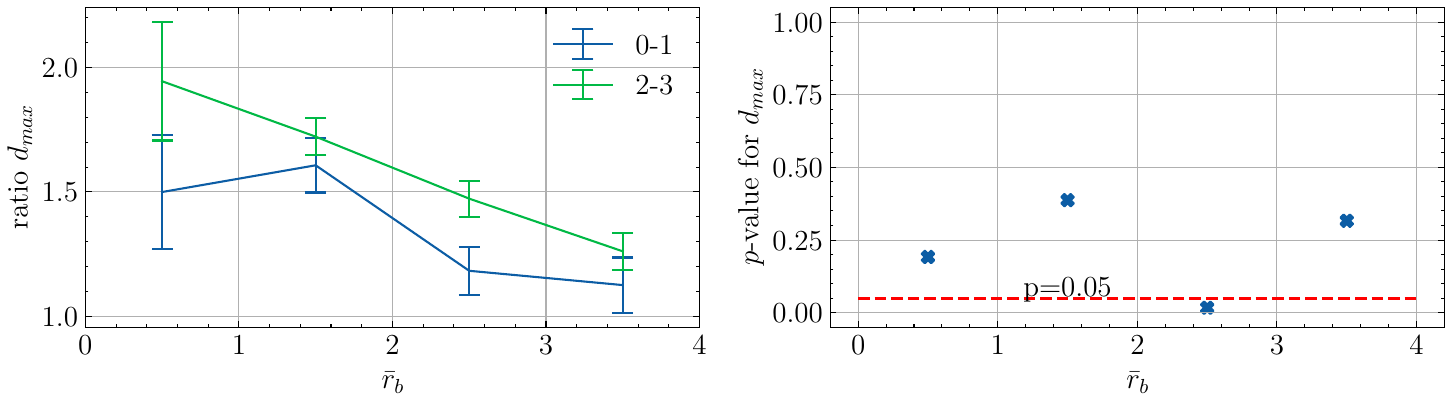}
        \caption{Individual}
        \label{fig:ratios_max_lateral_deviation_individuals_wrt_impact_parameter_01_23}
    \end{subfigure}
    \begin{subfigure}[t]{\textwidth}
        \centering
        \includegraphics[width=\textwidth]{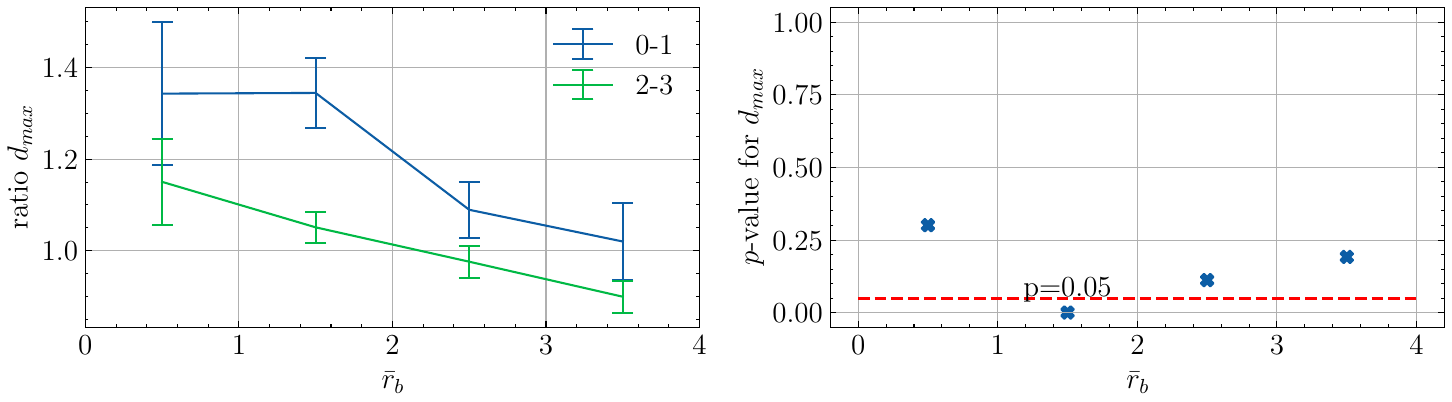}
        \caption{Dyad}
        \label{fig:ratios_max_lateral_deviation_groups_wrt_impact_parameter_01_23}
    \end{subfigure}
    \caption{Ratio of the value of the maximum lateral deviation $d_{max}$ in encounters to the undisturbed value for binned normalized impact parameter $\bar{r}_b$. The ratio are shown separately for encounters involving dyads with a low (0-1, in blue) and high (2-3, in green) level of interaction. The error bars represent the standard error of the mean. The $p$-values for the difference of means between 0-1 and 2-3 are also shown. The red dashed line represents the threshold $p=0.05$.}
    \label{fig:ratios_max_lateral_deviation_wrt_impact_parameter_01_23}
\end{figure}

\begin{figure}[htb]
    \centering
    \begin{subfigure}[t]{\textwidth}
        \centering
        \includegraphics[width=\textwidth]{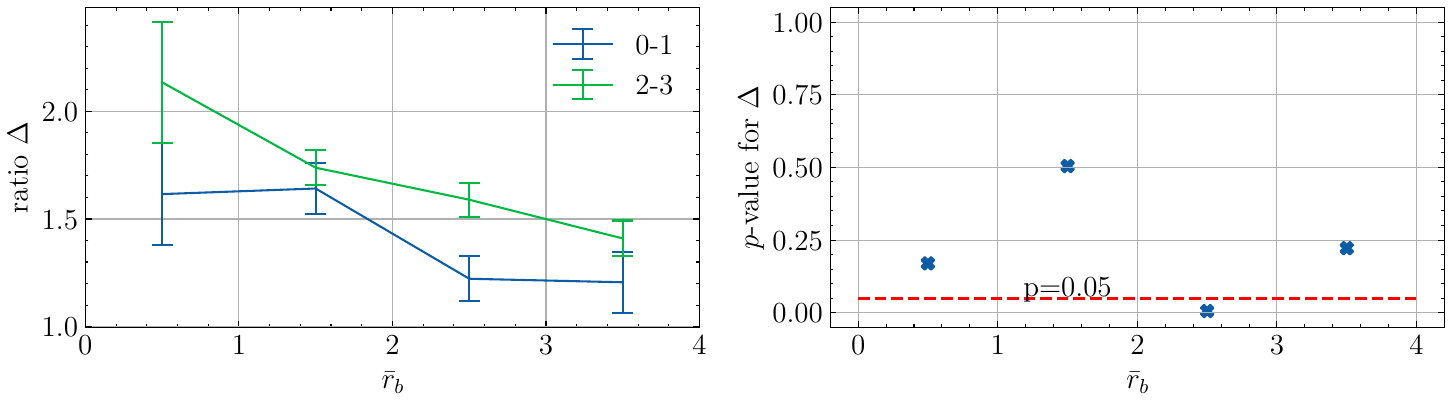}
        \caption{Individual}
        \label{fig:ratios_integral_deviation_individuals_wrt_impact_parameter_01_23}
    \end{subfigure}
    \begin{subfigure}[t]{\textwidth}
        \centering
        \includegraphics[width=\textwidth]{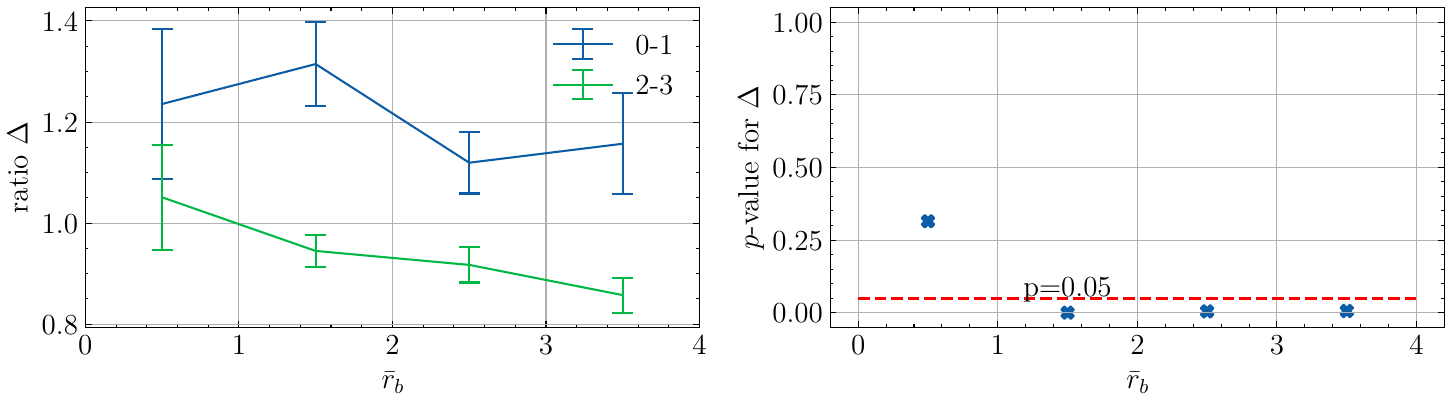}
        \caption{Dyad}
        \label{fig:ratios_integral_deviation_groups_wrt_impact_parameter_01_23}
    \end{subfigure}
    \caption{Ratio of the value of the integral of lateral deviation $\Delta$ in encounters to the undisturbed value for binned normalized impact parameter $\bar{r}_b$. The ratio are shown separately for encounters involving dyads with a low (0-1, in blue) and high (2-3, in green) level of interaction. The error bars represent the standard error of the mean. The $p$-values for the difference of means between 0-1 and 2-3 are also shown. The red dashed line represents the threshold $p=0.05$.}
    \label{fig:ratios_integral_deviation_wrt_impact_parameter_01_23}
\end{figure}

\begin{figure}[htb]
    \centering
    \begin{subfigure}[t]{\textwidth}
        \centering
        \includegraphics[width=\textwidth]{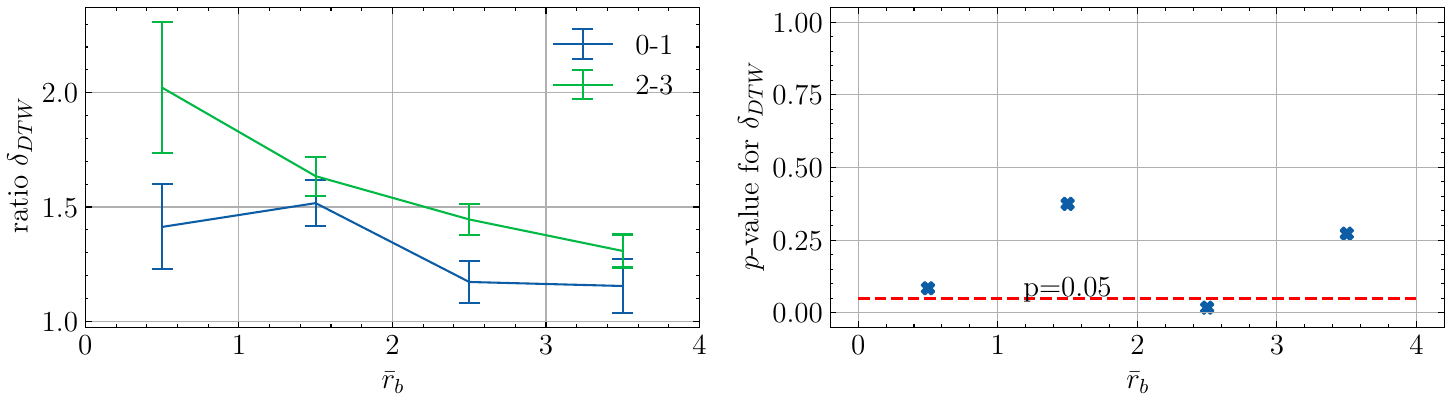}
        \caption{Individual}
        \label{fig:ratios_dynamic_time_warping_deviation_individuals_wrt_impact_parameter_01_23}
    \end{subfigure}
    \begin{subfigure}[t]{\textwidth}
        \centering
        \includegraphics[width=\textwidth]{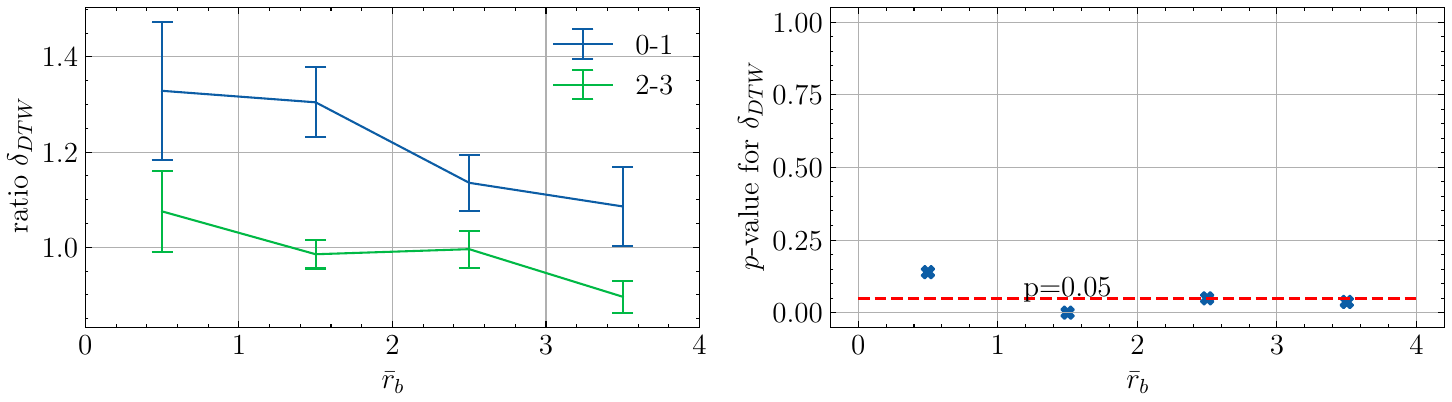}
        \caption{Dyad}
        \label{fig:ratios_dynamic_time_warping_deviation_groups_wrt_impact_parameter_01_23}
    \end{subfigure}
    \caption{Ratio of the value of the dynamic time warping deviation $\delta_{DTW}$ in encounters to the undisturbed value for binned normalized impact parameter $\bar{r}_b$. The ratio are shown separately for encounters involving dyads with a low (0-1, in blue) and high (2-3, in green) level of interaction. The error bars represent the standard error of the mean. The $p$-values for the difference of means between 0-1 and 2-3 are also shown. The red dashed line represents the threshold $p=0.05$.}
    \label{fig:ratios_dynamic_time_warping_deviation_wrt_impact_parameter_01_23}
\end{figure}

\begin{figure}[htb]
    \centering
    \begin{subfigure}[t]{\textwidth}
        \centering
        \includegraphics[width=\textwidth]{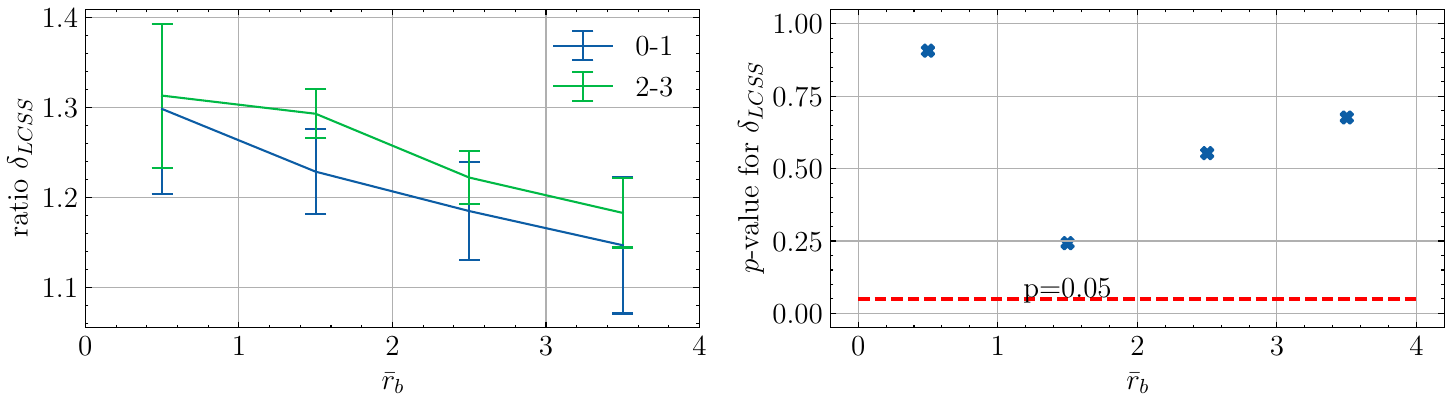}
        \caption{Individual}
        \label{fig:ratios_lcss_deviation_individuals_wrt_impact_parameter_01_23}
    \end{subfigure}
    \begin{subfigure}[t]{\textwidth}
        \centering
        \includegraphics[width=\textwidth]{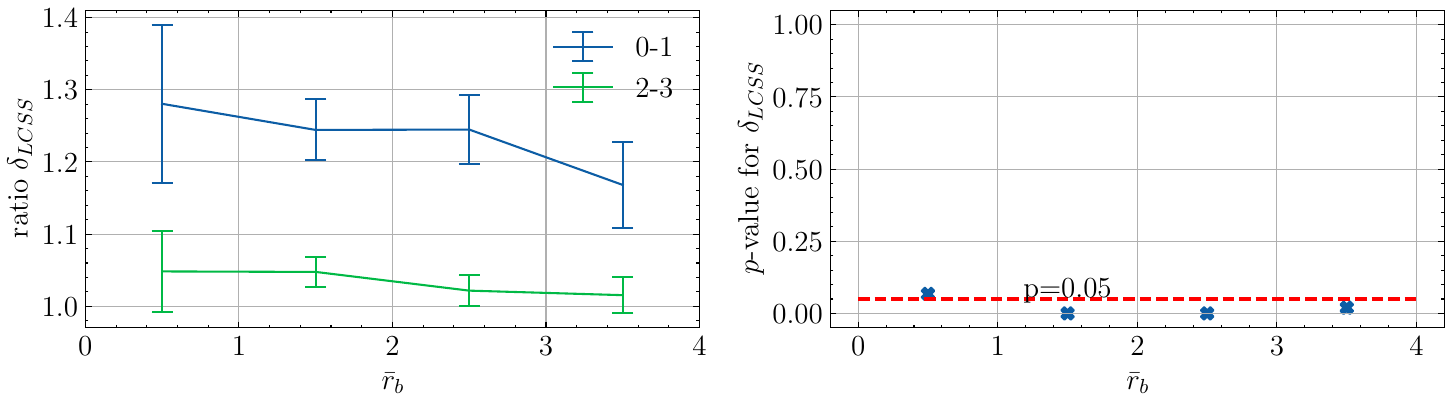}
        \caption{Dyad}
        \label{fig:ratios_lcss_deviation_groups_wrt_impact_parameter_01_23}
    \end{subfigure}
    \caption{Ratio of the value of the longest common subsequence deviation $\delta_{LCSS}$ in encounters to the undisturbed value for binned normalized impact parameter $\bar{r}_b$. The ratio are shown separately for encounters involving dyads with a low (0-1, in blue) and high (2-3, in green) level of interaction. The error bars represent the standard error of the mean. The $p$-values for the difference of means between 0-1 and 2-3 are also shown. The red dashed line represents the threshold $p=0.05$.}
    \label{fig:ratios_lcss_deviation_wrt_impact_parameter_01_23}
\end{figure}

\begin{figure}[htb]
    \centering
    \begin{subfigure}[t]{\textwidth}
        \centering
        \includegraphics[width=\textwidth]{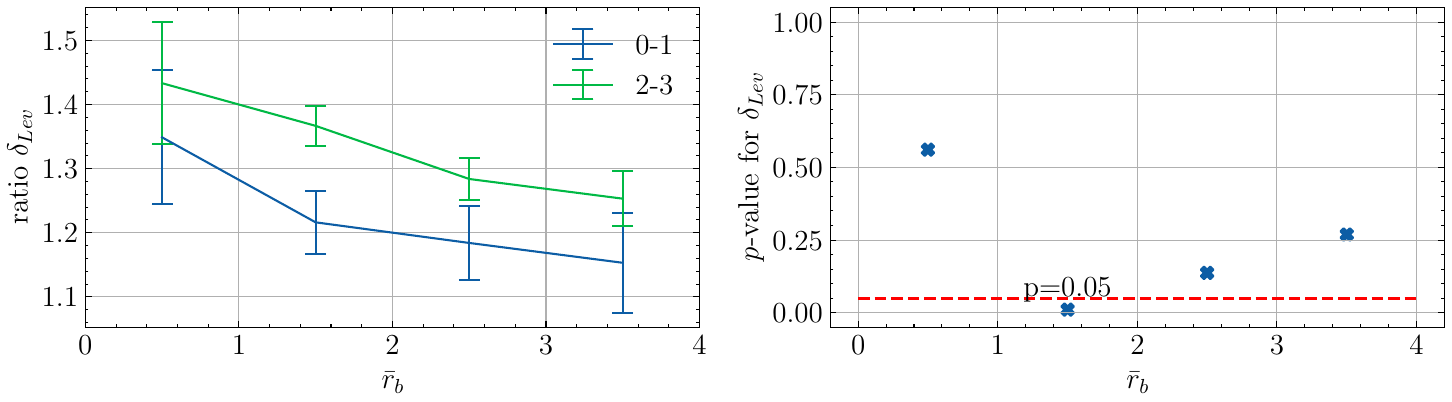}
        \caption{Individual}
        \label{fig:ratios_edit_distance_deviation_individuals_wrt_impact_parameter_01_23}
    \end{subfigure}
    \begin{subfigure}[t]{\textwidth}
        \centering
        \includegraphics[width=\textwidth]{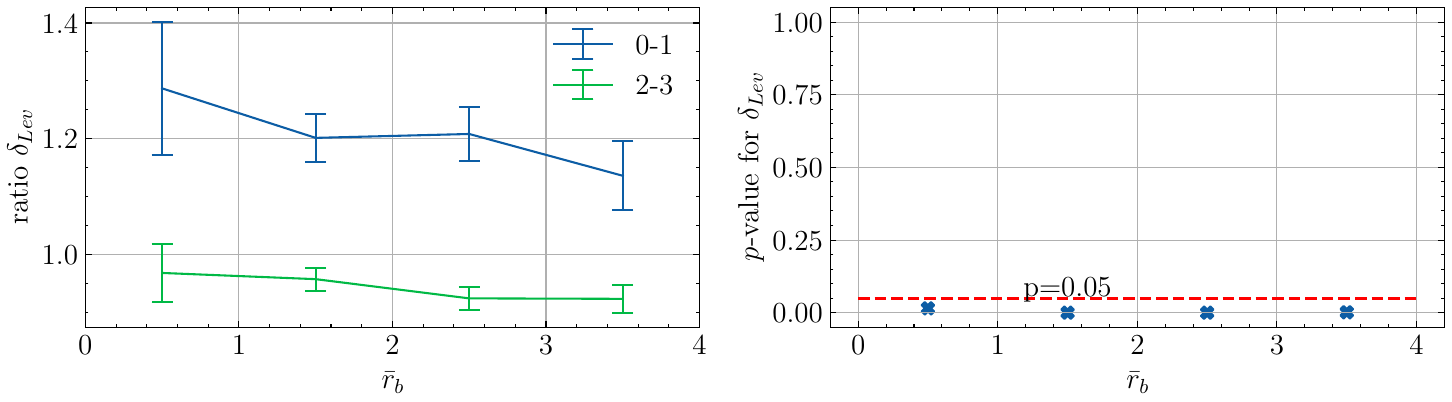}
        \caption{Dyad}
        \label{fig:ratios_edit_distance_deviation_groups_wrt_impact_parameter_01_23}
    \end{subfigure}
    \caption{Ratio of the value of the Levenshtein deviation $\delta_{Lev}$ in encounters to the undisturbed value for binned normalized impact parameter $\bar{r}_b$. The ratio are shown separately for encounters involving dyads with a low (0-1, in blue) and high (2-3, in green) level of interaction. The error bars represent the standard error of the mean. The $p$-values for the difference of means between 0-1 and 2-3 are also shown. The red dashed line represents the threshold $p=0.05$.}
    \label{fig:ratios_edit_distance_deviation_wrt_impact_parameter_01_23}
\end{figure}

\begin{figure}[htb]
    \centering
    \begin{subfigure}[t]{\textwidth}
        \centering
        \includegraphics[width=\textwidth]{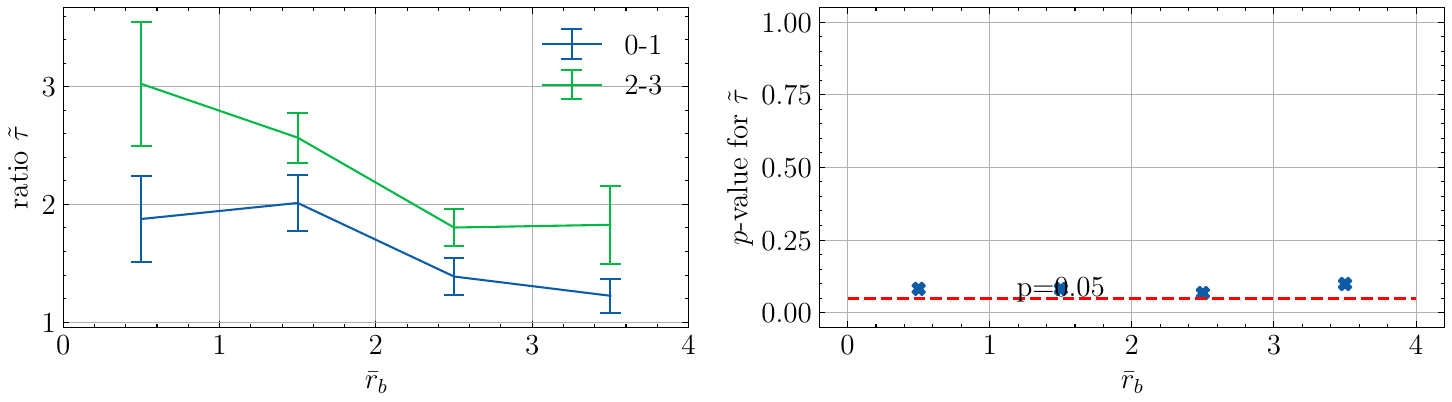}
        \caption{Individual}
        \label{fig:ratios_straightness_index_individuals_wrt_impact_parameter_01_23}
    \end{subfigure}
    \begin{subfigure}[t]{\textwidth}
        \centering
        \includegraphics[width=\textwidth]{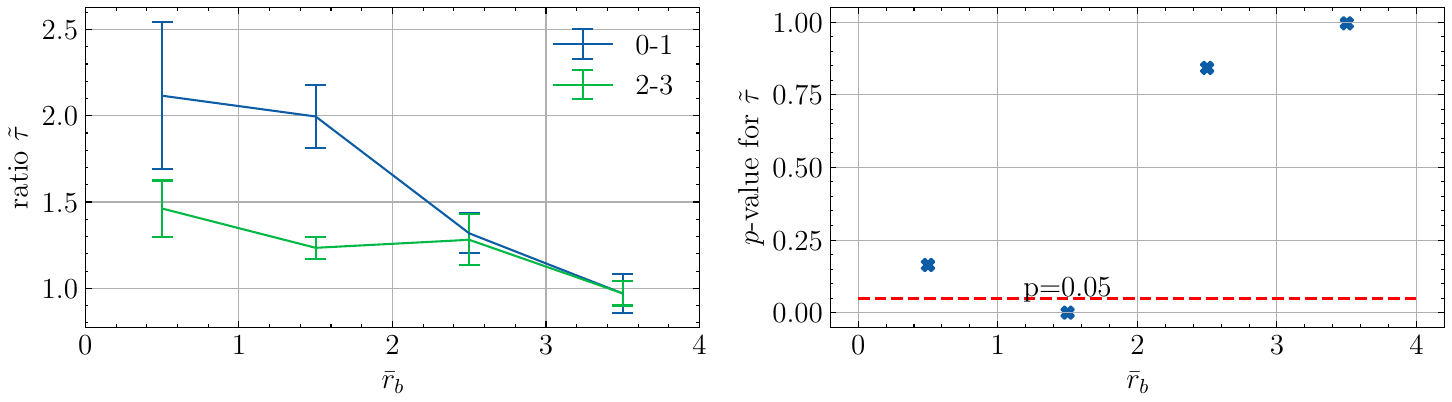}
        \caption{Dyad}
        \label{fig:ratios_straightness_index_groups_wrt_impact_parameter_01_23}
    \end{subfigure}
    \caption{Ratio of the value of the deviation index $\tilde{\tau}$ in encounters to the undisturbed value for binned normalized impact parameter $\bar{r}_b$. The ratio are shown separately for encounters involving dyads with a low (0-1, in blue) and high (2-3, in green) level of interaction. The error bars represent the standard error of the mean. The $p$-values for the difference of means between 0-1 and 2-3 are also shown. The red dashed line represents the threshold $p=0.05$.}
    \label{fig:ratios_straightness_index_wrt_impact_parameter_01_23}
\end{figure}

\begin{figure}[htb]
    \centering
    \begin{subfigure}[t]{\textwidth}
        \centering
        \includegraphics[width=\textwidth]{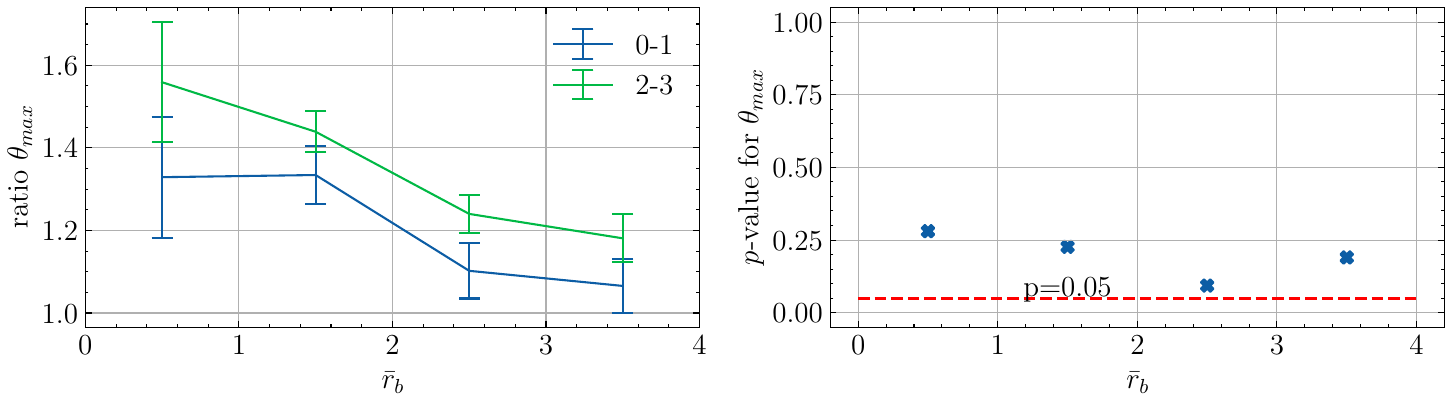}
        \caption{Individual}
        \label{fig:ratios_max_cumulative_turning_angle_individuals_wrt_impact_parameter_01_23}
    \end{subfigure}
    \begin{subfigure}[t]{\textwidth}
        \centering
        \includegraphics[width=\textwidth]{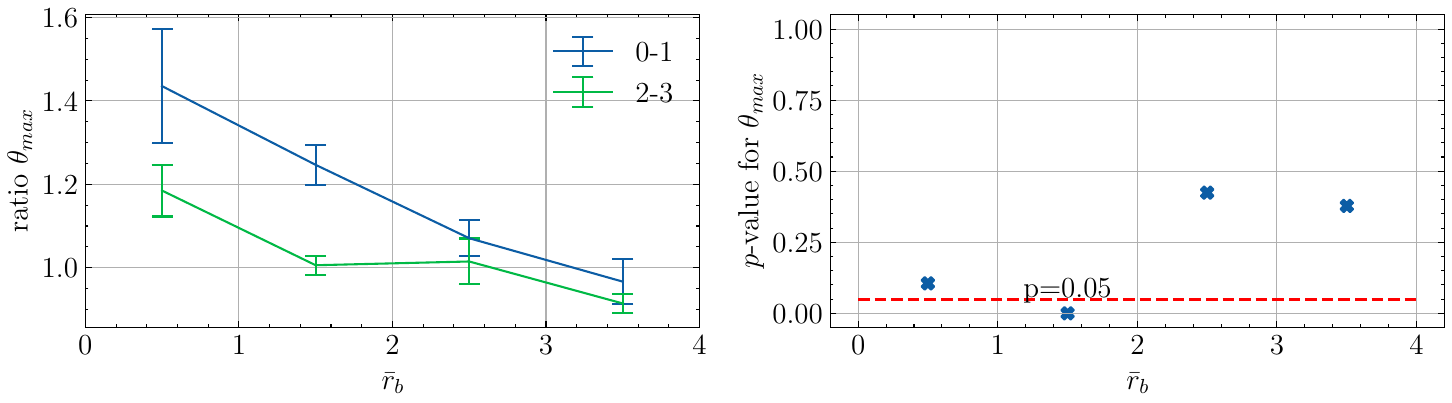}
        \caption{Dyad}
        \label{fig:ratios_max_cumulative_turning_angle_groups_wrt_impact_parameter_01_23}
    \end{subfigure}
    \caption{Ratio of the value of the maximum cumulative turning angle $\theta_{max}$ in encounters to the undisturbed value for binned normalized impact parameter $\bar{r}_b$. The ratio are shown separately for encounters involving dyads with a low (0-1, in blue) and high (2-3, in green) level of interaction. The error bars represent the standard error of the mean. The $p$-values for the difference of means between 0-1 and 2-3 are also shown. The red dashed line represents the threshold $p=0.05$.}
    \label{fig:ratios_max_cumulative_turning_angle_wrt_impact_parameter_01_23}
\end{figure}

\begin{figure}[htb]
    \centering
    \begin{subfigure}[t]{\textwidth}
        \centering
        \includegraphics[width=\textwidth]{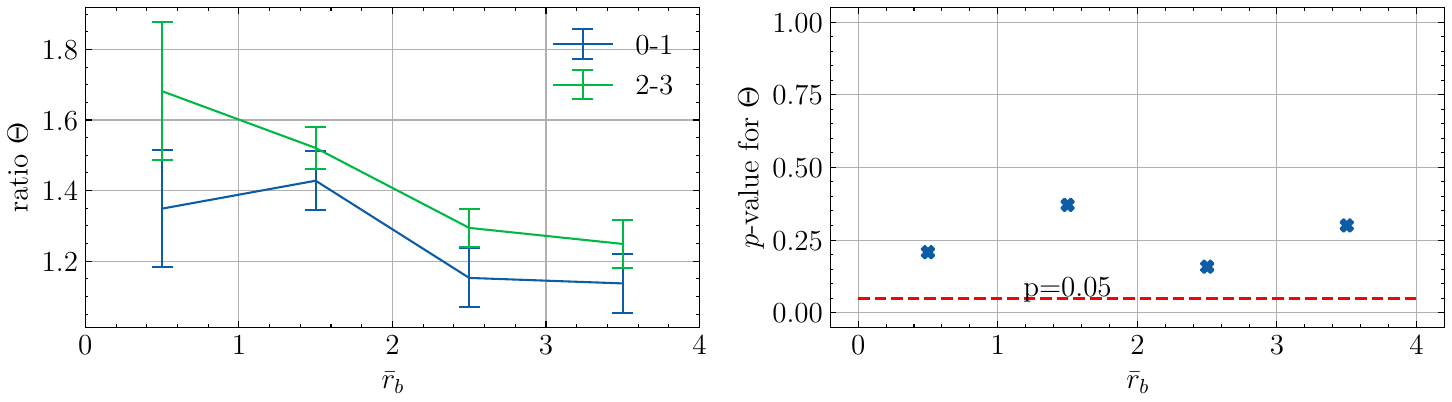}
        \caption{Individual}
        \label{fig:ratios_integral_cumulative_turning_angle_individuals_wrt_impact_parameter_01_23}
    \end{subfigure}
    \begin{subfigure}[t]{\textwidth}
        \centering
        \includegraphics[width=\textwidth]{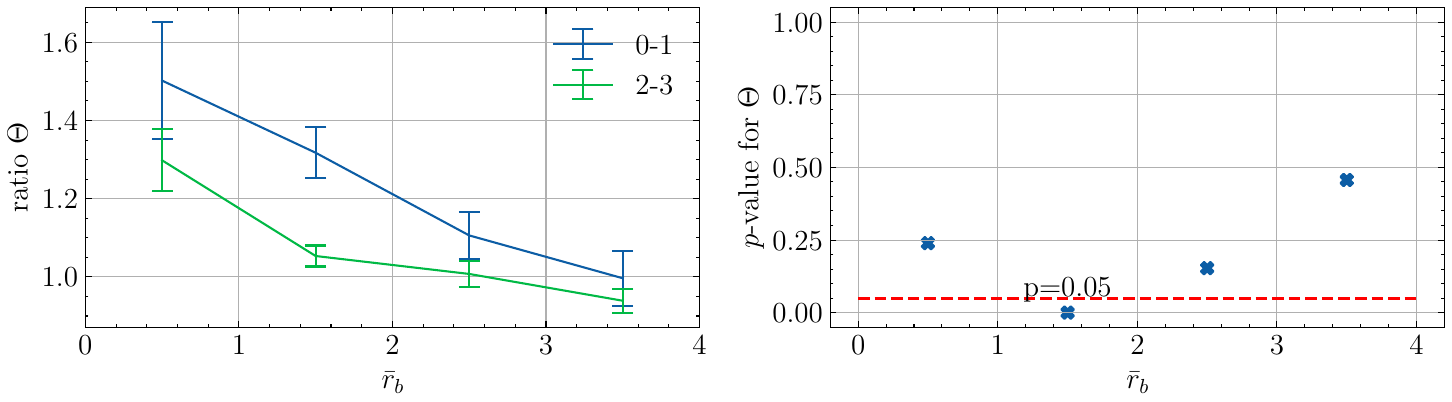}
        \caption{Dyad}
        \label{fig:ratios_integral_cumulative_turning_angle_groups_wrt_impact_parameter_01_23}
    \end{subfigure}
    \caption{Ratio of the value of the integral cumulative turning angle $\Theta$ in encounters to the undisturbed value for binned normalized impact parameter $\bar{r}_b$. The ratio are shown separately for encounters involving dyads with a low (0-1, in blue) and high (2-3, in green) level of interaction. The error bars represent the standard error of the mean. The $p$-values for the difference of means between 0-1 and 2-3 are also shown. The red dashed line represents the threshold $p=0.05$.}
    \label{fig:ratios_integral_cumulative_turning_angle_wrt_impact_parameter_01_23}
\end{figure}

\begin{figure}[htb]
    \centering
    \begin{subfigure}[t]{\textwidth}
        \centering
        \includegraphics[width=\textwidth]{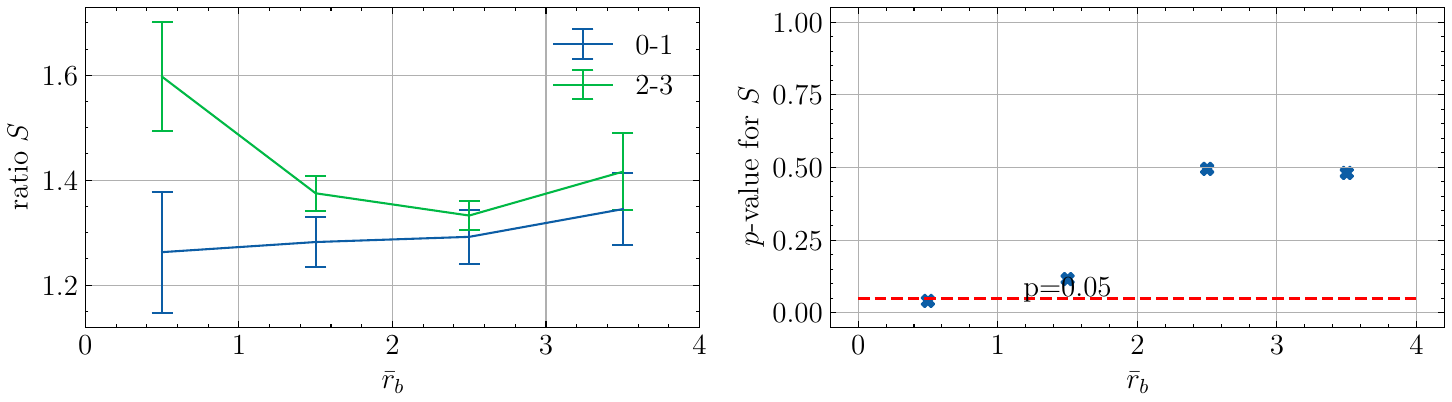}
        \caption{Individual}
        \label{fig:ratios_sinuosity_individuals_wrt_impact_parameter_01_23}
    \end{subfigure}
    \begin{subfigure}[t]{\textwidth}
        \centering
        \includegraphics[width=\textwidth]{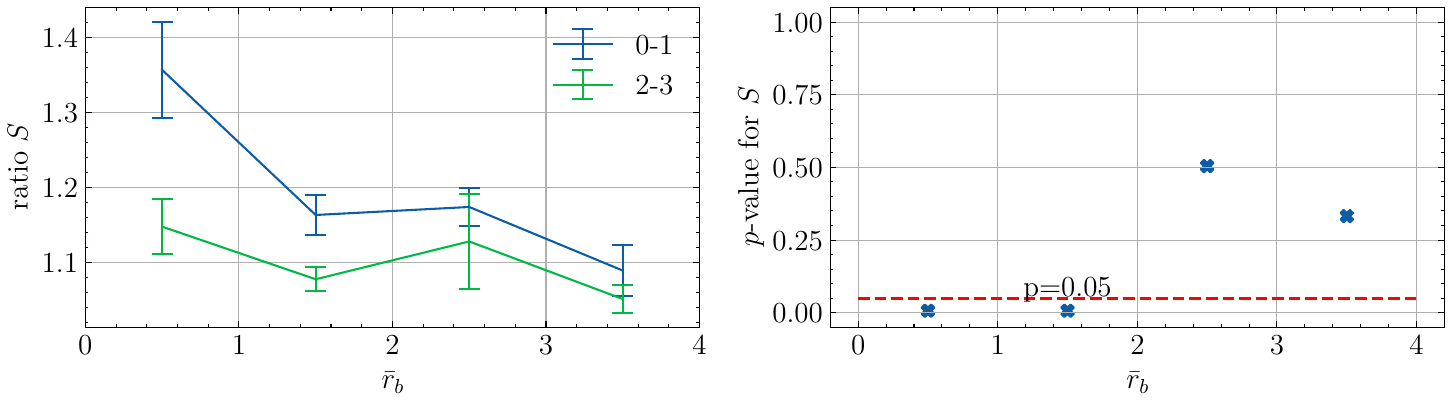}
        \caption{Dyad}
        \label{fig:ratios_sinuosity_groups_wrt_impact_parameter_01_23}
    \end{subfigure}
    \caption{Ratio of the value of the sinuosity $S$ in encounters to the undisturbed value for binned normalized impact parameter $\bar{r}_b$. The ratio are shown separately for encounters involving dyads with a low (0-1, in blue) and high (2-3, in green) level of interaction. The error bars represent the standard error of the mean. The $p$-values for the difference of means between 0-1 and 2-3 are also shown. The red dashed line represents the threshold $p=0.05$.}
    \label{fig:ratios_sinuosity_wrt_impact_parameter_01_23}
\end{figure}

\begin{figure}[htb]
    \centering
    \begin{subfigure}[t]{\textwidth}
        \centering
        \includegraphics[width=\textwidth]{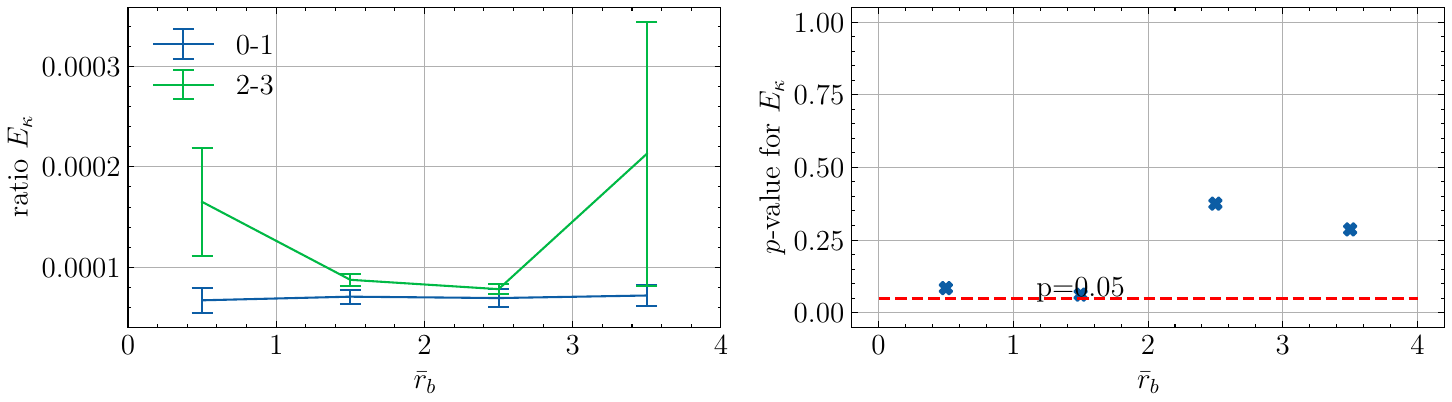}
        \caption{Individual}
        \label{fig:ratios_energy_curvature_individuals_wrt_impact_parameter_01_23}
    \end{subfigure}
    \begin{subfigure}[t]{\textwidth}
        \centering
        \includegraphics[width=\textwidth]{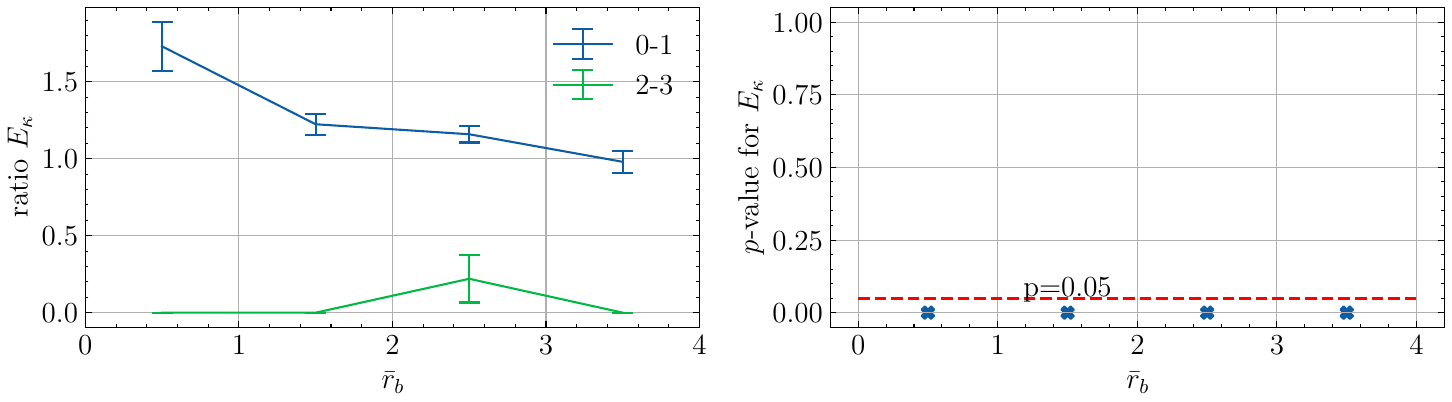}
        \caption{Dyad}
        \label{fig:ratios_energy_curvature_groups_wrt_impact_parameter_01_23}
    \end{subfigure}
    \caption{Ratio of the value of the energy curvature $E_{\kappa}$ in encounters to the undisturbed value for binned normalized impact parameter $\bar{r}_b$. The ratio are shown separately for encounters involving dyads with a low (0-1, in blue) and high (2-3, in green) level of interaction. The error bars represent the standard error of the mean. The $p$-values for the difference of means between 0-1 and 2-3 are also shown. The red dashed line represents the threshold $p=0.05$.}
    \label{fig:ratios_energy_curvature_wrt_impact_parameter_01_23}
\end{figure}

\begin{figure}[htb]
    \centering
    \begin{subfigure}[t]{\textwidth}
        \centering
        \includegraphics[width=\textwidth]{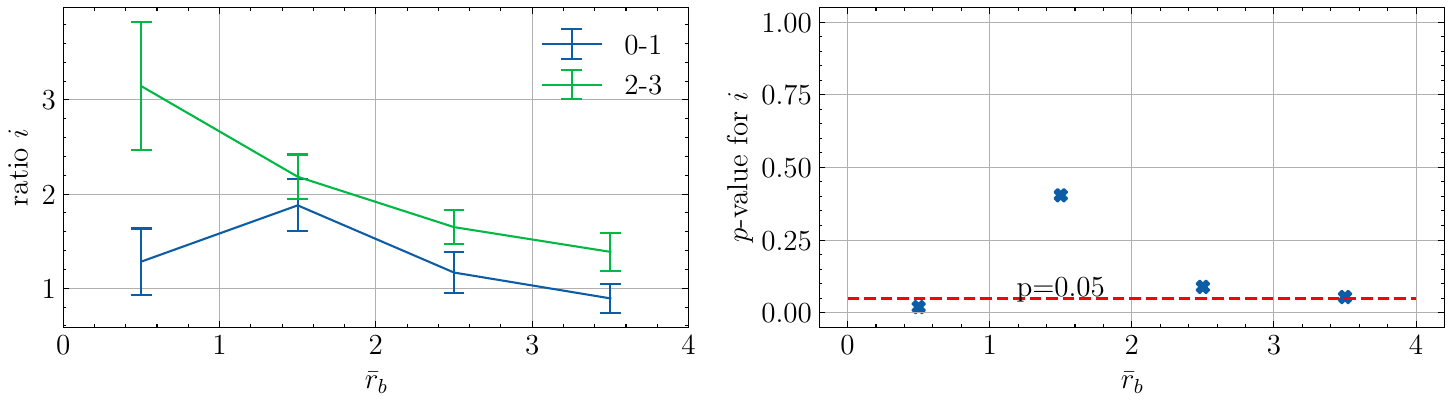}
        \caption{Individual}
        \label{fig:ratios_turn_intensity_individuals_wrt_impact_parameter_01_23}
    \end{subfigure}
    \begin{subfigure}[t]{\textwidth}
        \centering
        \includegraphics[width=\textwidth]{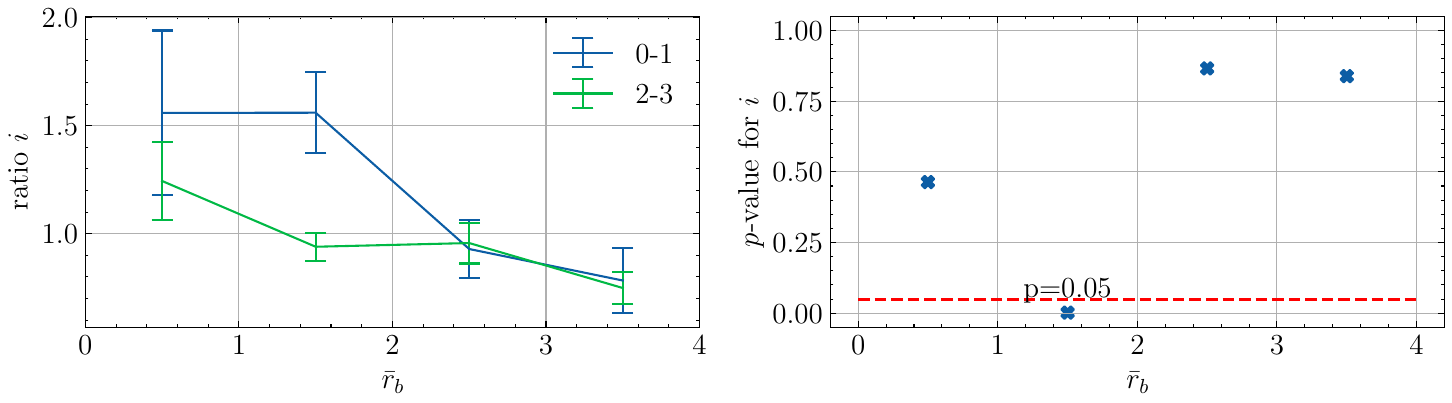}
        \caption{Dyad}
        \label{fig:ratios_turn_intensity_groups_wrt_impact_parameter_01_23}
    \end{subfigure}
    \caption{Ratio of the value of the turn intensity $i$ in encounters to the undisturbed value for binned normalized impact parameter $\bar{r}_b$. The ratio are shown separately for encounters involving dyads with a low (0-1, in blue) and high (2-3, in green) level of interaction. The error bars represent the standard error of the mean. The $p$-values for the difference of means between 0-1 and 2-3 are also shown. The red dashed line represents the threshold $p=0.05$.}
    \label{fig:ratios_turn_intensity_wrt_impact_parameter_01_23}
\end{figure}

% \bibliographystyle{abbrv}
% \bibliography{citations_v2z}

\printbibliography

\end{document}